\documentstyle[11pt,epsf]{article}

\advance \topmargin by -\headheight
\advance \topmargin by -\headsep

\evensidemargin \oddsidemargin
\def\np{Nucl. Phys.}
\def\pl{Phys. Lett.}
\def\prl{Phys. Rev. Lett.}
\def\pr{Phys. Rev.}

\def\mpl{Mod. Phys. Lett.}

\def\atmp{Adv. Theor. Math. Phys. }
\def\jhep{J. High Energy Phys.}
\def\ptp{Prog. Theor. Phys.}
\def\jgp{J. Geom. Phys.}
\def\atmp{Adv. Theor. Math. Phys.}

\begin{document}

\begin{titlepage}

\vskip4.5truecm
\rightline{PUTP-2240}
\smallskip
\begin{center} \Large \bf Holographic flavor in theories with eight supercharges\footnote{Electronic version of an article published as \textit{Holographic flavor in theories with eight supercharges}, IJMPA Vol. 22, pages 4717-4796 (2007). © [copyright World Scientific Publishing Company]}

\end{center}

\vskip 0.3truein
\begin{center}
Diego Rodr\'\i guez-G\'omez
%${}^{\,\dagger}$

\vspace{0.3in}

%${}^{\,\dagger}$
Joseph Henry Laboratories, Princeton University\\
Princeton NJ 08544, U.S.A.\\
\verb+drodrigu@princeton.edu+

\end{center}
\vskip.5truein

\begin{center}
\bf ABSTRACT
\end{center}
We review the holographic duals of gauge theories with eight supercharges obtained by adding very few flavors to pure supersymmetric Yang-Mills with sixteen supercharges. Assuming a brane-probe limit, the gravity duals are engineered in terms of probe branes (the so-called flavor brane) in the background of the color branes. Both types of branes intersect on a given subspace in which the matter is confined. The gauge theory dual is thus the corresponding flavoring of the gauge theory with sixteen supercharges. Those theories have in general a non-trivial phase structure; which is also captured in a beautiful way by the gravity dual. Along the lines of the gauge/gravity duality, we review also some of the  results on the meson spectrum in the different phases of the theories.

\end{titlepage}
\setcounter{footnote}{0}

\tableofcontents

\section{Introduction}

Gauge theories are the cornerstone of our current understanding of
Nature. The Standard Model is, with no doubt, the most successful
model of Nature we have so far constructed. It incorporates, under the
unified framework of Quantum Field Theory, the electroweak and the
strong interactions, being both gauge theories. However, there is yet
another force of Nature, gravity, which is left apart in this
scheme. String Theory is the most promising candidate for a unified
theory, in which gauge and gravity are two sides of the same
coin. Along this lines, the gauge/gravity correspondence \cite{jm}
(see \cite{MAGOO} for a very comprehensive review) has been a
breakthrough in our understanding of both gravity (and string theory)
and gauge field theories. This correspondence provides a closed string
description, based on classical supergravity, of the dynamics of gauge
theories at large 't Hooft coupling. It deeply relies on the dual
description of gravitational objects either as backgrounds on which
strings propagate; and as objects on its own right in the theory. The
most celebrated example considers the very special case of the $D3$
branes, which can be seen either as a certain supergravity background,
or as an object which carries a worldvolume gauge theory as the lowest
lying states. In a well-defined low energy limit, namely the {\it
decoupling limit}, changing from weak to strong coupling takes us from
one description to the other. This duality is indeed a holographic
duality \cite{Susskind,Witten,GKP}, since the weak coupling is
described in terms of a field theory living in four-dimensional
Minkowski, whereas the strong coupling is captured in terms of IIB
string theory propagating in ten-dimensional $AdS_5\times S^5$ (which
is the near horizon region of the $D3$ brane background, on which the
decoupling limit focuses). In a sense, it captures the original spirit
of string theory as an effective description of the strong coupling
regime of a gauge theory.

Many avenues of the gauge/gravity duality have been explored by
now. The dictionary between both sides has been established (see
\cite{MAGOO} and references therein), and many more examples
have been found (see also
\cite{MAGOO,Labc1,Labc2,Labc3,Labc4}). Technically, the
duality works in its most stelar way for $AdS$ backgrounds, whose
field theory duals are in terms of supersymmetric conformally
invariant field theories. This is inherited from the structure of the
$AdS$ space, which endows the holographic dual theory with a conformal
invariance. Restricting for a while to $3+1$ dimensions, in principle
one can find dualities for spaces of the form $AdS_5\times X$, as long
as $X$ is a five-dimensional Sasaki--Einstein manifold. This has been
done in the literature, where both the gravity and field theory sides
have been explicitly worked out, finding an amusing agreement (see
\cite{Labc1,Labc2,Labc3,Labc4} and references to those
papers). These backgrounds can be seen as the near horizon limit for
$D3$ branes at the tip of the Calabi--Yau cone whose base is the $X$
space.\footnote{Given that we are considering a CY, these theories will
preserve at least ${\cal{N}}=1$ in four dimensions.} This cone is, in
general, singular (although the near horizon removes this
singularity), and one can desingularize it by moving in the Kahler
moduli space resolving the singularities \cite{KW,KM}. This has also
been studied, leading to a deeper understanding of the interplay
geometry/gauge theory. However, understanding the breaking of
conformal invariance in this context remains as a major challenge,
since the ultimate challenge is to understand in a holographic way a
theory such as QCD. A major step was taken in \cite{KS},
where, by introducing fractional branes which in turn require
deforming the conifold (which amounts to moving along the complex
structure moduli space of the internal Calabi--Yau), the conformal
invariance was broken and the dual of a confining gauge theory was
found.

Going back to the original spirit of the gauge/gravity duality, one
could try to play the same game not just for the $D3$ brane, but for a
generic $Dp$ \cite{IMSY,BST}. In the general case the situation is
very different, since, once one finds the suitable holographic
coordinates, the gravity dual lives in a background which is not
$AdS$, but only conformally $AdS$. Since in addition in these
backgrounds the dilaton is not constant, the conformal invariance of
the dual theory is broken; which makes the duality somehow more
subtle, and valid just in a certain energy and parameter range. Since
the dilaton will be a function of the holographic coordinate; which in
turn has the interpretation of the energy scale in the dual theory,
generically we will have that, for some energy range, the gravity dual
opens up the M-theory circle. In a suitable parameter range, this
corresponds in the field theory to a strong coupling regime, which we
can surpass by uplifting the system to 11 dimensions. However, taking
into account all these subtleties, one can still formulate a
gauge/gravity duality for the generic case of $Dp$ branes.

In the dual field theories discussed, all fields are in the adjoint
representation. Clearly, a major issue is to introduce matter (quarks)
in the fundamental representation, and this will be precisely our main
interest. Our ultimate goal in this paper will be to understand the
dynamics of a certain class of gauge theories with flavors which admit
a gravity description. Those theories will arise as the flavoring of a
``bulk" Yang--Mills with 16 supercharges in $p+1$ dimensions. In order
to find the bulk theories, we will restrict from now on to the case in
which those $Dp$ branes live in ten-dimensional Minkowski
space. Therefore, the field theory description will be in terms of the
worldvolume theory on the branes; which is precisely the
aforementioned bulk theory. To be more precise, we will be interested
in adding fundamental matter to the gauge theories obtained from
dimensional reduction of the maximally supersymmetric Yang--Mills
theory in ten dimensions down to $p+1$. Indeed, we will consider
adding supersymmetrically $N_f$ hypermultiplets to those theories in
all the possible ways (namely confined to live inside a defect of the
various dimensionalities selected by supersymmetry).

Adding fundamental matter is equivalent to introducing open string
degrees of freedom to the supergravity side of the correspondence, and
can be achieved by adding $D$-branes to the supergravity background. A
first step towards the addition of an open string sector was taken in
\cite{Mateo,Mateo2,Mateo3,KR,KR2,KKW,KKW2}, where it was
suggested that one can have dynamical open string degrees of freedom
by introducing $N_f$ intersecting $Dq$ branes to the original $Dp$
branes. In the limit in which the number of $Dq$ branes is much
smaller than the number of $Dp$ branes, we can treat the system
effectively as $N_f$ probe branes in the background generated by the
$N_c$ $Dp$ branes. Thus, once we take the decoupling limit, this
background will reduce to the corresponding near horizon geometry of
the original $Dp$ branes, where the $Dq$ live embedded as probes.
Generically, the two types of branes overlap partially, which implies
that the additional $Dq$ branes create a defect on the worldvolume
theory of the $Dp$ branes. In the dual gauge theory description, the
extra branes give rise to additional matter, confined to live inside
the defect, which comes from the $Dp-Dq$ strings. When $q>p$, the
decoupling limit forces the $SU(N_f)$ gauge symmetry on the $Dq$ brane
to decouple. It then appears as a global flavor symmetry for the extra
matter, which is in the fundamental representation of the flavor
group; furnishing precisely the type of field theories which we wanted
to study. Although we will restrict to the aforementioned theories
(namely $p+1$ Yang--Mills with 16 supercharges containing a few
flavors confined in a half-BPS defect), this approach to the flavor
problem can be used in a generic way. In this context, the
fluctuations of the flavor branes should correspond to the mesons in
the dual gauge theory. The study of these mesons was started in
\cite{KMMW} for the $D7$ brane in the $AdS_5\times S^5$
geometry, and it was further extended to other flavor branes in
several backgrounds
\cite{Sonnen,Johana,Johana2,Johana3,KMMW-two,Carlos,Carlos2,Carlos3,Ouyang,Ouyang2,WH,flavoring,Hong,Evans,Evans2,Evans3,Ghoroku,Ghoroku1,Ghoroku2,Ghoroku3,melting,melting2,melting3,melting4,conifold,Kuper,Sakai,Sakai1,APR,AR,MT,Apreda:2006bu}
(for a review see \cite{R}). From the field theory point of
view, this approach is some sort of quenched approximation, since the
backreaction of the flavors on the color is not taken into account. It
is just since very recently that a full approach to the problem has
been considered with very interesting results (see
\cite{unquenched,unquenched2,unquenched3,unquenched4, rf60, rf61,unquenched5}).

Our purpose is to present a compilation of the accumulated results
which describe the gauge/gravity duality for the theories of
interest. We first start introducing the gauge/gravity duality which
will be the arena of our discussion. Inspired by the $AdS/CFT$
correspondence, whose biggest exponent is the $AdS_5/{\cal{N}}=4$
duality, we discuss a bit the duality for the rest of $Dp$ branes. An
exhaustive description of each case is, by far, out of the scope of
this paper, and we refer to the literature (in particular see
\cite{MAGOO} and references therein) for deeper
discussions. After introducing the gauge/gravity duality we turn to
the inclusion of fundamental matter along the lines of
\cite{Mateo,Mateo2,Mateo3,KR,KR2,KKW,KKW2}. The bottom line
is that, in the brane-probe approximation, the flavor is included as
probes in the color branes background, where we have to take the
gauge/gravity duality and go to the ``near horizon" region of the
space as dual of the gauge theory. However, the addition of the flavor
branes is somehow subtle. Since here we are mainly interested in
supersymmetric field theories, our first task will be to find the
supersymmetric embeddings for the probes; which will give rise to
three series of intersections characterized by the codimensionality of
the defect in the color branes: codimension 0, codimension 1 and
codimension 2. We will see in the next section that, as long as we do
not consider worldvolume gauge fields, the flavor branes do not couple
the background RR potential. Actually, in Sec.~2, we review all the
intersections in the Coulomb branch from the gravity side in a generic
way, paying a special attention to the $D3$ brane background for later
purposes. However, at this point, we preferred not to introduce yet
the full field theory analysis, and postpone it for later in order to
have a more unified picture. The fact of not considering worldvolme
gauge fields on the probe branes will have the consequence that this
brane embeddings correspond to the Coulomb branch of the theory; whose
properties, such as the meson spectrum, will be studied in Sec.~3 by
considering the fluctuations of the flavor branes. This was first
studied in \cite{KMMW} for the $D3-D7$ case and subsequently
extended to the other brane intersections in \cite{AR} and
\cite{MT} (for a review see \cite{R}).

We can have more involved situations in the field theory, such as
Higgs branches. We turn to them in Sec.~4. Since the $D3$ brane
background has special properties, such as the conformality of the
bulk ${\cal{N}}=4$ Yang--Mills theory and the fact that it is
$(3+1)$-dimensional, we will study the three intersections whose
background is that of the $D3$ brane in more detail using them as
examples for the rest of the intersections. Indeed, we will take
advantage of the gained perspective when studying the Coulomb branch
of the theory to discuss in detail, from the field theory point of
view, the dynamics of the systems. We will see that the field theory
results have a beautiful gravity counterpart. We start with the
codimension 0 defect. For the particular $D3-D7$ case, the Higgs phase
was first studied in \cite{EGG} (see also
\cite{Guralnik:2004ve} and \cite{ARR2}). It was proposed
in \cite{EGG} and \cite{Guralnik:2004ve} that, from the
point of view of the $D7$-brane, one can realize a (mixed
Coulomb--)Higgs phase of the $D3-D7$ system by switching on an
instanton configuration of the worldvolume gauge field of the
$D7$-brane. This instanton has the effect of separating some of the
color branes and dissolving them in the flavor ones since it couples
to the flavor branes the background potential. Heuristically this
explains why this corresponds to a Higgs branch. Since we are
separating some of the color branes, the gauge group is broken; and
the fact of dissolving (recombining) them with the flavor ones has the
effect of giving a nontrivial VEV for the quark fields, thus entering
into the Higgs branch. This picture will be universal for both the
codimension 0 and codimension 1 defects; and is shared by other
approaches to the same gauge theories (such as brane webs. For a
review see \cite{kutasov}. It was also suggested in
\cite{Aharony}). We will see that one can give a very explicit
realization of these ideas from the perspective of the ``separated
branes,'' which can be thought as moving in the background of the
rest. Because of the dielectric effect \cite{M}, they will polarize
into the effective flavor brane, giving a precise and beautiful
relation between the field theory and the gravity pictures.

We then turn to the codimension 1 defect. In this case we will study
in detail the $D3-D5$ intersection, which is dual to an
${\mathcal{N}}=1$ three-dimensional defect living in a bulk
${\mathcal{N}}=4$ four-dimensional gauge theory. The field theory was
extensively studied in \cite{WFO} and \cite{EGK}, as well
as some aspects of the brane construction in the Coulomb phase. The
corresponding Higgs phase for this intersection was discussed in
\cite{ARR}. On the field theory side the $D3-D5$ system
describes the dynamics of a $(2+1)$-dimensional defect containing
fundamental hypermultiplets living inside the $(3+1)$-dimensional
${\cal{N}}=4$ SYM. The meson spectra on the Coulomb branch was
extensively studied in \cite{AR}. In \cite{ARR} it
was found that, in the supergravity dual, the Higgs phase also
corresponds to adding magnetic worldvolume flux inside the flavor
$D5$-brane transverse to the $D3$-branes. This worldvolume gauge
field has the nontrivial effect of inducing $D3$-brane charge in the
$D5$-brane worldvolume (which reflects the recombination of some of
the color $D3$ with the flavor $D5$), which in turn suggests an
alternative microscopical description in terms of $D3$-branes expanded
to a $D5$-brane due to dielectric effect \cite{M} along the same lines
as in the $D3-D7$ case.\break Indeed, the vacuum conditions of the
dielectric theory can be mapped to the $F$ and $D$ flatness
constraints of the dual gauge theory, thus justifying the
identification with the Higgs phase, in very much of the same spirit
of what happened in the $D3-D7$ case. In this case, the Higgs vacua of
the field theory involve a nontrivial dependence of the defect fields
on the coordinate transverse to the defect. In the supergravity side
this is mapped to a bending of the flavor brane, which is actually
required by supersymmetry (see \cite{ST}). Moreover, in
\cite{ARR} the spectrum of transverse fluctuations was
computed in the Higgs phase, with the result that the discrete
spectrum is lost. The reason is that the IR theory is modified because
of the nontrivial profile of the flavor brane, so that in the Higgs
phase, instead of having an effective $AdS\times S$ worldvolume for
the flavor brane, one has Minkowski space, thus loosing the KK-scale
which would give rise to a discrete\break spectrum.

Lastly, we turn to the codimension 2 defect, which behaves rather
different from the other intersections. The defect conformal field
theory associated to the $D3-D3$ intersection was studied in \cite{CEGK}, where the corresponding fluctuation/operator
dictionary was established. The meson mass spectra of this system when
the two sets of $D3$-branes are separated was computed analytically in
\cite{AR}. In \cite{CEGK} the Higgs branch of the
$D3-D3$ system was identified as a particular holomorphic embedding of
the probe $D3$-brane in the $AdS_5\times S^5$ geometry, which was
shown to correspond to the vanishing of the $F$- and $D$-terms in the
dual superconformal field theory (see also
\cite{Erdmenger:2003kn} and \cite{Kirsch:2004km}). This
intersection behaves in a rather different way since the two
brane-stacks are of the same dimensionality. Indeed, in this case the
flavor symmetry will not decouple as local symmetry; and thus these
theories should be understood in a different way.

\section[Adding Matter to Gauge/Gravity Duality: BPS Intersections as\\ 
Holographic Flavor]{Adding Matter to Gauge/Gravity Duality: 
BPS Intersections as Holographic Flavor}	

As we said, a major challenge remains the addition of fundamental
matter to the gauge/gravity duality in a fully satisfactory manner. We
will consider a first approxi\-mation to the problem, in which we will
think of the flavors as coming from some brane probes in the
background of the branes generating the color degrees of
\hbox{freedom}. However, we first review the gauge/gravity correspondence for
theories with 16 supercharges. For further details we refer to the
original \cite{jm} and \cite{IMSY} and the review article
\cite{MAGOO}.

\subsection{An overview of gauge/gravity duality}

The most celebrated example of gauge/gravity duality is the $AdS/CFT$
correspondence, out of which the major example is the one relating
${\mathcal{N}}=4$ SYM theory in four-dimensional Minkowski to IIB
string theory on $AdS_5\times S^5$. A lot of effort has been put
towards understanding this duality and finding an explicit dictionary
between gravity and gauge theory. Also, by now, we have infinitely
many more \hbox{examples} of dualities between conformal field
theories with diverse supersym\-metries and IIB string theory on spaces
of the form $AdS_5\times L^{a,b,c}$. In addition, there are many other
examples in other dimensions, whose gravity dual involves various
$AdS$ spaces.

In general, the gauge/gravity duality relies on the dual description
of branes in a certain limit as supergravity backgrounds or as gauge
theories. In the very special case of the $D3$ brane, this duality can
be put forward in a very precise manner, and because of the very
special properties of the $D3$ brane background (in particular the
$AdS_5$ near horizon with constant dilaton), a precise $AdS/CFT$
duality can be stated. However, not without a number of subtleties,
one can, to some extend, adapt this correspondence to the generic case
of $Dp$ branes.

\subsubsection*{The most celebrated gauge/gravity duality\/$:$ 
$AdS/CFT$ for $D3$ branes}

Let us consider $N$ $D3$ branes in flat space. Since we want to use a
string theory picture, we need to keep the dilaton (or analogously
$e^{\Phi}$) small. However, for $D3$ branes, the dilaton is a
constant, so we simply have to ensure that the asymptotic value
$e^{\Phi_0}=g_s$ is small. Being massive objects, the $D3$ branes will
backreact on the geometry and generate an asymptotically flat space
with a horizon at $r=0$
\begin{equation}	%2.1
\label{space}
ds^2=f_3^{-\frac{1}{2}}\, dx_{1,3}^2 + f_3^{\frac{1}{2}}
(dr^2+r^2\,d\Omega_5^2)\,, \qquad f_3=1+\frac{R^4}{r^4}\,.
\end{equation}
The near horizon region reduces to the $AdS_5\times S^5$ geometry. The
size of the $AdS$ space is given by
\begin{equation}	%2.2
R^4=4\pi g_s Nl_s^4\,,
\end{equation}
and it can be thought as the size of the perturbation on the flat
space generated by the branes.

We will now take the so-called {\it decoupling limit\/} of\/
$l_s\rightarrow 0$ keeping fixed the energy of the
excitations. However, energies are measured at infinity, so the
precise relation between the proper energy $E_{\rm proper}$ of some
excitation and its energy measured at infinity $E$ is
\begin{equation}	%2.3
E=f_3^{-\frac{1}{4}}E_{\rm proper} = 
\bigg(1+\frac{R^4}{r^4}\bigg)^{-\frac{1}{4}} E_{\rm proper}\,.
\end{equation}
In the large $r$ asymptotically flat region we have $f_3\sim 1$, and
therefore the space (\ref{space}) reduces to ten-dimensional
Minkowski. Since $E=E_{\rm proper}$, just the massless excitations
(namely the supergravity multiplet) keep a finite energy and survive
the limit. On the other hand, in the near horizon region $r\sim 0$,
where
\begin{equation}	%2.4
f_3\sim \frac{R^4}{r^4}\,,
\end{equation}
we have that (\ref{space}) reduces to $AdS_5\times S^5$. Upon
redefining $r=R^2z$ we can write its metric as
\begin{equation}	%2.5
ds^2=R^2\bigg(z^2\,dx_{1,3}^2+\frac{1}{z^2}\,dz^2+d\Omega_5^2\bigg)\,.
\end{equation}
In addition, in this region we have $E=RzE_{\rm proper}$. Therefore,
all the excitations survive the limit since all of them appear
asymptotically as low energy modes. Amazingly, this two subsystems are
completely decoupled in this limit.\footnote{One can think of the $D3$
brane metric as the metric of a black $p$-brane in the extremal
limit. Then, it is possible to show in general that the absorption
cross-section of the black brane goes to zero as $l_s$ goes to zero
\cite{crosssection,crosssection2,crosssection3} (see also
\cite{MAGOO}), suggesting the true decoupling between the
near-horizon and the asymptotic region.} Therefore, we can think of
the system to be composed of IIB supergravity on ten-dimensional
Minkowski plus IIB string theory on $AdS_5\times S^5$.

In order to trust the description of branes as a supergravity
background, we need to have very small curvature in $l_s$ units. Since
the curvature is proportional to the inverse of the $AdS$ radius
${\mathcal{R}}\sim R^{-1}$, this amounts to require that $l_s
R^{-1}\ll 1$, so we need to require $g_sN$ to be large. In a sense, in
this limit we are regarding the branes as a delocalized perturbation
of the Minkowski space, and we are replacing them with the geometry
(plus RR 5-form flux) they source. Since $g_s$ should be small in
order to have a perturbative string description, it is clear that we
have to take $N$ to be large, so that $g_sN\gg 1$.\footnote{Actually,
if we consider $g_s$ large, the $D1$ string would become lighter than
the fundamental string, and thus, upon performing an S-duality, we
would be formally in the same situation.}

On the other hand, we can take the opposite limit, namely that in
which we regard the system as a stack of localized $D3$ branes in flat
Minkowski space. By taking the same limit as before, namely
$l_s\rightarrow 0$ with fixed energy for the excitations, we just keep
the low energy states; which in this case restrict to
${\mathcal{N}}=4$ Yang--Mills theory on the worldvolume of the branes
with a fixed and small Yang--Mills coupling $g_{\rm YM}^2=2\pi g_s$,
plus IIB supergravity in the bulk ten-dimensional Minkowski space. In
addition, both subsystems are decoupled, and therefore do not talk to
each other. Clearly, in order to trust this description, we must have
that, away from the branes, the space is not disturbed. In other
words, we have to ensure that the typical size of the perturbation of
flat space which the branes generate is localized in a small region in
$l_s$ units; so that we can think of the system as a localized
$D3$-brane stack in ten-dimensional Minkowski. This requires
$g_sN\sim\sqrt{\lambda}$ to be small (here $\lambda$ is the 't Hooft
coupling $\lambda=g_{\rm YM}^2N$). Meanwhile, in order to match the
dual description, $N$ has to be large. Thus, the dual $SU(N)$ theory
is taken at a small Yang--Mills coupling with large $N$, so that
$\lambda$ is small. This is the 't~Hooft limit, in which just the
planar sector of the gauge theory survives.

Given that we have two descriptions of the same system, and in both of
them there is a piece which is the same (namely IIB supergravity
excitations around ten-dimensional Minkowski space), it is natural to
conjecture following \cite{jm} that the remaining subsystems are also
equivalent, namely, that IIB strings on $AdS_5\times S^5$ are dual to
${\mathcal{N}}=4$ SYM.

Let us note that the $AdS/CFT$ duality is a strong/weak coupling
duality. The field theory approximation requires the 't~Hooft coupling
to be small, whereas in the supergravity side it must be large in
order to ensure small curvatures. Thus, increasing $g_{\rm YM}$ takes
us from a field theory description to a string theory description.

\subsubsection*{$AdS$ space, conformal invariance and holography}
 
From the point of view of the decoupling limit, in the gravity side
the fixed energies with $l_s\rightarrow 0$ are measured
asymptotically. This suggests that, in a sense, the dual gauge theory
lives in the boundary of the space. Given that the boundary is a lower
dimensional space but still the two descriptions carry the same
information, the gauge/gravity correspondence is a holographic
duality. One way to make this holography more explicit is by
considering the Euclidean version of $AdS_{p+1}$, which can be
considered as $R^{p+1}$ endowed with the following metric:
\begin{equation}	%2.6
ds^2=\frac{4dy_i\,dy_j\,\delta^{ij}}{(1-\vec{y}^2)^2}\,.
\end{equation}
This space has an $S^p$ boundary at $\vec{y}^2=1$ where the metric has
a double pole and blows up. Because of this double pole, naively one
can make sense of the metric just in the interior region. However, we
can extend the metric to the boundary provided we allow the metric on
the boundary to transform in such a way that it compensates the factor
which is blowing up. This endows the boundary with a conformal
structure responsible for the conformal invariance of the dual gauge
theory. For a wonderful explanation of the deep implications of these
facts see \cite{Witten}.

Rotating back to the Lorentz space, one can write the $AdS_{p+1}$
metric as
\begin{equation}	%2.7
ds^2=\frac{dx_{1,p}^3+du^2}{u^2}\,,
\end{equation}
which is related to the metric in (\ref{space}) as $u=z^{-1}$. In this
coordinates one can see that scale transformations $x^{\mu}\rightarrow
\lambda x^{\mu}$ are a symmetry only if $u\rightarrow \lambda
u$. Since scale transformations are linked with $u$ rescalings, it is
natural to interpret the radial coordinate in $AdS$ as the energy
scale of the theory.

The gauge/gravity correspondence should be provided with a dictionary
relating quantities computed in both sides of the duality
\cite{Witten,GKP,Igor} (see \cite{MAGOO,DF} or
\cite{holren} for reviews on this issues). Exploring this
dictionary is, by far, beyond the scope of this work. However, let us
mention that, in the gravity side, one expects to have supergravity
fluctuations propagating in the $AdS$ space. In general, those
fluctuations will be functions of the radial coordinate in $AdS$, and
we will typically have two types of behavior near the
boundary. Considering a scalar fluctuations for illustrative purposes,
if the fields are canonically normalized, the normalizable modes
behave at infinity as $\rho^{-\Delta}$, whereas the nonnormalizable
ones should behave as $\rho^{\Delta-d-1}$; being $\Delta$ the
conformal dimension of the field theory operator associated to the
super\-gravity fluctuation. In the case in which the modes are not
canonically normalized, the behavior of both types of modes is of the
form $\rho^{2a_1}=\rho^{-\Delta+\gamma}$ and
$\rho^{2a_2}=\rho^{\Delta-d-1+\gamma}$ for some $\gamma$. The standard
lore is that normalizable modes correspond to VEV's in the dual field
theory while nonnormalizable modes correspond to deformations of the
Lagrangian. The dictionary between the conformal dimension of the
associated operator and the behavior of the field near the boundary
is
\begin{equation}	%2.8
\label{Delta-hyper}
\Delta = {d+1\over 2} + a_2-a_1\,.
\end{equation}
Given this relation, by matching conformal dimension and quantum
numbers under global symmetries, one can relate a certain supergravity
field to a field theory operator.

\subsubsection*{Extending the correspondence to other $Dp$ branes}

We can try to play the same game for any other $Dp$ brane
\cite{IMSY}. The corresponding supergravity metric is of the form
\begin{equation}	%2.9
\label{spaceDp}
ds^2=f_p^{-\frac{1}{2}}dx_{1,p}+f_p^{\frac{1}{2}}
(dr^2+r^2d\Omega_{8-p})\,,\qquad f_p=1+\frac{R^{7-p}}{r^{7-p}}\,,
\end{equation}
where
\begin{equation}	%2.10
R^{7-p}=2^{5-p}\pi^{{5-p\over 2}}\Gamma
\bigg({7-p\over 2}\bigg)g_sNl_s^{7-p}\,.
\end{equation}
In this case, the dilaton is given by
\begin{equation}	%2.11
e^{\Phi}=\bigg(\frac{R^2}{r^2}\bigg)^{\frac{-(7-p)(p-3)}{8}} = 
(2\pi)^{2-p} g_{\rm YM}\bigg(\frac{c_p\hspace{0.8pt}g_{\rm YM}^2Nl_s^{2(7-p)}}
{r^{7-p}}\bigg)^{\frac{3-p}{4}}\,,
\end{equation}
where the Yang--Mills coupling is defined as
\begin{equation}	%2.12
\label{gYM}
g_{\rm YM}^2=(2\pi)^{p-2}g_sl_s^{p-3}\,,
\end{equation}
being $g_s$ the asymptotic value of the dilaton. Also, we have grouped
the numerical coefficients in $c_p$:
\begin{equation}	%2.13
c_p=2^{6-2p}\pi^{\frac{9-3p}{2}}\Gamma\bigg(\frac{7-p}{2}\bigg)\,.
\end{equation}
Note that for every $p\ne 3$ the dilaton will be a function of the
radial coordinate. Soon we will see the important implications of this
fact.

We will take the decoupling limit $l_s\rightarrow 0$ while keeping
finite energy excitations (measured at infinity). In order to do that,
we have to introduce a new variable $z$ ($u=rl_s^{-2}$)
\begin{equation}	%2.14
\label{z}
z=\frac{(5-p)u^{\frac{5-p}{2}}}{2\sqrt{c_p\hspace{0.8pt}g_{\rm YM}^2N}}\,.
\end{equation}
In terms of $z$ the background metric in the near-horizon
region\footnote{For $p\ne 3$ the space has a singularity at $r=0$
rather than a horizon. One can define \cite{BST} a new metric in the
``dual frame" $ds_{\rm dual}=(e^{\Phi}N)^{\frac{2}{p-7}}ds^2$ which,
instead of a singularity, has a horizon at $r=0$; and therefore the
near-horizon limit makes sense. Writing the metric in terms of $z$
puts the dual frame metric as $AdS$; yielding to the metric
(\ref{spaceDp}) when going back to the string frame.} reads
\jot=7pt
\begin{eqnarray}	%2.15
\label{NHspaceDp}
ds^2 &=& \alpha'\bigg(\frac{2}{5-p}\bigg)^{\frac{7-p}{5-p}}
\big(c_p\hspace{0.8pt}g_{\rm YM}^2N\big)^{\frac{1}{5-p}}z^{-\frac{3-p}{5-p}} 
\bigg\{z^2dx^2_{1,p}+\frac{1}{z^2}dz^2 + 
\frac{(5-p)^2}{4}d\Omega_{8-p}^2\bigg\}\,, \nonumber \\[-4pt]
\end{eqnarray}
which is conformally $AdS$. Given that in $AdS$ the rescalings in the
Minkowski space are linked to rescalings in the radial coordinate, it
is natural to identify $z$ with the energy scale of the dual
theory. For $p=3$ we see that, like the dilaton, the conformal factor
relating the $Dp$ background to $AdS$ becomes a constant; and,
therefore, rescalings are a true symmetry, which manifests in the
conformality of the dual theory. However, for generic $p$, the scale
transformation is no longer a true symmetry; which reflects the fact
that the dual field theory will not be conformal. Note in addition
that none of these manipulations are well-defined for $p=5$.

In order to proceed further, it is useful to take a little jump ahead
and notice that, since the worldvolume low energy theory on the $Dp$
will not be conformal, it must be defined at a given energy
scale. Since the Yang--Mills coupling is dimen\-sionful, we will have
an effective dimensionless coupling at an energy scale $\mu$, which,
by dimensional analysis, must be given by
\begin{equation}	%2.16
g_{\rm eff}^2=g_{\rm YM}^2N\mu^{p-3}\,.
\label{geff}
\end{equation}

As we have noticed, for generic $p$ the dilaton will be a function of
the radial coordinate, which means that the string theory description
ceases to be valid at some point and we need a nonperturbative
completion in terms of an uplift to M-theory. In order to avoid this,
and trust the string theory description, one has to require that
$e^{\Phi}\ll 1$:
\begin{equation}	%2.17
e^{\Phi}\sim g_{\rm eff}^{\frac{7-p}{2}}N^{-1}\ll 1\,.
\end{equation}
In addition, in terms of $g_{\rm eff}$, the curvature of the $Dp$
background in $l_s$ units is proportional to $1/g_{\rm eff}$, so in
order to trust the supergravity approximation we have to take $g_{\rm
eff}\gg 1$. Both things can be combined into
\begin{equation}	%2.18
1\ll g_{\rm eff}\ll N^{\frac{2}{7-p}}\,,
\end{equation}
which defines the range of validity of the gravity approximation in
terms of a string theory background. However, there is a parameter
range in which $e^{\Phi}\sim 1$, in which one would start resolving
the M-theory circle. In this case, one could continue the gravity
description by uplifting to M-theory. As long as the curvature of the
11-dimensional background is kept small, it is possible to give an
M-theory description in this regime.

On the other hand, we can consider the $Dp$ branes as a localized
stack in Minkowski space. In the very same limit as before, we would
have that it the system decouples into the bulk supergravity plus the
worldvolume gauge theory. Since the effective dimensionless parameter
we will use as expansion parameter is $g_{\rm eff}$, we can control
this approximation for $g_{\rm eff}\ll 1$; where the energy scale
$\mu$ is set by $z$. Since $g_{\rm eff}$ is a function of the scale,
for fixed $g_{\rm YM}^2N$ at some scale we will fail to have a
controlled field theory approximation. However, it is possible to find
an energy range in which the gauge theory fails to be weakly coupled,
demanding some nonperturbative completion. We can find this completion
in terms of the M-theory uplift, in which, in the suitable parameter
range, we can match the gravity dual in terms of an 11-dimensional
supergravity description.

Naively, thinking just as for the $D3$, we would be tempted to
conjecture that the near-horizon background (\ref{spaceDp}) captures,
for the above range of parameters, the physics of the system,
meanwhile when $g_{\rm eff}$ is small it is the corresponding gauge
theory the one capturing the physics. However, a careful analysis case
by case should be done, since it is not obvious at all (indeed for the
$D6$ it is false) that the open and closed string modes (namely
asymptotic region and near horizon) really decouple. In addition, as
we have pointed out, all the manipulations above are not well defined
for $p=5$, where a careful analysis yields to a dual description in
terms of a little string theory. Analyzing each case is beyond our
scope, and we refer to the comprehensive review \cite{MAGOO}
and references therein. However, taking into account these subtleties,
we can still play the same game for a generic $Dp$-brane. This way, we
can obtain a dual description, valid in general in some energy regime
and in a different corner in parameter space, of the SYM field theory
on the worldvolume of a $Dp$ brane in terms of the near horizon of the
background corresponding to the $Dp$. The field theory can be obtained
as the dimensional reduction of the (maximally) supersymmetric $SU(N)$
Yang--Mills theory in ten dimensions down to $p+1$ dimensions.

It is clear then that the field theory dual to any $Dp$-brane stack
will contain just adjoint matter corresponding to the transverse
scalars to the $Dp$'s which host the SYM theory. More explicitly, when
taking the $l_s\rightarrow 0$ limit while regarding the system as
localized branes in flat space, what survives from the open string
sector attached to the branes are precisely the states that are
necessary to generate the $SU(N)$ non-Abelian gauge theory, in which
the scalar fields are in the adjoint representation and have the
interpretation of the transverse positions to the branes. Since this
branes generate a pure glue theory, we will call this branes ``color"
branes.

\subsection{Flavoring the gauge/gravity duality}

It is of obvious interest bringing fundamental matter into the
game. The key idea of
\cite{Mateo,Mateo2,Mateo3,KR,KR2}. References~\cite{KKW}
and \cite{KKW2} is to add extra $N_f$ ``flavor" branes to the $N_c$
color ones giving rise to a new sector of strings stretching between
the two stacks. Thus, the idea is to use the gauge/gravity duality
above for this extended system exactly as we did in the case of just
one stack of color branes.

In this case, the field theory description will come up from analyzing
the low energy limit of the brane system when thought as localized
intersection of two stacks in the ambient flat Minkowski space. This
intersection will contain three open string sectors: the $Dp_1-Dp_1$
strings, which will give rise to the corresponding $SU(N_c)$ SYM on
the worldvolume of the $Dp_1$; the $Dp_2-Dp_2$ strings, which will
give rise to the corresponding $SU(N_f)$ SYM on the worldvolume of the
$Dp_2$; and the $Dp_1-Dp_2$ strings giving rise to some extra matter
transforming in the $(\mathbf{N_c},\mathbf{N_f})$ and confined to the
common intersection between the two stacks. Note that, in general, we
can consider a transverse separation between the two stacks, which
corresponds to a minimum length for the $Dp_1-Dp_2$ strings. Since the
mass of an open string is proportional to its length, we have that the
separation of the color and flavor branes amounts, in the field
theory, to introduce a mass scale for the quarks confined to the
intersection. In this common intersection, which will be seen as a
defect in the worldvolume of both stacks of branes, there is a
$SU(N_c)\times SU(N_f)$ gauge theory. In the decoupling limit, the low
energy description of the whole system will be in terms of
supergravity plus a field theory which schematically reads
\begin{equation}	%2.19
S = \int d^{\hspace{1pt}p_1+1}x\, L_{Dp_1}+\int d^{\hspace{1pt}p_2+1}x\,L_{Dp_2} + 
\int d^{\hspace{1pt}p_2-p_1}x\, L_{\rm defect}\,.
\end{equation}
In each of the two stacks of branes, the strength of the gauge
couplings will be governed by the tension as $g_{p}^{-2}\sim
T_{D_p}$. Actually, the kinetic terms for the gauge fields read
\begin{eqnarray}	%2.20
S_{\rm gauge\;kinetic} &=& \int d^{\hspace{1pt}p_1+1}x\, 
\frac{1}{4\pi g^2_{p_1}}\,F_{Dp_1,SU(N_c)}^2 \nonumber \\
&&{}+ \int d^{\hspace{1pt}p_2+1}x\,\frac{1}{4\pi g^2_{p_2}}F_{Dp_2,SU(N_f)}^2\,.
\end{eqnarray}
The relation between the gauge couplings $g_{p_1}$, $g_{p_2}$ is given
by
\begin{equation}	%2.21
\frac{T_{p_1}}{T_{p_2}}=(2\pi l_s)^{{p_2}-{p_1}} \leadsto 
\frac{g^2_{p_2}}{g^2_{p_1}}=(2\pi l_s)^{{p_2}-{p_1}}\,.
\end{equation}
As in the $Dp$ brane case, we will keep fixed the Yang--Mills coupling
$g_{p_1}$ on the $Dp_1$. Then, if $p_2>p_1$, in the low energy limit
$l_s\rightarrow 0$:
\begin{equation}	%2.22
g_{p_2}\rightarrow 0\,.
\end{equation}
Since we have that the $SU(N_f)$ gauge group has a vanishing coupling
constant, its kinetic term should vanish, so it decouples as a local
symmetry; leaving as remnant just the global $SU(N_f)$ rotations for
the matter confined to the common intersection. Then, the effective
field theory description is a $SU(N_c)$ pure SYM theory in $p_1+1$
dimensions containing a lower-dimensional defect in which matter
transforming under a $SU(N_f)$ flavor symmetry lives.\footnote{Hence
the name {\it defect field theories} for these gauge theories with
matter confined to a lower-dimensional defect.} Since we will restrict
ourselves to supersymmetric intersections, this defect field theory
will preserve eight supersymmetries.

Exactly as in the $Dp$ case, in order to ensure the validity of the
approximation, we have to consider small dilaton to trust the string
description, and ensure that the effective scale-dependent Yang--Mills
coupling constant is small enough so as to trust the perturbative
Yang--Mills.

On the other hand, we can also think that the branes backreact the
geometry in a given way; and consider the suitable curvature range so
as to think of the brane system as the geometry it
backreacts. However, in this case the full backreacted solution will
be quite complicated, and it is just since very recent that some
results have appeared along these lines (see
\cite{unquenched,unquenched2,unquenched3,unquenched4, rf60, rf61,unquenched5}). In
order to simplify the problem, we can take the limit in which we have
much more color branes than flavor branes $N_f\ll N_c$, so that we can
think that the effect of the flavor branes is negligible compared to
the effect of the color branes, in some sort of quenched
approximation. More explicitly, the mass of the color branes will be
given by $m_c=N_cT_{p_1}Vol_{Dp_1}$, meanwhile the mass of the
flavor ones will be $m_f=N_fT_{p_2}Vol_{Dp_2}$. Then, in order
to ensure the validity of the approximation, we have to require that
the mass of the flavor branes is negligible compared to the mass of
the color branes, so that we can approximate well enough the full
background with the one sourced by the color branes in which the
flavors move as probes. This amounts to impose $m_1\gg m_2$, which
requires $N_c\gg N_f V_{\perp}$, where $V_{\perp}$ is the volume of
the transverse coordinates to the $Dp_1$ contained in the $Dp_2$
measured in $l_s$ units. Since this will go to infinity because the
branes wrap a noncompact cycle, we need to take $N_c$ going to
infinity in a sufficiently rapid manner to ensure the brane-probe
approximation. In addition, the curvature of the background should be
small in $l_s$ units while keeping small dilaton. Therefore, under
these circumstances, a good approximation to the supergravity
description is to consider the flavor branes as probe branes in the
background of the color ones. Upon taking the decoupling limit, which
amounts to consider just the near horizon region of the corresponding
$Dp_1$ brane background, we will have the gravity dual for the defect
field theory in this quenched approach.

\subsubsection{Supersymmetric brane intersections}

Since we will be considering just supersymmetric field theories, our
first task will be to find the corresponding supersymmetric
intersections. A particular feature of a supersymmetric brane system
is that there is no force between the branes. This requires that the
potential for the separation between the branes vanishes. This is the
so-called no-force condition. Following \cite{AR}, we will
make use of this condition to find supersymmetric intersections by
considering a stack of flavor probe branes in the background of the
color ones at a distance $L$ in the transverse space; and then
imposing the no-force condition, which amounts to demand that the
transverse distance is a flat direction in the brane--brane
potential. However, an explicit check of the supersymmetry can be
given (see for example \cite{intersections} and
\cite{intersections2} for a discussion at the level of the
supergravity backgrounds).

The energy density $E$ of the probe is determined by its
Dirac--Born--Infeld plus Chern--Simons action. The latter would
involve the coupling of the probe brane to the background RR
potential. However, in the cases at hand, this coupling is easily seen
to be zero as long as we do not consider a nonzero worldvolume vector
field. For the moment, we will restrict to these cases, which will
correspond to the Coulomb branch of the theory; meanwhile the
inclusion of the worldvolume field will correspond to the Higgs
branch. Then, for static configurations as those we are considering
here, we have that the energy is minus the DBI action, which in this
case reads
\begin{equation}	%2.23
\label{DBI}
S=S_{\rm DBI}=-T_p\int d^{\hspace{1pt}p+1}x\,\sqrt{-\det{\cal{G}}}\,,
\end{equation}
where ${\cal{G}}$ stands for the induced metric on the worldvolume of
the brane, which is the pull-back of the target space metric.

Let us rewrite the near-horizon limit of (\ref{spaceDp}) as
\begin{equation}	%2.24
ds^2 = \biggl[{r^2\over R^2}\biggr]^{\alpha}dx^2_{1,p_1} + 
\biggl[{R^2\over r^2}\biggr]^{\alpha}\,d\vec y\cdot d\vec y\,.
\label{metric}
\end{equation}
Here $dx^2_{1,p_1}$ denotes the $(p_1+1)$-dimensional
Minkowski\vspace*{2pt} metric, while $\vec y = (y^1,\ldots,
y^{9-p_1})$ and $r^2=\vec y\cdot \vec y$. As we know, in these
backgrounds there is also a dilaton given by
\begin{equation}	%2.25
e^{-\phi(r)} = \biggl[{R^2\over r^2}\biggr]^{\gamma}\,,
\label{dilaton}
\end{equation}
where the exponents $\gamma,\alpha$ are given by
\begin{equation}	%2.26
\alpha = {7-p_1\over 4}\,, \qquad \gamma = {(7-p_1)(p_1-3)\over 8}\,.
\label{Dpgammas}
\end{equation}

\begin{figure}[th]
\centerline{\hskip -.8in \epsffile{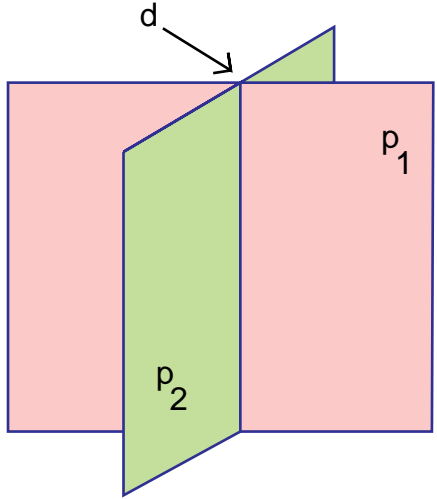}}
\vspace*{8pt}
\caption{An orthogonal intersection of a $p_1$- and a $p_2$-brane along
$d$ spatial directions.\protect
\label{fig1}}
\end{figure}

Let us now add a probe $Dp_2$-brane sharing $d$ common spatial
directions with the $p_1$-brane. The corresponding
orthogonal%\forcebreak{}
 intersection will be denoted as $(d|p_1\perp
p_2)$ and is depicted in Fig.~\ref{fig1}. We will assume that the
probe is extended along the directions $(x^1,\ldots,x^d, y^1,\ldots,
y^{p_2-d})$ and we will denote by $\vec z$ the set of $y$ coordinates
transverse to the probe. Notice that $|\vec z|$ represents the
separation of the branes along the directions transverse to both
background and probe branes.

We will consider a static configuration in which the probe is located
at a constant value of $|\vec z|$, namely at $|\vec z|=L$. The
induced metric on the probe worldvolume for such a static
configuration will be $ds^2_{I}={\cal G}_{ab}\,d\xi^a d\xi^b$ with
$\xi^a$ being a set of worldvolume coordinates. In what follows we
shall take these coordinates as the common Cartesian coordinates
$x^0,\ldots, x^d$, together with the spherical coordinates of the
$y^1,\ldots, y^{p_2-d}$ hyperplane. Assuming that $p_2-d\ge 2$, we
will represent the line element of this hyperplane as 
\begin{equation}	%2.27
(dy^1)^2 + \cdots + (dy^{p_2-d})^2 = 
d\rho^2 + \rho^2\, d\Omega^2_{p_2-d-1}\,,
\label{spherical}
\end{equation}
where $d\Omega^2_{p_2-d-1}$ is the line element of a unit
$(p_2-d-1)$-sphere. It is now straightforward to verify that the
induced metric $ds^2_{I}$ can be written as
\begin{equation}	%2.28
ds^2_{I} = \biggl[{\rho^2+L^2\over R^2}\biggr]^{\alpha}dx^2_{1,d} + 
\biggl[{R^2\over \rho^2+L^2}\biggr]^{\alpha}
(d\rho^2 + \rho^2 d\Omega^2_{p_2-d-1})\,.
\label{indmetric}
\end{equation}

Using Eqs.~(\ref{dilaton}) and (\ref{indmetric}), we have that the
energy for the probe-brane is
\begin{equation}	%2.29
E = \biggl[{\rho^2+L^2\over R^2}\biggr]^{{\alpha\over 2}
(d+1) - {\alpha\over 2}(p_2-d)-\gamma}\rho^{p_2-d-1}\sqrt{\det \tilde g}\,,
\label{staticH}
\end{equation}
where $\tilde g$ is the metric of the unit $(p_2-d-1)$-sphere.

Since we are interested in studying BPS intersections, we have to
impose that the branes do not exert any force among each other. This
no-force condition requires the energy to be independent of the
distance $L$ between the branes which, in view of the right-hand side
of Eq.~(\ref{staticH}), is only possible if the number $d$ of common
dimensions is related to the total dimensionality $p_2$ of the probe
as
\begin{equation}	%2.30
d = {p_2\over 2} + {2\gamma-\alpha\over 2\alpha}\,.
\label{BPSrule}
\end{equation}

Using (\ref{Dpgammas}), we get the following relation between $d$ and
$p_2$:
\begin{equation}	%2.31
d={p_2+p_1-4\over 2}\,.
\label{Common}
\end{equation}
However, as the brane of the background and the probe should live in
the same theory, $p_2-p_1$ should be even. Since $d\le p$, we are left
with the following three possibilities $p_2=p_1$, $p_1+2$, $p_1+4$,
for which Eq.~(\ref{Common}) gives $d=p-2$, \hbox{$p-1$,} $p$
respectively. Thus, we get the following well-known set of orthogonal
BPS intersections of $D$-branes:
\begin{equation}	%2.32
\label{series}
(p\hspace{1pt}|\hspace{1pt}Dp\perp D(p+4))\,,\qquad 
(p-1\hspace{1pt}|\hspace{1pt}Dp\perp D(p+2))\,,\qquad 
(p-2\hspace{1pt}|\hspace{1pt}Dp\perp Dp)\,. \ \
\end{equation}
The cases of (\ref{series}) give rise to three series of gauge/gravity
dualities for gauge \hbox{theories} containing flavors confined to a
defect; which in each case is of codimension 0, codimension 1 and
codimension 2 in the ambient gauge theory. The gravity description is
nothing but the $Dp_1$ background in which we embed the corresponding
$Dp_2$ as probes, meanwhile for the first two series the field theory
dual is the dimensional reduction of ${\mathcal{N}}=1$ $SU(N_c)$ SYM
in ten dimensions down to $d+1$ dimensions where we have to consider a
1/2 supersymmetric defect of the corresponding dimensionality
containing fundamental matter with a global $SU(N_f)$ symmetry. Given
that the system is supersymmetric, it is possible to reconstruct the
field theory in each case based on supersymmetry (and global
symmetries) argu\-ments. However, we will postpone a detailed
field-theoretic analysis to Sec.~4 in order to give a more unified
presentation of the different branches of the gauge theory.

The last intersection type deserves a particular comment, since it is
somehow special. Given that both intersecting branes are of the same
dimensionality, none of the local dynamics of the branes decouples
from the system. Therefore, in the intersection, we have a product
gauge group under which the matter is in the bifundamental
representation, thus being not a flavoring of the ``bulk" gauge theory
in a proper sense.

\section{The Coulomb Branch of the Gauge Theories}

In the last section we argued that it is possible to find a gravity
dual for defect field theories in terms of an intersection of branes
in which the matter lives confined. In the brane-probe approximation,
the gravity description corresponds to consider very few flavor branes
suitably embedded in the near-horizon region of the background sourced
by the (infinite) color branes, which partially overlap. On the other
hand, the field theory dual corresponds to the gauge theory on the
color branes plus the matter confined in the common intersection. As
we anticipated, the color sector comes from the dimensional reduction
down to the color brane worldvolume of the ${\cal{N}}=1$
ten-dimensional Yang--Mills theory. The matter comes from the open
string sector connecting both brane stacks. Since both stacks will be
separated a distance $L$, those strings will have a minimal length
given by this $L$, so they will give rise to a massive matter sector
with mass
\begin{equation}	%3.1
m=\frac{L}{2\pi l_s^2}\,.
\end{equation}

We have discussed the flavor probe branes with vanishing worldvolume
vector field, which in particular had the consequence of not coupling
the background RR potential to the probe worldvolume. Generically, a
worldvolume vector field on a given brane has the effect of dissolving
lower dimensional branes. This can be argumented through the
Chern--Simons coupling
\begin{equation}	%3.2
S_{\rm CS}=T_p\int P[e^BC^{(n)}]e^{2\pi l_s^2F}\,,
\end{equation}
where it should be understood that a form of suitable dimensions
should be constructed inside the integral. This way, one can couple
$C^{(n)}$ to a $n+2$ brane through $F$ and to a $n+4$ brane through
$F\wedge F$; meaning that both the $n+2$ and $n+4$ branes have
$n$-brane charge dissolved. In the cases at hand, the RR potential
will be that sourced by the color branes, so by turning on the
worldvolume vector field we will be able to dissolve some of the color
branes in the flavor ones. This way one would separate some of the
branes of the color stack. But in the original picture, the stack of
$N_c$ color branes gives rise to a $SU(N_c)$ gauge theory, so
separating the color branes amounts to break the gauge group down to
some subgroup with a pattern given by the separation of the branes;
which corresponds to a higgsing the gauge group. However, the
separated branes are being dissolved in the flavor stack, which in
turn requires to give some VEV's to some of the open string fields. In
particular, as we argued, we have to give a VEV to the worldvolume
vector field. Since generically open string fields correspond to
matter fields in the field theory, one would expect that, in addition
to the breaking of the gauge group, some nontrivial quark VEV's are
generated, which should correspond to the Higgs branch of the theory.

On the other hand, taking the worldvolume gauge field to zero amounts
to keep all the color branes agrupated in a single stack without
recombining with the flavor ones.\footnote{More precisely, not
recombined with the flavors, since indeed, in general, they will be
separated. Actually, as we will discuss, in a generic point of the
Coulomb branch moduli space we will have a broken gauge group
corresponding to moving the color branes without dissolving them in
the flavor ones.} In particular, this means that we will not give any
VEV to any open string field (matter sector in the field theory side),
so this should correspond to the Coulomb phase of the gauge
theory. Thus, the precise statement of the gauge/gravity duality is
that the background of color branes with very few probe flavor branes
with vanishing worldvolume gauge field corresponds to the Coulomb
phase of the dual defect field theory.

\subsection{Fluctuations as mesons}

Since the probe branes we are considering in the background of the
color ones have the interpretation of flavors in the field theory,
their fluctuations will correspond to the possible excitations in the
dual defect theory. In particular, if one finds a discrete spectrum
for those fluctuations, this will correspond to the bound state
spectrum of quarks (a.k.a. mesons) of the field theory. For the class
of theories of interest, the meson spectrum study was initiated in
\cite{KMMW} for the $D3-D7$ case, where the whole meson
spectrum (in the Coulomb branch) was studied. However, a lot of work
has been devoted towards studying the meson spectrum in various
theories (in various contexts in several backgrounds; see for example
\cite{KMMW,Sonnen,Johana,Johana2,Johana3,KMMW-two,Carlos,Carlos2,Carlos3,Ouyang,Ouyang2,WH,flavoring,Hong,Evans,Evans2,Evans3,Ghoroku,Ghoroku1,Ghoroku2,Ghoroku3,melting,melting2,melting3,melting4,conifold,Kuper,Sakai,Sakai1,APR,AR,MT,Apreda:2006bu}).

Let us study the fluctuations around the flavor brane embeddings
discussed above. Without loss of generality we can take, for a generic
$Dp_1-Dp_2$ intersection, $z^1=L$, $z^m=0$ ($m>1$) as the unperturbed
configuration. Among all the fluctuations, we will restrict for
simplicity to the transverse scalar modes $\chi$, which are of the
type:
\begin{equation}	%3.3
z^1=L+\chi^1\,, \qquad z^m=\chi^m (m>1)\,.
\label{perturbation}
\end{equation}
By analyzing the whole set of fluctuations, one can see that
restricting to these modes is a consistent truncation (see
\cite{AR} and \cite{R}). The dynamics of the fluctuations
is determined by the Dirac--Born--Infeld Lagrangian (\ref{DBI}). By
expanding this action and keeping up to second order terms, one can
see that the relevant Lagrangian for the fluctuations is
\begin{equation}	%3.4
{\cal L} = -{1\over 2}\,\rho^{p_2-d-1}\sqrt{\det\tilde g}
\biggl[{R^2\over \rho^2+L^2}\biggr]^{{7-p_1\over 4}}
{\cal G}^{ab}\partial_{a}\chi^m\partial_{b}\chi^m\,,
\label{fluct-lag-general}
\end{equation}
where ${\cal G}^{ab}$ is the (inverse of the) metric
(\ref{indmetric}). The equations of motion derived from ${\cal L}$
are
\begin{equation}	%3.5
\partial_{a}\Bigg[{\rho^{p_2-d-1}\sqrt{\det \tilde g}\over 
(\rho^2+L^2)^{{7-p_1\over 4}}}\,{\cal G}^{ab}\partial_{b}\chi\Bigg] = 0\,,
\label{eom-general}
\end{equation}
where we have dropped the index $m$ in the $\chi$'s. Using the
explicit form of the metric elements ${\cal G}^{ab}$,
Eq.~(\ref{eom-general}) can be written as the following differential
equation:
\begin{equation}	%3.6
{R^{7-p_1}\over (\rho^2+L^2)^{{7-p_1\over 2}}}\,
\partial^{\mu}\partial_{\mu}\,\chi + {1\over \rho^{p_2-d-1}}\,
\partial_{\rho}(\rho^{p_2-d-1}\partial_{\rho}\chi) + 
{1\over\rho^2}\,\nabla^i\nabla_i\chi=0\,,
\label{eom-separated}
\end{equation}
where the index $\mu$ corresponds to the directions $x^{\mu}=(t,
x^1,\ldots, x^d)$ and $\nabla_i$ is the covariant derivative on the
$(p_2-d-1)$-sphere. To solve this equation, let us separate variables
as
\begin{equation}	%3.7
\chi = \xi(\rho)e^{ikx}Y^l(S^{p_2-d-1})\,,
\label{sepvar}
\end{equation}
where the product $kx$ is performed with the flat Minkowski metric and
$Y^l(S^{p_2-d-1})$ are scalar spherical harmonics on the
$(p_2-d-1)$-dimensional sphere, which satisfy
\begin{equation}	%3.8
\nabla^i\nabla_i Y^l(S^{p_2-d-1}) = -l(l+p_2-d-2)Y^l(S^{p_2-d-1})\,.
\label{casimir}
\end{equation}
If we redefine the variables as
\begin{equation}	%3.9
\varrho = {\rho\over L}\,, \qquad \bar M^2 = -R^{7-p_1}L^{p_1-5}k^2\,,
\label{newvariables}
\end{equation}
the differential equation (\ref{eom-separated}) becomes
\begin{equation}	%3.10
\partial_{\varrho}(\varrho^{p_2-d-1}\partial_{\varrho}%\ko
\xi) + 
\bigg[\bar M^2\,{\varrho^{p_2-d-1}\over (1+\varrho^2)^{{7-p_1\over 2}}} - 
l(l+p_2-d-2)\varrho^{p_2-d-3}\bigg]\xi = 0\,.
\label{fluc}
\end{equation}

For generic $p_1$, $p_2$, (\ref{fluc}) has no simple analytic
solution. However, a numerical analysis can be carried
out. Generically, one obtains a discrete mass spectrum, being the
masses of the mesons proportional to $L\sim m_q$. They will also
depend on the inverse effective coupling, which is to be expected
since the bound states should be a nonperturbative effect in the field
theory. Remarkably, it turns out that in the $p_1=3$ case the
differential equation (\ref{fluc}) can be solved in terms of
hypergeometric functions and the spectrum of values of $\bar M$ can be
found exactly. Indeed, in this case, the fluctuations we have studied
are a subset of those in \cite{KMMW}; where the whole spectrum
was studied. We now turn to a more detailed analysis of this case.

\subsubsection{$AdS_5\times S^5$ background}

It is of particular interest the case of a bulk $(3+1)$-dimensional
theory corresponding to a stack of color $D3$ branes. As we mentioned,
this case was exhaustively studied in \cite{KMMW}. For
$p_1=3$, in the limit we are considering in which the flavor branes
are treated as probes, the decoupling limit works as in the usual
$AdS/CFT$ case. Thus, once we take the decoupling limit, the gravity
description is in terms of the corresponding embedding of the flavor
branes in the near horizon of the $D3$ background, which is
$AdS_5\times S^5$, being the possible flavorings
\begin{equation}	%3.11
(3\hspace{1pt}|\hspace{1pt}D3\perp D7)\,,\qquad 
(2\hspace{1pt}|\hspace{1pt}D3\perp D5)\,,\qquad (1\hspace{1pt}|\hspace{1pt}D3\perp D3)\,.
\label{AdSdefects}
\end{equation}
Let us now introduce the quantity $\lambda$, related to the rescaled
mass $\bar M$ as
\begin{equation}	%3.12
\bar M^2=4\lambda (\lambda+1)\,.
\label{Mlambda}
\end{equation}
Then, the solution of (\ref{fluc}) for $p_1=3$ that is regular when
$\varrho\to 0$ is
\begin{equation}	%3.13
\xi(\varrho) = \varrho^l(\varrho^2+1)^{-\lambda}
F\bigg({-%\ko
}\lambda, -\lambda+l-1+{p_2-d\over 2}; 
l+{p_2-d\over 2};-\varrho^2\bigg)\,.
\label{scalarhyper}
\end{equation}
We also want that $\xi$ vanishes when $\varrho\to\infty$. A way to
ensure this is by imposing that
\begin{equation}	%3.14
-\lambda+l-1+{p_2-d\over 2} = -n\,, \quad n=0,1,2,\ldots\,.
\label{quant}
\end{equation}
When the quantization condition (\ref{quant}) is imposed, the series
defining the hypergeometric function in (\ref{scalarhyper}) truncates,
and the highest power of $\varrho$ is $(\varrho^2)^n$. As a
consequence $\xi$ vanishes as $\varrho^{-(l+p_2-d-2)}$ when
$\varrho\to\infty$. Moreover, the quantization condition
(\ref{quant}) of the values of $\lambda$ implies that the allowed
values of $\bar M^2$ are
\begin{equation}	%3.15
\bar M^2 = 4\bigg(n+l-1+{p_2-d\over 2}\bigg)
\bigg(n+l+{p_2-d\over 2}\bigg)\,.
\label{exactM}
\end{equation}
Notice that, for the three cases in (\ref{AdSdefects}), $p_2=2d+1$ for
$d=3,2,1$. By using this relation between $p_2$ and $d$, one can
rewrite the mass spectra (\ref{exactM}) of scalar fluctuations for the
intersections (\ref{AdSdefects}), corresponding to the meson spectrum
in the Coulomb branch, as
\begin{equation}	%3.16
M = {2L\over R^2}\sqrt{\bigg(n + l + {d-1\over 2}\bigg)
\bigg(n + l + {d+1\over 2}\bigg)}\,,
\label{generalMS}
\end{equation}
where $M^2=-k^2$ and we have taken into account that, in this case,
$\bar M^2=-R^4L^{-2}k^2$ (see Eq.~(\ref{newvariables})).

It drops from (\ref{generalMS}) that the mass gap is proportional to
$L$; which in turn is related to the quark mass. Indeed, in terms of
field theory quantities we have that (\ref{generalMS}) reads
\begin{equation}	%3.17
\label{mass}
M = {2\pi\over\sqrt{\lambda}}\,m_q\sqrt{2\bigg(n + l + {d-1\over 2}\bigg)
\bigg(n + l + {d+1\over 2}\bigg)}\,,
\end{equation}
so for zero quark mass the generated mass gap vanishes. Moreover, the
induced metric on the probe (\ref{indmetric}) reduces to
\begin{equation}	%3.18
ds^2_{I} = {\rho^2+L^2\over R^2}\,dx_{1,d}^2 + 
{R^2\over \rho^2+L^2}\,d\rho^2 + 
R^2\,{\rho^2\over \rho^2+L^2}\,d\Omega^2_d\,.
\label{AdSindmetric}
\end{equation}
In the so-called conformal case, corresponding to $L=0$,
(\ref{AdSindmetric}) reduces to $AdS_{d+2}\times S^d$, and therefore
the dual theory is indeed a conformal field theory. Since a conformal
theory does not generate a dynamical mass scale, and given that in the
massless case there is no ``tree-level" scale, the mass gap in the
spectrum should vanish in the $L\sim m_q\rightarrow 0$ limit, as we
found in (\ref{mass}). The defect conformal field theory is engineered
as a defect which lives immersed in a ${\mathcal{N}}=4$ theory. This
bulk theory enjoys a conformal symmetry, which naively one would
expect to be broken by the addition of extra matter in the defect,
even in the case in which the extra matter is massless. However, the
conformal symmetry will appear just in the large $N_c$ and small $N_f$
limit, since there the would-be ADS (Affleck, Dine, Seiberg)
superpotential vanishes.

In the massive case corresponding to $L\ne 0$, the bare mass of the
quarks sets a scale which explicitly breaks the conformal symmetry. In
turn, this is responsible for the appearance of the mass gap, since
the meson masses are proportional to the scale $L$ corresponding to
$m_q$. However, the same asymptotic $AdS$ metric is achieved in the
ultraviolet limit $\rho\to\infty$. This $\varrho\to\infty$ limit is
simply the high energy regime of the theory, where the mass of the
quarks, which are proportional to the brane separation $L$, can be
ignored and the theory becomes conformal. Therefore, one expects that
the dual theory enjoys a conformal symmetry in the UV, and it is only
in the IR, when the quark masses are relevant, that this conformal
symmetry is broken and a mass gap with a discrete spectrum is
generated.

The field theory lives in the boundary of the space, which corresponds
to the large $\rho$ (namely UV) region. Since the UV regime is
insensitive of the possible IR breaking of the conformal invariance,
the $\varrho\to\infty$ behavior of the fluctuations, even in the $L\ne
0$ case, should provide us with information about the conformal
dimension $\Delta$ of the corresponding operator. Taking into account
that for large $\varrho$ the hypergeometric function behaves as
\begin{equation}	%3.19
F(a_1, a_2;b;-\varrho^2) \approx c_1 \varrho^{-2a_1} + 
c_2\varrho^{-2a_2}\,, \quad (\varrho\to \infty)\,,
\label{asymptotic-hyper}
\end{equation}
we have that $a_1=-\lambda$ and $a_2=-\lambda+l+{d-1\over 2}$. Using
(\ref{Delta-hyper}) in this case, we get the following value for the
dimension of the operator associated to the scalar fluctuations:
\begin{equation}	%3.20
\Delta = l+d\,.
\label{generalDeltaS}
\end{equation}

It turns out that the mass spectra of all the Born--Infeld modes (and
not only those reviewed here that correspond to the transverse
scalars) can be computed analytically as in \cite{KMMW} (see
also \cite{AR,R} and \cite{MT}). This full set of
fluctuation modes can be accommodated in multiplets, with the mass
spectra displaying the expected degeneracy. The dual operators in the
gauge theory side can be matched with the fluctuations by looking at
the UV dimensions and at the R-charge quantum numbers. Generically,
the dual fields are bilinear in the fundamental fields and contain the
powers of the adjoint fields needed to construct the appropriate
representation of the R-charge symmetry.

\section{Higgsing the Theories}

So far, in our gravity approximation to the systems we are interested
in, we have not considered the effect of the RR gauge field in the
worldvolume of the probe flavor branes. As we argued, the reason is
that we considered a vanishing worldvolume gauge field on the probe
branes. In turn, this field is required in order to couple such RR
background potential to the worldvolume theory. However, as we argued,
we have reasons to believe that the inclusion of this field will
correspond to the Higgs branch of the theory. As we already described,
the reason is that, by means of this field, we can dissolve some of
the color branes in the flavor ones. Heuristically, as already
suggested in \cite{Aharony}, this would represent separating
some of the color branes and therefore breaking the gauge group. But
moreover, those color branes are being dissolved in the flavor ones,
which gives some VEV's to some open string fields, which in turn
should correspond to quark VEV's. Thus, on very general grounds, one
would expect that the dual field theory is in the Higgs branch. In
this section we will explicitly see how it is the case. Actually, by
considering as examples the flavorings of ${\mathcal{N}}=4$, we will
explicitly discuss the field theories and explore how they contain
different branches apart from the Coulomb phase already discussed.

\subsection{The codimension zero defect}

We will first study the codimension 0 intersection. This corresponds
to the $Dp-Dp+4$ intersection where the flavors fill the whole bulk
where the gauge theory lives:
\begin{eqnarray*}	
\arraycolsep5pt\begin{array}{rccccccccccl}
& 1 & \cdots & p & p+1 & p+2 & p+3 & p+4 & p+5 & \cdots & 9 & \\[2pt]
Dp: & \times & \cdots & \times & - & - & - & - & - & \cdots & - & \\[2pt]
D(p+4): & \times & \cdots & \times & \times & \times
& \times & \times & - & \cdots & - &
\end{array}
\label{DpDp+4intersection}
\end{eqnarray*}

From the gravity side, all the cases behave in a similar way. However,
we will examine the $p=3$ case, where a detailed description of both
the field theory and gravity side will be given. The other cases will
behave in a similar way, and indeed, when computing the fluctuations
giving rise to the mesons, we will treat in a unified way all the
dimensionalities.

\subsubsection{A case study I\/$:$ the $D3-D7$ intersection}

Let us start considering the $D3-D7$ intersection first studied in
\cite{EGG} and further analyzed in \cite{ARR2}. It can
be seen that the dual gauge theory is a ${\mathcal{N}}=2$ SYM theory
in $3+1$ dimensions obtained by adding $N_f$ ${\mathcal{N}}=2$
fundamental hypermultiplets to the ${\mathcal{N}}=4$ $SYM$ theory, in
which, as we know, the transverse separation of the branes gives a
bare mass to the quarks. The Lagrangian is given by \cite{EGG}
\begin{eqnarray}	%4.1
\label{LFT}
{\cal L} &=& \tau \int d^{\hspace{0.5pt}2} \theta\, d^{\hspace{0.5pt}2} \bar\theta
\big(Tr\big(\Phi_I^{\dagger}e^V \Phi_I e^{-V}\big) + 
Q_i^\dagger e^V Q^i + \tilde Q_i e^{-V} \tilde Q^{i\dagger}\big) \nonumber \\
&&{}+ \tau \int d^{\hspace{0.5pt}2} \theta(Tr ({\cal W}^\alpha{\cal W}_\alpha) + W) +
\tau \int d^{\hspace{0.5pt}2} \bar \theta(Tr (\bar{\cal W}_{\dot \alpha}
\bar{\cal W}^{\dot\alpha}) + \bar W)\,,
\end{eqnarray}
where the superpotential is
\begin{equation}	%4.2
W = \tilde{Q}_i(m+\Phi_3)Q^i+{1\over 3}\,\epsilon^{IJK}
Tr[\Phi_I\Phi_J\Phi_K]\,.
\end{equation}
In Eq.~(\ref{LFT}) we are working in ${\cal N}=1$ language, where
$Q_i$, ($\tilde{Q}_i$) $i=1,\ldots, N_f$ are the chiral (antichiral)
superfields in the hypermultiplet, while $\Phi_I$ are the adjoint
scalars of ${\mathcal{N}}=4$ SYM once complexified as
$\Phi_1=X^1+iX^{2}$, $\Phi_2=X^3+iX^{4}$ and $\Phi_3=X^5+iX^{6}$;
where $X^I$ ($I=1,\ldots,6$) is the scalar which corresponds to the
direction $I+3$ in the array (\ref{DpDp+4intersection}). It is worth
mentioning that an identity matrix in color space is to be understood
to multiply the mass parameter of the quarks $m$.

We are interested in the classical SUSY vacua of this theory, which
can be obtained by imposing the corresponding $D$- and $F$-flatness
conditions following from the Lagrangian (\ref{LFT}). The vanishing of
the $F$-terms corresponding to the quark hypermultiplets amounts to
set:
\begin{equation}	%4.3
\tilde{Q}_i(\Phi_3+m)=0\,, \quad (\Phi_3+m)Q^i=0\, .
\label{Phi3}
\end{equation}
In turn, the vanishing of the $F$-terms associated to the adjoint
scalars gives rise to
\begin{equation}	%4.4
[\Phi_1,\Phi_3] = [\Phi_2,\Phi_3]=0\,,
\end{equation}
together with the equation
\begin{equation}	%4.5
\label{F3}
Q^i\tilde{Q}_i+[\Phi_1,\Phi_2]=0\,.
\end{equation}
In (\ref{F3}) $Q^i\tilde{Q}_i$ denotes a matrix in color space of
components $Q^i_{\alpha}\tilde{Q}_i^{\beta}$.

In addition to the $F$-flatness condition, we also have to impose the
$D$-flatness
\begin{equation}	%4.6
\label{D}
|Q^i|^2-|\tilde{Q}_i|^2+[\Phi_1,\Phi_1^{\dagger}] + 
[\Phi_2,\Phi_2^{\dagger}]=0\,.
\end{equation}
Note that $|Q^i|^2=Q^i_{\alpha}(Q^i_{\beta})^{\dagger}$ is also a
matrix in color space, as well as the commutator of the $\Phi$
fields, since they are in the adjoint representation of the $SU(N)$
gauge group.

Let us start considering the Coulomb branch of the theory, which
corresponds to setting all the $Q$, $\tilde{Q}$ are zero. This forces
to take the $\Phi_i$ as commuting matrices. In general, they will be
of the form
\begin{displaymath}
\Phi_i=\left(
\begin{array}{c c c c c}
 m_i^1 & 0 & \cdots & 0 & 0 \\ 0 & m_i^2 & \cdots & 0 & 0\\ \vdots & \vdots & \ddots & \vdots & \vdots \\ 0 & 0 &  \cdots & m_i^{N_c-1} & 0 \\ 0 & 0 & \cdots & 0 &  m_i^{N_c}
\end{array}
\right)\ .
\end{displaymath}
begin the $m_i^{\alpha}$ some constants. Motion along the Coulomb
branch amounts to consider different $m_i^{\alpha}$. Since the
eigenvalues of the adjoint fields have the interpretation of the
transverse coordinates to the color branes, changing the
$m_i^{\alpha}$ represents slight separations of the background $D3$;
so moving along the Coulomb branch amounts to changing the different
relative positions of each color brane. Actually, this \hbox{implies}
that in a generic point of the Coulomb branch we will have a broken
gauge group according to the particular brane separation pattern.

However, as shown in \cite{ARR2}, from (\ref{Phi3}) it is
clear that whenever some eigen\-values of the $\Phi_3$ matrix are set to
$-m$ we have the possibility of developing a nonzero value for the
$Q$, $\tilde{Q}$ in the corresponding entry of the matrix while still
satisfying the $F$-term flatness condition. Since we are giving a VEV
to some quark fields, we are entering the Higgs branch of the
theory. Indeed, in order to give this nonzero VEV's, we had to choose
in a given way the $\Phi_3$ eigenvalues, breaking the gauge group down
to some subgroup by moving in the Coulomb branch. Therefore, as it is
well-known, we see that we must to go to a particular point of the
Coulomb branch (namely that with some entries of $\Phi_3$ set to $-m$)
to have the possibility to develop a nonzero VEV for the quarks and
enter Higgs branch (see for example \cite{kutasov} for an
argumentation of this in a different context which we will shortly
review below). In general, we can go to the Higgs branch considering a
solution to (\ref{Phi3}) as
\begin{eqnarray}	%4.7
\Phi_3 = \left(\arraycolsep5pt\begin{array}{@{\hspace{1pt}}cccccc@{\hspace{1pt}}}
\tilde{m}_1& & & & & \\
&\ddots & & & \\
& & \tilde{m}_{N-k} & & & \\
& & & -m & & \\
& & & & \ddots & \\
& & & & & -m
\end{array}\right)\,,
\label{Phi3sol}
\end{eqnarray}
where the number of $m$'s is $k$. In order to have $\Phi_3$ in the Lie
algebra of $SU(N)$, one must have $\sum_{j=1}^{N-k}
\tilde{m}_j=km$. This choice of $\Phi_3$ lead us to take $Q^i$ and
$\tilde{Q}_i$ as
\begin{eqnarray}	%4.8
\tilde{Q}_i = (0 \cdots 0,\tilde{q}^1_i,\ldots,\tilde{q}^{k}_{i})\,,
\qquad Q^i = \left(\begin{array}{c} 
0 \\ 
\vdots \\ 
0 \\ 
q^i_1 \\ 
\vdots \\ 
q^i_{k}\end{array}\right)\,.
\label{QQtilde-sol}
\end{eqnarray}
Indeed, it is trivial to check that the values of $\Phi_3$,
$\tilde{Q}_i$ and $Q^i$ displayed in Eqs.~(\ref{Phi3sol}) and
(\ref{QQtilde-sol}) solve Eq.~(\ref{Phi3}). Since the quark VEV in
this solution has some components which are zero and others that are
different from zero, this choice of vacuum leads to a so-called mixed
Coulomb--Higgs phase.

For a vacuum election as in Eq.~(\ref{QQtilde-sol}) we can restrict
ourselves to the lower $k\times k$ matrix block, and we can write
Eq.~(\ref{F3}) as
\begin{equation}	%4.9
\label{F}
q^i\tilde{q}_i+[\Phi_1,\Phi_2]=0\,,
\end{equation}
where now, and it what follows, it is understood that $\Phi_1$ and
$\Phi_2$ are $k\times k$ matrices. In addition, we can write the
$D$-term restricted to just this $k\times k$ subspace inside the color
space as
\begin{equation}	%4.10
\label{Dpart}
|q^i|^2-|\tilde{q}_i|^2+[\Phi_1,\Phi_1^{\dagger}] + 
[\Phi_2,\Phi_2^{\dagger}]=0\,.
\end{equation}

The constraints (\ref{F}) and (\ref{Dpart}), together with the
condition $[\Phi^I,\Phi^3]=0$, define the mixed Coulomb--Higgs phase
of the theory.

\subsubsection*{Gravity dual of the mixed Coulomb--Higgs phase}
\label{MacroD3D7}

An important point coming from the field theory discussion is that in
order to enter the Higgs branch of the theory we need to go to a
particular point in the Coulomb branch. Given that the eigenvalues of
the adjoint fields represent the positions of the individual branes of
the color stack, this particular point on the Coulomb branch allowing
for the Higgs branch should correspond to having some of the color
branes away from the others and coincident in the $(8,9)$ directions,
which the $\Phi_3$ field represent. This branes precisely sit at a
distance $m$ to the origin, where the flavor $D7$ sits in the brane
construction. It is then natural to guess that these $D3$ are
dissolved as instantons in the flavor $D7$, with the quark VEV's being
the responsibles of the dissolution. Since as we argued dissolving
lower-dimensional branes is done via a nontrivial configuration of the
worldvolume gauge field, on the worldvolume of the flavor $D7$ there
should be an instantonic magnetic field. Indeed, as it is well-known,
there is a one-to-one correspondence between the Higgs phase of
${\cal{N}}=2$ gauge theories and the moduli space of instantons
\cite{MRD1,MRD2,W}. This comes from the fact that the $F$- and
$D$-flatness conditions can be directly mapped into the ADHM equations
(see \cite{Tong} and \cite{Tong2} for reviews). Because of
this map, we can identify the Higgs phase of the gauge theory with the
space of four-dimensional instantons; which, in the context at hand,
can be understood in terms of the instantonic worldvolume vector field
necessary to dissolve $D3$ inside the $D7$. Actually, this provides a
natural interpretation of the Higgs phase-ADHM equations map.

To summarize, the gravity dual of the Higgs phase of the field theory
above is realized in terms of a $D7$ brane with dissolved $D3$
representing the separation and further dissolution of some of the
color branes in the flavor ones. Therefore, once we take the
decoupling limit, the gravity dual of the field theory in the previous
subsection corresponds to the near-horizon of the color branes where
we should embed the flavor ones as probes. In the case at hand, the
corresponding background is $AdS_5\times S^5$, which also includes a
4-form RR potential given by
\begin{equation}	%4.11
C^{(4)} = \bigg(\frac{r^2}{R^2}\bigg)^2\ dx^0\wedge \cdots\wedge dx^3\,.
\label{C4}
\end{equation}
Then, in order to couple (\ref{C4}) to the worldvolume of the flavor
brane, we see that indeed we have to include the instantonic
worldvolume gauge field along the coordinates in the $D7$ transverse
to the $D3$. Let us concrete, and write the $AdS_5\times S^5$
background in a system of coordinates suitable for our purposes. Let
$\vec y = (y^1,\ldots, y^4)$ be the coordinates along the directions
$4,\ldots,7$ in the array (\ref{DpDp+4intersection}) and let us denote
by $\rho$ the length of $\vec y$ (i.e. $\rho^2 = \vec y\cdot\vec y$).
Moreover, we will call $\vec z=(z^1,z^2)$ the coordinates $8,9$ of
(\ref{DpDp+4intersection}). Notice that $\vec z$ is a vector in the
directions which are orthogonal to both stacks of D-branes. Clearly,
$r^2 = \rho^2 + \vec{z}^{2}$, so the metric can be written as
\begin{equation}	%4.12
\label{bckgr}
ds^2 = \frac{\rho^2+\vec{z}^{2}}{R^2}\,dx^2_{1,3} + 
\frac{R^2}{\rho^2+\vec{z}^{2}}\,(d\vec{y}^{2} + d\vec{z}^{2})\,.
\end{equation}

The DBI action for a stack of $N_f$ $D7$-branes with worldvolume
gauge field is given by
\begin{eqnarray}	%4.13
S_{\rm DBI}^{D7} = -T_7 \int d^8\xi\,e^{-\phi}
Str\big\{\sqrt{-\det(g+F)}\big\}\,,
\label{DBI-D7-general}
\end{eqnarray}
where $\xi^a$ is a system of worldvolume coordinates, $\phi$ is the
dilaton, $g$ is the induced metric and $F$ is the field strength of
the $SU(N_f)$ worldvolume gauge group.\footnote{Notice that, with our
notations, $F_{ab}$ is dimensionless and, therefore, the relation
between $F_{ab}$ and the gauge potential $A$ is
$F_{ab}=\partial_aA_b-\partial_bA_a+{1\over 2\pi\alpha'}[A_a,A_b]$,
whereas the gauge covariant derivative is $D_a=\partial_a+{1\over
2\pi\alpha'}A_a$.} Let us assume that we take $\xi^a = (x^{\mu},y^i)$
as worldvolume coordinates and that we consider a $D7$-brane embedding
in which $|\vec{z}|=L$, where $L$ represents the constant transverse
separation between the two stacks of $D3$- and $D7$-branes. Notice
that this transverse separation will give a mass $L/2\pi\alpha'$ to
the $D3-D7$ strings, which corresponds to the quark mass in the field
theory dual. For an embedding with $|\hspace{0.5pt}\vec z\hspace{0.5pt}|=L$, the
induced metric takes the form
\begin{eqnarray}	%4.14
g_{x^{\mu} x^{\nu}} = {\rho^2 + L^2\over R^2}\,\eta_{\mu\nu}\,,
\qquad g_{y^{i} y^{j}} = {R^2\over \rho^2 + L^2}\,\delta_{ij}\,.
\label{inducedgD3-D7}
\end{eqnarray}
Let us now assume that the worldvolume field strength $F$ has nonzero
entries only along the directions of the $y^i$ coordinates and let us
denote $F_{y^iy^j}$ simply by $F_{ij}$. Then, after using
Eq.~(\ref{inducedgD3-D7}) and the fact that the dilaton is trivial for
the%\break
$AdS_5\times S^5$ background, the DBI action
(\ref{DBI-D7-general}) takes the form
\begin{equation}	%4.15
S_{\rm DBI}^{D7} = -T_{7}\int d^{\hspace{0.5pt}4}x\,d^{\hspace{0.5pt}4}y Str
\Bigg\{\sqrt{\det\bigg(\delta_{ij} + 
\bigg(\frac{\rho^2 + L^2}{R^2}\bigg)F_{ij}\bigg)}\,\Bigg\}\,.
\label{DBI-D3D7-reduced}
\end{equation}
The matrix appearing on the r.h.s. of Eq.~(\ref{DBI-D3D7-reduced}) is
a $4\times 4$ matrix whose entries are $SU(N_f)$ matrices. However,
inside the symmetrized trace such matrices can be considered as
commutative numbers. Actually, we will evaluate the determinant in
(\ref{DBI-D3D7-reduced}) by means of the following identity. Let
$M_{ij}=-M_{ji}$ be a $4\times 4$ antisymmetric matrix. Then, one can
check that
\begin{eqnarray}	%4.16
\det (1 + M) = 1 + {1\over 2}\,M^2 + {1\over 16}\,({}^*MM)^2\,,
\label{matrix-identity}
\end{eqnarray}
where $M^2$ and ${}^*MM$ are defined as follows:
\begin{eqnarray}	%4.17
M^2 \equiv M_{ij} M_{ij}\,, \qquad {}^*M M \equiv {}^*M_{ij}M_{ij}\,,
\label{MM}
\end{eqnarray}
and ${}^*M$ is defined as the following matrix:
\begin{eqnarray}	%4.18
{}^*M_{ij} = {1\over 2}\,\epsilon_{ijkl}M_{kl}\,.
\label{*M}
\end{eqnarray}
When the $M_{ij}$ matrix is self-dual (i.e. when ${}^*M=M$), the
three terms on the r.h.s. of (\ref{matrix-identity}) build up a
perfect square:
\begin{eqnarray}	%4.19
\det (1 + M)|_{self-dual} = \bigg(1 + {1\over 4}\,M^2\bigg)^2\,.
\label{detM(s-d)}
\end{eqnarray}

Let us consider a configuration in which the worldvolume gauge field
is self-dual in the internal $R^4$ of the worldvolume spanned by the
$y^i$ coordinates which, as one can check, satisfies the equations of
motion of the $D7$-brane probe. For such an instantonic gauge
configuration ${}^*F=F$, where ${}^*F$ is defined following
Eq.~(\ref{*M}). Using the expression in Eq.~(\ref{detM(s-d)}), we can
write
\begin{eqnarray}	%4.20
S_{\rm DBI}^{D7}\hbox{(self-dual)} = 
-T_7\int d^{\hspace{0.5pt}4}x\,d^{\hspace{0.5pt}4}y Str\bigg\{1 + \frac{1}{4}
\bigg(\frac{\rho^2 + L^2}{R^2}\bigg)^{2}{}^*FF\bigg\}\,.
\label{SDBI-sd}
\end{eqnarray}

In turn, the WZ piece of the worldvolume action reduces in this case
to
\begin{eqnarray}	%4.21
S_{\rm WZ}^{D7} = {T_{7}\over 2}\int
Str\big[P\big[C^{(4)}\big] \wedge F\wedge F\big]\,,
\label{WZactionD7}
\end{eqnarray}
By using the same set of coordinates as in (\ref{DBI-D3D7-reduced}),
and the explicit expression of $C^{(4)}$ (see Eq.~(\ref{C4})), one
can rewrite $S_{\rm WZ}^{D7}$ as
\begin{equation}	%4.22
\label{WZ}
S_{\rm WZ}^{D7} = T_{7}\int d^{\hspace{0.5pt}4}x\,d^{\hspace{0.5pt}4}y Str 
\bigg\{\frac{1}{4}\bigg(\frac{\rho^2 + L^2}{R^2}\bigg)^{2}{}^*FF\bigg\}\,.
\end{equation}

Remarkably, once we assume the instantonic character of $F$, the WZ
term partially cancels the DBI giving
\begin{eqnarray}	%4.23
S^{D7}\hbox{(self-dual)} = -T_7 \int d^{\hspace{0.5pt}4}x\,d^{\hspace{0.5pt}4}y
Str[1] = -T_7 N_f \int d^{\hspace{0.5pt}4}x\,d^{\hspace{0.5pt}4}y\,.
\label{totalaction}
\end{eqnarray}
Notice that in the total action (\ref{totalaction}) the transverse
distance $L$ does not appear. This ``no-force" condition is an
explicit manifestation of the SUSY of the system. Indeed, the fact
that the DBI action is a square root of a perfect square is required
for supersymmetry, and actually can be regarded as the saturation of a
BPS bound. Furthermore, had we changed the sign of the WZ by
considering an antibrane rather than a brane, we would have had an
explicit appearance of $L$, breaking the no-force condition since the
system would be nonsupersymmetric.

In order to get a proper interpretation of the role of the instantonic
gauge field on the $D7$-brane probe, let us recall that for self-dual
configurations the integral of the Pontryagin density ${\cal P}(y)$ is
quantized for topological reasons. Actually, with our present
normalization of $F$, ${\cal P}(y)$ is given by
\begin{eqnarray}	%4.24
{\cal P}(y) \equiv \frac{1}{16\pi^2}\,{1\over (2\pi\alpha')^2}Tr[{}^*FF]\,,
\label{Poyntriagin}
\end{eqnarray}
and, if $k\in Z$ is the instanton number, one has
\begin{equation}	%4.25
\int d^{\hspace{0.5pt}4}y\,{\cal P}(y) = k\,.
\label{instanton-number}
\end{equation}
A worldvolume gauge field satisfying (\ref{instanton-number}) is
inducing $k$ units of $D3$-brane charge into the $D7$-brane
worldvolume along the subspace spanned by the Minkowski coordinates
$x^{\mu}$. To verify this fact, let us rewrite the WZ action
(\ref{WZactionD7}) of the $D7$-brane~as
\begin{equation}	%4.26
S_{\rm WZ}^{D7} = {T_{7}\over 4} \int d^{\hspace{0.5pt}4}x\,d^{\hspace{0.5pt}4}y\,
C_{x^0x^1x^2x^3}^{(4)} Tr[{}^*FF] = T_3\int
d^{\hspace{0.5pt}4}x\,d^{\hspace{0.5pt}4}y\,C_{x^0x^1x^2x^3}^{(4)}{\cal P}(y)\,,
\label{D3induced}
\end{equation}
where we have used (\ref{Poyntriagin}) and the relation
$T_3=(2\pi)^4\,(\alpha')^2\,T_7$ between the tensions of the $D3$- and
$D7$-branes. If $C_{x^0x^1x^2x^3}^{(4)}$ does not depend on the
coordinate $y$, we can integrate over $y$ by using
Eq.~(\ref{instanton-number}), namely
\begin{eqnarray}	%4.27
S_{\rm WZ}^{D7} = k T_3 \int d^{\hspace{0.5pt}4}x\,C_{x^0x^1x^2x^3}^{(4)}\,.
\label{D3charge}
\end{eqnarray}
Equation~(\ref{D3charge}) shows that the coupling of the $D7$-brane
with $k$ instantons in the worldvolume to the RR potential $C^{(4)}$
of the background is identical to the one corresponding to $k$
$D3$-branes, as claimed above. It is worth to remark here that the
existence of these instanton configurations relies on the fact that we
are considering $N_f>1$ flavor $D7$ branes, i.e. that we have a
non-Abelian worldvolume gauge theory.

\subsubsection*{Recovering the field theory picture from the 
microscopical interpretation of the $D3-D7$ intersection with flux}

The fact that the $D7$-branes carry $k$ dissolved $D3$-branes on them
opens up the possibility of a new perspective on the system, which
could be regarded not just from the point of view of the $D7$-branes
with dissolved $D3$s, but also from the point of view of the dissolved
$D3$-branes which expand due to dielectric effect \cite{M} to a
transverse fuzzy $R^4$ (see the appendix for a very short
introduction to the dielectric effect). From this point of view, the
$D7$ appears as an effective description, which we will call
``macroscopic," while the picture in terms of blown-up $D3$ will be
called ``microscopic." Going back to the field theory description, the
fact that, once we are in the adequate point in the Coulomb branch, we
enter the fields $\Phi_1, \Phi_2$ are matrix-valued suggest precisely
that we can think those $D3$ at $m$ distance to expand dielectrically
to an effective $D7$ brane. To see this, let us assume that we have a
stack of $k$ $D3$-branes in the background given by
(\ref{bckgr}). These $D3$-branes are extended along the four Minkowski
coordinates $x^{\mu}$, whereas the transverse coordinates $\vec{y}$
and $\vec{z}$ must be regarded as the matrix scalar fields $Y^i$ and
$Z^j$, taking values in the adjoint representation of
$SU(k)$. Actually, we will assume in what follows that the $Z^j$
scalars are Abelian, as it corresponds to a configuration in which the
$D3$-branes are localized (i.e. not polarized) in the space
transverse to the $D7$-brane.

The dynamics of a stack of coincident $D3$-branes is determined by the
Myers dielectric action \cite{M} (see Appendix), which is the sum of a
Dirac--Born--Infeld and a Wess--Zumino part:
\begin{eqnarray}	%4.28
\label{Myers}
S_{D3} = S_{\rm DBI}^{D3} + S_{\rm WZ}^{D3}\,.
\end{eqnarray}
For the background we are considering the DBI action is
\begin{eqnarray}	%4.29
\label{MyersDBI}
S_{\rm DBI}^{D3} = -T_3 \int d^{\hspace{0.5pt}4}\xi Str
\Big[\sqrt{-\det[P[G+G(Q^{-1}-\delta)G]_{ab}]}\sqrt{\det Q}\,\Big]\,,
\end{eqnarray}
%\vskip-\lastskip
%\pagebreak

\noindent
In Eq.~(\ref{MyersDBI}) $G$ is the background metric, ${\bf
Str}(\cdots)$ represents the symmetrized trace over the $SU(k)$
indices and $Q$ is a matrix which depends on the commutator of the
transverse scalars (see below). The WZ term for the $D3$-brane in the
$AdS_5\times S^5$ background under consideration is
\begin{eqnarray}	%4.30
\label{couplingD3D7}
S_{\rm WZ}^{D3} = T_{3} \int d^{\hspace{0.5pt}4}\xi Str\big[P\big[C^{(4)}\big]\big]\,.
\end{eqnarray}
As we are assuming that only the $Y$ scalars are noncommutative, the
only elements of the matrix $Q$ appearing in (\ref{MyersDBI}) that
differ from those of the unit matrix are given by
\begin{eqnarray}	%4.31
Q_{y^iy^j} = \delta_{ij} + {i\over 2\pi\alpha'}[Y^i,Y^k]G_{y^ky^j}\,.
\end{eqnarray}
By using the explicit form of the metric elements along the $y$
coordinates (see Eq.~(\ref{bckgr})), one can rewrite $Q_{ij}$ as
\begin{eqnarray}	%4.32
Q_{y^iy^j} = \delta_{ij} + {i\over 2\pi\alpha'}\,
{R^2\over \hat r^{2}}\,[Y^i,Y^j]\,,
\end{eqnarray}
where $\hat r^{\,2}$ is the matrix
\begin{eqnarray}	%4.33
\hat r^{2} = (Y^i)^2 + Z^2\,.
\end{eqnarray}
Let us now define the matrix $\theta_{ij}$ as
\begin{eqnarray}	%4.34
i\theta_{ij} \equiv {1\over 2\pi\alpha'}\,[Y^i,Y^j]\,.
\label{thetaij}
\end{eqnarray}
It follows from this definition that $\theta_{ij}$ is antisymmetric in
the $i,j$ indices and, as an $SU(k)$ matrix, is Hermitian:
\begin{eqnarray}	%4.35
\theta_{ij} = -\theta_{ji}\,, \qquad \theta_{ij}^{\dagger} = \theta_{ij}\,.
\end{eqnarray}
The algebra given by (\ref{thetaij}) defines a fuzzy $R^4$ (see for
example \cite{ND}). The appearance of this algebra should be
expected, since from the macroscopical picture we expect the $D3$ to
polarize into a transverse $R^4$ giving rise to the effective $D7$.

Moreover, in terms of $\theta_{ij}$, the matrix $Q_{ij}$ can be
written as
\begin{eqnarray}	%4.36
Q_{y^iy^j} = \delta_{ij} - {R^2\over \hat r^{2}}\theta_{ij}\,.
\end{eqnarray}
Using these definitions, we can write the DBI action (\ref{MyersDBI})
for the dielectric $D3$-brane in the $AdS_5\times S^5$ background as
\begin{eqnarray}	%4.37
S_{\rm DBI}^{D3} = -T_3 \int d^{\hspace{0.5pt}4}x\ \, {\bf Str}  
\Bigg[\bigg({\hat r^{2}\over R^2}\bigg)^2\sqrt{\det\bigg(\delta_{ij} - 
{R^2\over \hat r^{2}}\theta_{ij}\bigg)}\,\Bigg]\,,
\label{DBI-dielectricD3}
\end{eqnarray}
%\vskip-\lastskip
%\pagebreak

\noindent
where we have chosen the Minkowski coordinates $x^{\mu}$ as our set of
worldvolume coordinates for the dielectric $D3$-brane. Similarly, the
WZ term can be written as
\begin{eqnarray}	%4.38
S_{\rm WZ}^{D3} = T_{3} \int d^{\hspace{0.5pt}4}x\ \, {\bf Str}
\bigg[\bigg({\hat r^{2}\over R^2}\bigg)^2\bigg]\,.
\label{SWZ-D3}
\end{eqnarray}
Let us now assume that the matrices $\theta_{ij}$ are self-dual with
respect to the $ij$ indices, i.e. that ${}^*\theta = \theta$. Notice
that, in terms of the original matrices $Y^i$, this is equivalent to
the condition
\begin{equation}	%4.39
\label{selfduality}
[Y^i,Y^j]=\frac{1}{2}\epsilon_{ijkl}[Y^k,Y^l]\,. 
\end{equation}

Moreover, the self-duality condition implies that there are three
independent $\theta_{ij}$ matrices, namely
\begin{eqnarray}	%4.40
\theta_{12} = \theta_{34}\,,\qquad \theta_{13} = \theta_{42}\,,
\qquad \theta_{14} = \theta_{23}\,.
\label{sd-explicit}
\end{eqnarray}

The description of the $D3-D7$ system from the perspective of the
color $D3$-branes should match the field theory analysis performed at
the beginning of this section. In particular, the $D$- and
$F$-flatness conditions of the adjoint fields in the Coulomb--Higgs
phase of the ${\cal N}=2$ SYM with flavor should be the same as the
ones satisfied by the transverse scalars of the dielectric
$D3$-brane. Let us define the~following complex combinations of the
$Y^i$ matrices:
\begin{eqnarray}	%4.41
2\pi\alpha' \Phi_1 \equiv {Y^1+iY^2\over \sqrt{2}}\,,
\qquad 2\pi\alpha' \Phi_2 \equiv {Y^3+iY^4\over \sqrt{2}}\,,
\label{Phi12}
\end{eqnarray}
where we have introduced the factor $2\pi\alpha'$ to take into account
the standard relation between coordinates and scalar fields in string
theory. We are going to \hbox{identify} $\Phi_1$ and $\Phi_2$ with the
adjoint scalars of the field theory side. From the definitions
(\ref{thetaij}) and (\ref{Phi12}) and the self-duality condition
(\ref{sd-explicit}), it is straightforward to\break check that 
\begin{eqnarray}	%4.42
&&[\,\Phi_1\,,\Phi_2\,]\,=\,-{\theta_{23}\over 2\pi\alpha'}
\,+\,i{\theta_{13}\over 2\pi\alpha'}
\,\,,\\
&&[\,\Phi_1\,,\Phi_1^{\dagger}\,]\,=\,[\,\Phi_2\,,\Phi_2^{\dagger}\,]\,=\,
{\theta_{12}\over 2\pi\alpha'}\,\,.
\end{eqnarray}
By comparing with the results of the field theory analysis
(Eqs.~(\ref{F}) and (\ref{D})), we get the following identifications
between the $\theta$'s and the vacuum expectation values of the matter
fields:
\begin{eqnarray}	%4.43
q^i\tilde{q}_i = {\theta_{23} \over 2\pi\alpha'} - i
{\theta_{13}\over 2\pi\alpha'}\,, \qquad
|\tilde{q}_i|^2 - |q^i|^2 = {\theta_{12}\over \pi\alpha'}\,.
\label{q-theta}
\end{eqnarray}
\vskip-\lastskip
\pagebreak

\noindent
Moreover, from the point of view of this dielectric description, the
$\Phi_3$ field in the field theory is proportional to
$Z^1+iZ^2$. Since the stack of branes is localized in that directions,
$Z^1$ and $Z^2$ are Abelian and clearly we have that
$[\Phi_1,\Phi_3]=[\Phi_2,\Phi_3]=0$, thus matching the last
$F$-flatness condition for the adjoint field $\Phi_3$.

It is also interesting to relate the present microscopic description
of the $D3-D7$ intersection, in terms of a stack of dielectric
$D3$-branes, to the macroscopic description, in terms of the flavor
$D7$-branes. With this purpose in mind, let us compare the actions of
the $D3$- and $D7$-branes. First of all, we notice that, when the
matrix $\theta$ is self-dual, we can use Eq.~(\ref{detM(s-d)}) and
write the DBI action (\ref{DBI-dielectricD3}) as
\begin{eqnarray}	%4.44
S_{\rm DBI}^{D3}\hbox{(self-dual)} = -T_3\int d^{\hspace{0.5pt}4}x\ \, {\bf Str}
\bigg[\bigg({\hat r^{2}\over R^2}\bigg)^2 + {1\over 4}\,\theta^2\bigg]\,.
\label{SD3-sd}
\end{eqnarray}
Moreover, by inspecting Eqs.~(\ref{SWZ-D3}) and (\ref{SD3-sd}) we
discover that the WZ action cancels against the first term of the
right-hand side of (\ref{SD3-sd}), in complete analogy to what happens
to the $D7$-brane. Thus, one has
\begin{equation}	%4.45
S^{D3}\hbox{(self-dual)} = -{T_3\over 4}\int d^{\hspace{0.5pt}4}x\, {\bf Str}
[\theta^2] = - \pi^2T_7(2\pi\alpha')^2\int d^{\hspace{0.5pt}4}x\, {\bf Str}[\theta^2]\,,
\label{completeactionD3}
\end{equation}
where, in the last step, we have rewritten the result in terms of the
tension of the $D7$-brane. Moreover, an important piece of information
is obtained by comparing the WZ terms of the $D7$- and $D3$-branes
(Eqs.~(\ref{D3induced}) and (\ref{SWZ-D3})). Actually, from this
comparison we can establish a map between matrices in the $D3$-brane
description and functions of the $y$ coordinates in the $D7$-brane
approach. Indeed, let us suppose that $\hat f$ is a $k\times k$ matrix
and let us call $f(y)$ the function to which $\hat f$ is mapped. It
follows from the identification between the $D3$- and $D7$-brane WZ
actions that the mapping rule is
\begin{equation}	%4.46
\, {\bf Str}[\hat f] \Rightarrow \int d^{\hspace{0.5pt}4}y\, {\cal P}(y)f(y)\,,
\label{micro-macro}
\end{equation}
where the kernel ${\cal P}(y)$ on the r.h.s. of (\ref{micro-macro}) is
the Pontryagin density defined in Eq.~(\ref{Poyntriagin}). Actually,
the comparison between both WZ actions tells us that the matrix $\hat
r^2$ is mapped to the function $\vec y^{\,2}+\vec z^{\,2}$. Notice
also that, when $\hat f$ is the unit $k\times k$ matrix and $f(y)=1$,
both sides of (\ref{micro-macro}) are equal to the instanton number
$k$ (see Eq.~(\ref{instanton-number})). Another interesting
information comes by comparing the complete actions of the $D3$- and
$D7$-branes. It is clear from (\ref{completeactionD3}) and
(\ref{totalaction}) that
\begin{eqnarray}	%4.47
(2\pi\alpha')^2\, {\bf Str}[\theta^2] \Rightarrow 
\int d^{\hspace{0.5pt}4}y\,{N_f\over \pi^2}\,.
\label{theta-map}
\end{eqnarray}
By comparing Eq.~(\ref{theta-map}) with the general relation
(\ref{micro-macro}), one gets the function that corresponds to the
matrix $\theta^2$, namely
\begin{eqnarray}	%4.48
(2\pi\alpha')^2 \theta^2 \Rightarrow {N_f\over \pi^2{\cal P}(y)}\,.
\label{theta-instanton}
\end{eqnarray}
Notice that $\theta^2$ is a measure of the noncommutativity of the
adjoint scalars in the dielectric approach, i.e. is a quantity that
characterizes the fuzziness of the space transverse to the
$D3$-branes. Equation~(\ref{theta-instanton}) is telling us that this
fuzziness is related to the (inverse of the) Pontryagin density for
the macroscopic $D7$-branes. Actually, this identification is
reminiscent of the one found in \cite{SW} between the
noncommutative parameter and the NSNS $B$-field in the string theory
realization of noncommutative geometry. Interestingly, in our case the
commutator matrix $\theta$ is related to the VEV of the matter fields
$q$ and $\tilde q$ through the $F$- and $D$-flatness conditions
(\ref{F}) and (\ref{D}). Notice that Eq.~(\ref{theta-instanton})
implies that the quark VEV is somehow related to the instanton density
on the flavor brane. In order to make this correspondence more
precise, let us consider the one-instanton configuration of the
$N_f=2$ gauge theory on the $D7$-brane worldvolume. In the so-called
singular gauge, the $SU(2)$ gauge field is given by
\begin{equation}	%4.49
{A_i\over 2\pi\alpha'} = 2i\Lambda^2\frac{\bar{\sigma}_{ij}y^j}
{\rho^2(\rho^2+\Lambda^2)}\,,
\label{instanton-potential}
\end{equation}
where $\rho^2=\vec y\cdot\vec y$, $\Lambda$ is a constant (the
instanton size) and the matrices $\bar{\sigma}_{ij}$ are defined as
\begin{eqnarray}	%4.50
\bar{\sigma}_{ij} = {1\over 4}\,
(\bar\sigma_i\sigma_j - \bar\sigma_j\sigma_i)\,,\quad
\sigma_i = (i\vec \tau,1_{2\times 2})\,,\quad 
\bar\sigma_i = \sigma_i^{\dagger} = (-i\vec \tau,1_{2\times 2})\,.\quad
\label{sigmaij}
\end{eqnarray}
In (\ref{sigmaij}) the $\vec \tau$'s are the Pauli matrices. Notice
that we are using a convention in which the $SU(2)$ generators are
Hermitian as a consequence of the relation
$\bar{\sigma}_{ij}^{\dagger} = -\bar{\sigma}_{ij}$. The non-Abelian
field strength $F_{ij}$ for the gauge potential $A_i$ in
(\ref{instanton-potential}) can be easily computed, with the result
\begin{eqnarray}	%4.51
{F_{ij}\over 2\pi\alpha'} = -{4i\Lambda^2\over (\rho^2 + \Lambda^2)^2}\,
\bar{\sigma}_{ij} - {8i\Lambda^2\over \rho^2(\rho^2 + \Lambda^2)^2}
(y^i\bar{\sigma}_{jk} - y^j\bar{\sigma}_{ik})y^k\,.
\end{eqnarray}
Using the fact that the matrices $\bar{\sigma}_{ij}$ are
anti-self-dual one readily verifies that $F_{ij}$ is
self-dual. Moreover, one can prove that
\begin{eqnarray}	%4.52
{F_{ij} F_{ij}\over (2\pi\alpha')^2} = 
{48 \Lambda^4\over (\rho^2 + \Lambda^2)^4}\,,
\label{F2-inst}
\end{eqnarray}
which gives rise to the following instanton density:
\begin{eqnarray}	%4.53
{\cal P}(y) = {6\over \pi^2}\,{\Lambda^4\over (\rho^2 + \Lambda^2)^4}\,.
\end{eqnarray}
As a check one can verify that Eq.~(\ref{instanton-number}) is
satisfied with $k=1$.

Let us now use this result in (\ref{theta-instanton}) to get some
qualitative understanding of the relation between the Higgs mechanism
in field theory and the instanton density in its holographic
description. For simplicity we will assume that all quark VEV's are
proportional to some scale $v$, i.e. that
\begin{eqnarray}	%4.54
q, \ \tilde q \sim v\,.
\end{eqnarray}
Then, it follows from (\ref{q-theta}) that
\begin{eqnarray}	%4.55
\theta \sim \alpha' v^2\,,
\end{eqnarray}
and, by plugging this result in (\ref{theta-instanton}) one arrives at
the interesting relation
\begin{equation}	%4.56
v\sim \frac{\rho^2+\Lambda^2}{\alpha'\Lambda}\, .
\label{holoVEV}
\end{equation}
Equation~(\ref{holoVEV}) should be understood in the holographic
sense, i.e. $\rho$ should be regarded as the energy scale of the
gauge theory. Actually, in the far IR ($\rho\approx 0$) the relation
(\ref{holoVEV}) reduces to
\begin{equation}	%4.57
v \sim \frac{\Lambda}{\alpha'}\,,
\label{v-Lambda}
\end{equation}
which, up to numerical factors, is precisely the relation between the
quark VEV and the instanton size that has been obtained in
\cite{EGG}. Let us now consider the full expression
(\ref{holoVEV}) for $v$. For any finite nonzero $\rho$ the quark VEV
$v$ is nonzero. Indeed, in both the large and small instanton limits
$v$ goes to infinity. However, in the far IR a subtlety arises, since
there the quark VEV goes to zero in the small instanton limit. This
region should be clearly singular, because a zero quark VEV would mean
to unhiggs the theory, which would lead to the appearance of extra
light degrees of freedom. This will have interesting consequences in
the meson spectrum of the theories.

Finally, let us notice that the dielectric effect considered here is
not triggered by the influence of any external field other than the
metric background. This explicitly shows up in (\ref{couplingD3D7}),
where the CS coupling in the $D3$ worldvolume is the sum of the
individual CS of each brane composing the stack, with no need of the
non-Abelian character of the stack. In this sense it is an example of
a purely gravitational dielectric effect, as in
\cite{gravdielec} and \cite{gravdielec2}.

\subsubsection*{Another UV completion}

It is interesting to compare the results we have presented with other
ways of embedding the same field theory in string theory. It is well
known that the ${\mathcal{N}}=2$ field theory dual to the $D3-D7$
intersection can be engineered in a different way by means of a web of
branes (for a detailed review of these issues see
\cite{kutasov}). Consider the following configuration in the
IIA theory:
\begin{eqnarray*}
\arraycolsep5pt\begin{array}{rcccccccccl}
& 1 & 2 & 3 & 4 & 5 & 6 & 7 & 8 & 9 & \\[2pt]
NS5: & \times & \times & \times & - & \times & \times & - & - & - \\[2pt]
NS5': & \times & \times & \times & - & \times & \times & - & - & - & \\[2pt]
D4: & \times & \times & \times & \times & - & - & - & - & - &
\end{array}
\end{eqnarray*}
Here the $D4$ branes are suspended between the parallel $NS$ and $NS'$
a distance $l$. Since the $NS$ branes are very massive objects, the
low energy description is in terms of the worldvolume gauge theory on
the $D4$. For energies below $l$, the theory is effectively
$(3+1)$-dimensional, and reduces to a ${\mathcal{N}}=2$ pure gauge
theory, whose gauge coupling is given by $g\sim l^{-1}$. The positions
of the $D4$ branes in the $5,6$ directions parametrize the Coulomb
branch moduli space. When all the $D4$ coincide, the gauge group is
$SU(N_c)$, while when separating them we break it in a pattern given
by the separation.

One can add flavors to this theory by adding a new sector of $D4$
branes and ending on a $D6$ brane perpendicular to the other $NS$:
\begin{eqnarray*}	
\arraycolsep5pt\begin{array}{rcccccccccl}
& 1 & 2 & 3 & 4 & 5 & 6 & 7 & 8 & 9 & \\[2pt]
NS5: & \times & \times & \times & - & \times & \times & - & - & -\\[2pt]
NS5': & \times & \times & \times & - & \times & \times & - & - & - & \\[2pt]
D4_c: & \times & \times & \times & \times & - & - & - & - & - & \\[2pt]
D6: & \times & \times & \times & - & - & - & \times & \times & \times & \\[2pt]
D4_f: & \times & \times & \times & \times & - & - & - & - & - &
\end{array}
\end{eqnarray*}

\begin{figure}[th]
\centerline{\hskip -.8in \epsffile{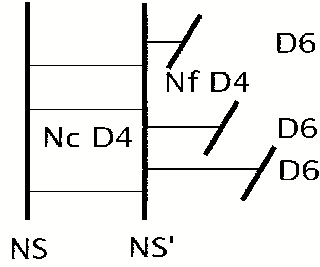}}
\vspace*{8pt}
\caption{Web of branes giving rise to the theory of interest.\protect
\label{fig2}}
\end{figure}

The low energy description is nothing but the same field theory given
by (\ref{LFT}). However, the construction is different, and it
corresponds to a different UV completion to the one so far
considered. However, this construction gives a very nice intuition of
what is going on. The matter sector comes from the open string sector
connecting the color and the flavor $D4$, and therefore, the masses
are given by the separation between two $D4$ at each side of the
$NS'$. As in the unflavored case, the positions of the $D4$ correspond
to the eigenvalues of the adjoint fields. Therefore, motion along the
Coulomb branch corresponds to moving the $D4$ inside the $NS$. If all
of the $D4$ coincide, we clearly have an unbroken $SU(N_c)$, while
separating the branes breaks the gauge group.

When two $D6$ are at the same point in $5,6$, we have the possibility
of breaking the $D4$ connecting the $NS'$ and the farest $D6$ in a
piece between the $NS'$ and the nearest $D6$ and another piece between
the $D6$, which can freely move in $7,8,9$. This excites some open
string fields giving VEV to the quark hypermultiplets. Note that
having two $D6$ at the same point corresponds to having two quark
hypermultiplets with the same mass. However, this is way we would have
the nearest $D6$ with two $D4$ connecting it to the $NS'$, which, by
the so-called {\it $s$-rule} \cite{HW}, is not supersymmetric. In
order to solve this, we can bring one of the color $D4$ and reconnect
one of those $D4$ with it, so each $NS-D6$ is connected by a single
$D4$. This corresponds to the Higgs branch of the theory.

\begin{figure}[h]
\vspace{1cm}
\centerline{\hskip -.8in \epsffile{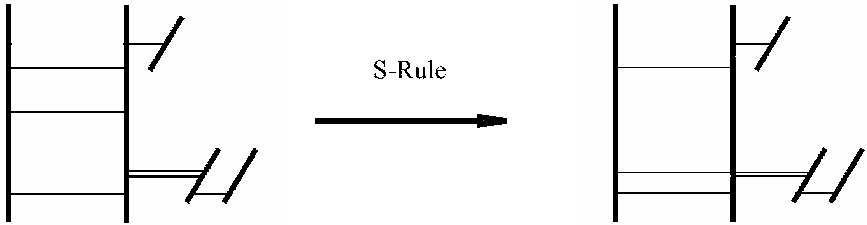}}
\vspace*{8pt}
\caption{The brane moving between $D6$ excites some VEV for the quarks. Using the s-rule demands picking a particular point in the Coulomb branch moduli space.}\protect
\end{figure}

In this picture, it is also clear that one has to go to a particular
point of the Coulomb branch in which one of the color $D4$ is aligned
with a flavor $D4$ so that we can have the recombination allowing for
a supersymmetric Higgs branch. If we denote by $a^{\alpha}$ the
positions of the color $D4$ and $m_i$ the positions of the flavor
$D4$, the effective masses of the quarks are given by
$M_i^{\alpha}=|a^{\alpha}-m_i|$. Consider first the case in which we
have the full unbroken gauge group, and assume all the quark masses
equal $m_i=m$. Then we can choose $a^{\alpha}=0$ and
$M_i^{\alpha}=m$. This corresponds to having all the $D6$ at the same
point in $5,6$ at a distance $m$ of the stack containing all the color
$D4$. We can cut $k$ of the flavor $D4$ ending on a far brane and
force them to end on a nearer brane. Then, we can move the resulting
$D4$ pieces between the $D6$ giving VEV to some quarks. However, this
would break the s-rule, so we need to pick $k$ of the color $D4$ and
recombine them with the broken $D4$ so that each $NS$ is connected to
each $D6$ with just one $D4$ according to the s-rule. But this will
break the gauge group, since we are moving away $k$ color
branes. Furthermore, in order to do this, we have to align the $D4$
branes, so we have to put the $k$ $D4$ at $a=m$ each aligned with a
different cut flavor brane, in very much the same spirit as in the
picture we found in our holographic setup, in which we also had to
move $k$ of the color branes to recombine them with the flavor
ones. It is interesting to note that also in this brane web setup the
Higgs branch is realized by means of the recombination of the color
branes with the flavor ones.

\subsubsection{Fluctuations in $Dp-D(p+4)$ with flux}

So far we have seen how we can realize the Higgs branch of the theory
in the gravity side by means of dissolving some of the color branes in
the flavor ones; providing an explicit route between field theory and
gravity by means of the dielectric description. We will now turn back
to the macroscopical description, and we will consider fluctuations
around the instanton configuration; which will correspond to the meson
spectrum in the Higgs branch of the gauge theory.

Since we have a similar situation for all the $Dp-D(p+4)$
intersections, namely a one to one correspondence between the Higgs
phase of the corresponding field theory and the moduli space of
instantons in four dimensions, in this section we will work with the
general $Dp-D(p+4)$ system. Both the macroscopic and the microscopic
analysis of the previous section can be extended in a straightforward
manner to the general case, so we will first briefly sketch the
macroscopical computation to set notations, and turn to the
fluctuations.

The metric background (\ref{NHspaceDp}) contains a dilaton given by
(\ref{dilaton}) and a RR 4-form potential given by
\begin{equation}	%4.58
\label{DpRR}
C^{(p+1)}=\bigg(\frac{r^2}{R^2}\bigg)^{\alpha}dx^0\wedge\cdots\wedge dx^p\,,
\end{equation}
where $\alpha$ is that in (\ref{Dpgammas}).

We will separate again the $\vec{r}$ coordinates in two sets, namely
$\vec r=(\vec y, \vec z)$, where $\vec y$ has four components, and we
will denote $\rho^2=\vec y\cdot\vec y$. As $r^2=\rho^2+\vec{z}^{2}$,
the metric can be written as
\begin{equation}	%4.59
\label{bckgrDp}
ds^2 = \bigg(\frac{\rho^2+\vec{z}^{2}}{R^2}\bigg)^{\alpha}dx_{1,p}^2 + 
\bigg(\frac{R^2}{\rho^2+\vec{z}^{2}}\bigg)^{\alpha}
(d\vec{y}^{2}+d\vec{z}^{2})\,.
\end{equation}
In this background we will consider a stack of $N_f$ $D(p+4)$-branes
extended along $(x^{\mu}, \vec y)$ at fixed distance $L$ in the
transverse space spanned by the $\vec z$ coordinates (i.e. with
$|\hspace{0.5pt}\vec z\hspace{0.5pt}|=L$). If $\xi^a=(x^{\mu}, \vec y)$ are the
worldvolume coordinates, the action of a probe $D(p+4)$-brane is
\begin{eqnarray}	%4.60
S^{D(p+4)} &=& -T_{p+4}\int d^{\hspace{1pt}p+5}\xi\, e^{-\phi}Str
\big\{\sqrt{-\det(g + F)}\big\} \nonumber \\
&&{} + {T_{p+4}\over 2}\int Str\big\{P(C^{(p+1)})\wedge F\wedge F\big\}\,,
\label{SD(p+4)}
\end{eqnarray}
where $g$ is the induced metric and $F$ is the $SU(N_f)$ worldvolume
gauge field strength. In order to write $g$ more compactly, let us
define the function $h$ as follows:
\begin{eqnarray}	%4.61
h(\rho) \equiv \bigg(\frac{R^2}{\rho^2+L^{2}}\bigg)^{\alpha}\,.
\label{h}
\end{eqnarray}
Then, one can write the nonvanishing elements of the induced metric
as
\begin{eqnarray}	%4.62
g_{x^{\mu}x^{\nu}} = {\eta_{\mu\nu}\over h}\,,
\qquad g_{y^{i}y^{j}} = h\delta_{ij}\,.
\end{eqnarray}
Let us now assume that the only nonvanishing components of the
worldvolume gauge field $F$ are those along the $y^i$
coordinates. Following the same steps as in (\ref{MacroD3D7}), the
action for the $D(p+4)$-brane probe can be written as
\begin{eqnarray}	%4.63
&&S^{D(p+4)}\,=\,-T_{p+4}\int\, d^4x\,d^4y\,\,
{\rm Str}\,\Bigg\{
\sqrt{\,1+{1\over 2}\,\Bigg(\frac{\rho^2\,+\,L^2}{R^2}\Bigg)^{2\alpha} F^2
+\frac{1}{16}\Bigg(\frac{\rho^2\,+\,L^2}{R^2}\Bigg)^{4\alpha}
\Big(\,{}^*FF\,\Big)^2}\,-\,\nonumber\\
&&\qquad\qquad\qquad\qquad\qquad\qquad\qquad\qquad
\,\,-\frac{1}{4}\Bigg(\,\frac{\rho^2\,+\,L^2}{R^2}\,\Bigg)^{2\alpha}
\,\,{}^*FF
\Bigg\}\,,
\label{DBI-DpDp+4-explicit}
\end{eqnarray}
where $F^2$ and ${}^*FF$ are defined as in Eqs.~(\ref{MM}) and
(\ref{*M}). If, in addition, $F_{ij}$ is self-dual, one can check that
the equations of motion of the gauge field are satisfied and,
actually, there is a cancellation between the DBI and WZ parts of the
action (\ref{DBI-DpDp+4-explicit}) generalizing (\ref{totalaction}),
namely
\begin{equation}	%4.64
S^{D(p+4)}\hbox{(self-dual)} =-T_{p+4}\int Str[1] = 
-N_{f}T_{p+4}\int d^{\hspace{1pt}p+1}x\int d^{\hspace{0.5pt}4}y\,.
\end{equation}

We turn now to the analysis of the fluctuations around the self-dual
configuration and the computation of the corresponding meson spectrum
for this fluxed $Dp-D(p+4)$ intersection. Since the main particularity
of this embedding corresponding to the Higgs branch is the presence of
the worldvolume gauge field, following \cite{EGG} and
\cite{ARR2} we will focus on its fluctuations, for which we will
write
\begin{eqnarray}	%4.65
A=A^{\rm inst}+a\,,
\end{eqnarray}
where $A^{\rm inst}$ is the gauge potential corresponding to a
self-dual gauge field strength $F^{\rm inst}$ and $a$ is the
fluctuation. The total field strength $F$ reads
\begin{equation}	%4.66
F_{ab}=F^{\rm inst}_{ab} + f_{ab}\,,
\end{equation}
with $f_{ab}$ being given by
\begin{eqnarray}	%4.67
f_{ab} = \partial_{a}a_{b}-\partial_{b}a_{a}+{1\over 2\pi\alpha'}\,
[A^{\rm inst}_{a},a_{b}] + {1\over 2\pi\alpha'}\,
[a_{a},A^{\rm inst}_{b}] + {1\over 2\pi\alpha'}\,[a_{a},a_{b}]\,,\qquad
\end{eqnarray}
where the indices $a$, $b$ run now over all the worldvolume
directions. Next, let us expand the action (\ref{SD(p+4)}) in powers
of the field $a$ up to second order. With this purpose in mind, we
rewrite the square root in the DBI action as
\begin{eqnarray}	%4.68
\sqrt{-\det(g + F^{\rm inst} + f)} = 
\sqrt{-\det(g + F^{\rm inst})} \sqrt{\det(1 + X)}\,,
\label{DetX}
\end{eqnarray}
where $X$ is the matrix
\begin{eqnarray}	%4.69
X \equiv (g + F^{\rm inst})^{-1}f\,.
\label{defX}
\end{eqnarray}
We will expand the r.h.s. of (\ref{DetX}) in powers of $X$ by using
the equation
\begin{eqnarray}	%4.70
\label{expansion}
\sqrt{\det(1+X)} = 1 + {1\over 2}Tr X - {1\over 4}Tr X^2 + 
{1\over 8}(Tr X)^2 + {\mathcal{O}}(X^3)\,.
\end{eqnarray}
In our case, let us denote by ${\cal G}$ and ${\cal J}$ to the
symmetric and antisymmetric part of the inverse of $X$, i.e.
\begin{eqnarray}	%4.71
X\,=\,\Big(\,g\,+\,F^{inst}\,\Big)^{-1}\,=\,{\cal G}\,+\,{\cal J}\, ,\qquad {\cal G}^{\mu\nu}\,=\,h\,\,\eta^{\mu\nu}\,,
\qquad
{\cal G}^{ij}\,=\,{h\over H}\,\delta_{ij}\,, \qquad{\cal J}^{ij}\,=\,-{F^{inst}_{ij}\over H}\,;
\end{eqnarray}
where $h$ has been defined in (\ref{h}) and the function $H$ is given
by
\begin{eqnarray}	%4.72
H \equiv h^2 + {1\over 4}\,(F^{\rm inst})^2\,.
\label{H-def}
\end{eqnarray}
The symmetric part ${\cal{G}}$ behaves as an ``open string metric,"
and it carries combined information from the worldvolume gauge field
and metric.

By using these results we get, after a straightforward computation,
the action up to quadratic order in the fluctuations; namely
\begin{eqnarray}	%4.73
S^{D(p+4)} &=& -T_{p+4}\int Str
\bigg\{1+\frac{H}{4}f_{\mu\nu} f^{\mu\nu}+\frac{1}{2}\,
f_{i\mu}f^{i\mu}+\frac{1}{4H}f_{ij}f^{ij} \nonumber \\
&&{}+\frac{1}{8h^2H}(F^{ij}f_{ij})^2-\frac{1}{4h^2H}\,
F^{ij}F^{kl}f_{jk}f_{li} - \frac{1}{8h^2}\,
f_{ij}f_{kl}\epsilon^{ijkl}\bigg\}\,,\qquad\ 
\end{eqnarray}
where we are dropping the superscript in the instanton field strength.

From now on we will assume again that $N_f=2$ and that the unperturbed
configuration is the one-instanton $SU(2)$ gauge field written in
Eq.~(\ref{instanton-potential}). Moreover, we will focus on the subset
of fluctuations for which $a_i=0$, i.e. on those for which the
fluctuation field $a$ has nonvanishing components only along the
Minkowski directions. However, we should impose this ansatz at the
level of the equations of motion in order to ensure the consistency of
the truncation. Let us consider first the equation of motion for
$a_i$, which after imposing $a_i=0$ reduces to
\begin{eqnarray}	%4.74
D_i \partial^{\mu} a_{\mu} = 0\,.
\label{ai-eom}
\end{eqnarray}
Moreover, the equation for $a_{\mu}$ when $a_i=0$ becomes
\begin{eqnarray}	%4.75
H D^{\mu} f_{\mu\nu} + D^i f_{i\nu} = 0\,,
\label{amu-eq}
\end{eqnarray}
where now $H$ is given in (\ref{H-def}), with $(F^{\rm inst})^2$ as in
(\ref{F2-inst}). Equation~(\ref{ai-eom}) is solved by requiring
\begin{eqnarray}	%4.76
\partial^{\mu}a_{\mu} = 0\,.
\label{transv}
\end{eqnarray}
Using this result, Eq.~(\ref{amu-eq}) can be written as
\begin{eqnarray}	%4.77
&&H\partial^{\mu}\partial_{\mu} a_{\nu} + \partial_i\partial_i a_{\nu} +
\partial^i\bigg[{A_i\over 2\pi\alpha'},a_{\nu}\bigg] \nonumber \\
&&\qquad{} + \bigg[{A_i\over 2\pi\alpha'},\partial_ia_{\nu}\bigg] + 
\bigg[{A_i\over 2\pi\alpha'},
\bigg[{A_i\over 2\pi\alpha'},a_\nu\bigg]\bigg] = 0\,.
\label{fluc-eom-DpDp+4}
\end{eqnarray}
Let us now adopt the following ansatz for $a_{\mu}$:
\begin{equation}	%4.78
a_{\mu}^{(l)} = \xi_{\mu}(k)f(\rho)e^{ik_{\mu}x^{\mu}}\tau^l\,,
\label{amu-ansatz}
\end{equation}
where $\tau^l$ is a Pauli matrix. This ansatz solves
Eq.~(\ref{transv}) provided the following transversality condition is
fulfilled:
\begin{eqnarray}	%4.79
k^{\mu}\xi_{\mu} = 0\,.
\end{eqnarray}
Moreover, one can check that, for this ansatz, one has
\begin{eqnarray}	%4.80
&& \partial^i\,\big[\,A_i\,,\,a_{\nu}^{(l)}\,\big]\,=\,
\big[\,A_i\,,\partial_ia_{\nu}^{(l)}\,\big]\,=\,0\,\,,\\
&&\Big[\,{A_i\over 2\pi\alpha'}\,,\Big[\,{A_i\over 2\pi\alpha'}\,,a_\nu^{(l)}
\Big]\Big]\,=\,-\,\frac{8\Lambda^4}{\rho^2(\rho^2+\Lambda^2)^2}\,\,
\xi_{\nu}(k)\,f(\rho)\,\,e^{ik_{\mu}x^{\mu}}\,\tau^l\,\,.
\end{eqnarray}
Let us now use these results in Eq.~(\ref{fluc-eom-DpDp+4}). Denoting
$M^2=-k^2$ (which will be identified with the mass of the meson in the
dual field theory) and using Eq.~(\ref{F2-inst}) to compute the
function $H$ (see Eq.~(\ref{H-def})), one readily reduces
(\ref{fluc-eom-DpDp+4}) to the following second-order differential
equation for the function $f(\rho)$ of the ansatz (\ref{amu-ansatz}):
\begin{eqnarray}	%4.81
\label{e1}
\bigg[\frac{R^{4\alpha}M^2}{(\rho^2{%\ko
+%\ko
}L^2)^{2\alpha}}
\bigg(1{%\ko
+%\ko
}\frac{12(2\pi\alpha')^2\Lambda^4}{R^{4\alpha}}
\frac{(\rho^2{%\ko
+%\ko
}L^2)^{2\alpha}}{(\rho^2{%\ko
+%\ko
}\Lambda^2)^4}\bigg)  
{%\ko
-%\ko
}\frac{8\Lambda^4}{\rho^2(y^2{%\ko
+%\ko
}\Lambda^2)^2} {%\ko
+%\ko
} 
\frac{1}{\rho^3}\partial_\rho(\rho^3\partial_\rho)\bigg]f &=& 0\,. \nonumber \\[-4pt]
\end{eqnarray}
In order to analyze Eq.~(\ref{e1}), let us introduce a new radial
variable $\varrho$ and a reduced mass $\bar M$, which are related to
$\rho$ and $M$ as
\begin{eqnarray}	%4.82
\rho=L\varrho\,, \qquad \bar{M}^2=R^{7-p}L^{p-5}M^2\,.
\end{eqnarray}
Moreover, following \cite{MT}, it is interesting to rewrite
the fluctuation equation in terms of field theory
quantities. Accordingly, let us introduce the quark mass $m_q$ and its
VEV $v$ as follows:
\begin{eqnarray}	%4.83
m_q = {L\over 2\pi\alpha'}\,, \qquad
v = {\Lambda\over 2\pi\alpha'}\,.
\end{eqnarray}
Notice that the relation between $v$ and the instanton size $\Lambda$
is consistent with our analysis in (\ref{microD3D7}) (see
Eq.~(\ref{v-Lambda})) and with the proposal of \cite{EGG}.

We can reexpress the equation for the fluctuations in terms of field
theory quantities by means of (\ref{gYM}) and (\ref{geff}), giving
\begin{eqnarray}	%4.84
\label{fluctuationsDp+4}
&&\left[\frac{\bar{M}^2}{(1+\varrho^2)^{2\alpha}}
\left(1+c_{p}(v,m_q)\,\frac{(1+\rho^2)^{2\alpha}}
{\Big(\varrho^2+\big(\frac{v}{m_q}\big)^2\Big)^4}\right)\right. \nonumber \\
&&\left.\qquad{}-\bigg(\frac{v}{m_q}\bigg)^4
\frac{8}{\varrho^2\Big(\varrho^2+\big(\frac{v}{m_q}\big)^2\Big)^2} +
\frac{1}{\rho^3}\partial_{\varrho}(\varrho^3\partial_{\varrho})\right]f=0\,,
\end{eqnarray}
where $c_{p}(v,m_q)$ is defined as
\begin{eqnarray}	%4.85
c_{p}(v,m_q) \equiv \frac{12\cdot 2^{p-2}
\pi^{\frac{p+1}{2}}}{\Gamma\big(\frac{7-p}{2}\big)}\,
{v^4\over g_{\rm eff}^2(m_q)m_q^4}\,.
\label{cp}
\end{eqnarray}
Notice that everything conspires to absorb the powers of $\alpha'$ and
give rise to the effective coupling at the quark mass in
$c_{p}(v,m_q)$.

Equation (\ref{fluctuationsDp+4}) differs in the $\bar{M}$ term from
the one obtained in \cite{EGG}, where the term proportional to
$c_{p}(v,m_q)$ is absent. We would like to point out that in order to
arrive to (\ref{fluctuationsDp+4}) we expanded up to quadratic order
in the fluctuations and we have kept all orders in the instanton
field. The extra factor compared to that in \cite{EGG} comes
from the fact that, for a self-dual worldvolume gauge field, the
unperturbed DBI action actually contains the square root of a perfect
square, which can be evaluated exactly and shows up in the Lagrangian
of the fluctuations. This extra term is proportional to the inverse
of the effective Yang--Mills coupling. In order to ensure the validity
of the DBI approximation, we should have slowly varying gauge fields,
which further imposes that $F\wedge F$ should be much smaller than
$\alpha'$. Also, to trust the supergravity approximation, the
effective Yang--Mills coupling should be large, which would suggest
that the effect of this term is indeed negligible. However, in the
region of small $\frac{v}{m_q}$ the full term is actually dominating
in the IR region and determines the meson spectrum.

We have postponed the detailed analysis of the meson spectrum to the
\hbox{appendix}. Let us mention that this meson spectrum is discrete, i.e. we
have a mass gap for the mesons proportional to the quark mass also in
this Higgs branch of the gauge theory. Interestingly, as one can see
in the Appendix, the meson spectrum exhibits a so-called {\it spectral
flow}. When the instanton size is varied, the masses of the mesons
change in a very similar way as if the quantum numbers, instead of the
instanton size, were changing. This is the so-called {\it spectral
flow} phenomenon, which was first suggested in
\cite{EGG}. However, the small instanton region is somehow
singular, and indeed one encounters that the masses appear to go to
zero. In this singular point, the approximation is no longer valid,
since the $F^2$ terms becomes basically a delta function. Indeed, at
this point one should expect some sort of singular behavior, since
there the VEV of the quark fields vanishes and the gauge group
unhiggses; and one would expect that at this point new light degrees
of freedom would enter the low energy description.

Let us now study the dependence of the mass gap as a function of the
quark mass $m_q$ and the quark VEV $v$. First of all, we notice that
the relation between the reduced mass $\bar M$ and the mass $M$ can be
rewritten in terms of the quark mass $m_q$ and the dimensionless
coupling constant $g_{\rm eff}(m_q)$ as
\begin{eqnarray}	%4.86
M \propto {m_q\over g_{\rm eff}(m_q)}\,\bar M\,.
\label{M-barM}
\end{eqnarray}
For large $v$ the reduced mass $\bar M$ tends to a value independent
of both $m_q$ and $v$. Thus, the meson mass $M$ depends only on $m_q$
in a holographic way, namely
\begin{eqnarray}	%4.87
M \sim {m_q\over g_{\rm eff}(m_q)} \quad (v\to\infty)\,.
\end{eqnarray}
Notice that this dependence on $m_q$ and $v$ is exactly the same as in
the unbroken symmetry case, although the numerical coefficient is
different from that found in \cite{AR} and \cite{MT}. On
the contrary, for small $v$, after combining Eq.~(\ref{M-barM}) with
the WKB result (\ref{WKB-smallv}), we get that the mass gap depends
linearly on $v$ and is independent on the quark mass $m_q$:
\begin{eqnarray}	%4.88
M\sim v \quad (v\to 0)\,,
\end{eqnarray}
and, in particular, the mass gap disappears in the limit $v\to 0$,
which corresponds to having a zero size instanton.

\subsection{The codimension one defect}\label{DpDp+2section}

We will now consider the intersection of $Dp$- and $D(p+2)$-branes
according to the array:
\begin{eqnarray*}	
\arraycolsep5pt\begin{array}{rcccccccccl}
& 1 & \cdots & p-1 & p & p+1 & p+2 & p+3 & \cdots & 9 & \\[2pt]
Dp: & \times & \cdots & \times & \times & - & - & - & \cdots & - &\\[2pt]
D(p+2): & \times & \cdots & \times & - 
& \times & \times & \times & \cdots &- &\end{array}
\label{DpDp+2intersection}
\end{eqnarray*}
Clearly, this defines a codimension 1 defect in the bulk
$p$-dimensional gauge theory where the matter is confined.

Exactly as in the codimension 0 defect, from the gravity point of
view, also in this case all the dimensionalities behave in a similar
way. However, we will study in a detailed way the $p=3$ case, where
both the field theory and gravity descriptions will be analyzed. Since
all the other dimensionalities behave similarly, when computing the
spectrum we will do it in a unified manner for all $p$.

\subsubsection{A case study II\/$:$ the $D3-D5$ intersection}

Since the dual field theory to the $D3-D5$ intersection is somehow
less familiar, we will start with the gravity description in order to
gain some intuition on the dynamics of the system. The study of the
Coulomb branch of this system was initiated in \cite{WFO} and
further pursued in \cite{AR} and \cite{ARR}; where an
interpretation of the gravity dual of the Higgs branch was given. In
Sec.~3 we discussed the Coulomb branch, and we now turn to the Higgs
branch of the system. For that, let us split the six transverse
coordinates $\vec y$ to the color $D3$ branes in two sets of three
elements, according to the $D3-D5$ intersection represented by the
array (\ref{DpDp+2intersection}). The coordinates $(y^1,y^2 ,y^3)$ are
those which are parallel to the $D5$-brane worldvolume in
(\ref{DpDp+2intersection}). It is convenient to go to spherical
coordinates as $(dy^1)^2+(dy^2)^2+(dy^3)^2 = d\rho^2 + \rho^2
d\Omega_2^2$, where $d\Omega_2^2$ is the line element of a unit
two-sphere. Moreover, let us denote by $\vec z = (z^1,z^2 ,z^3) =
(y^4,y^5 ,y^6)$ the coordinates transverse to both the $D3$- and
$D5$-branes. Clearly, $r^2 = \rho^2+\vec z^{2}$, so the background
$AdS_5\times S^5$ metric can be written as
\begin{eqnarray}	%4.89
ds^2 = {\rho^2+\vec z^{2}\over R^2}\,dx_{1,3}^2 + 
{R^2\over \rho^2+\vec z^{2}}
(d\rho^2 + \rho^2\, d\Omega_2^2 + d\vec z\cdot d\vec z)\,.
\label{polarbackmetric}
\end{eqnarray}

Since our probe flavor branes will be $D5$ branes partially
overlapping the $D3$, in order to couple the background 4-form
potential we have to turn on a nonzero worldvolume magnetic field. In
this case, the action to be considered is
\begin{eqnarray}	%4.90
S_{D5} = -T_{5}\int d^{\hspace{0.5pt}6}\xi\sqrt{-\det (g+F)} + 
T_{5}\int d^{\hspace{0.5pt}6}\xi\,P[C^{(4)}]\wedge F\,,
\label{DBI-D5}
\end{eqnarray}
where $g$ is the pullback of the metric (\ref{polarbackmetric}), $F$
is the strength of the Abelian worldvolume gauge field and $\xi^a$
$(a=0,\ldots, 5)$ are a set of worldvolume coordinates. In what
follows we will use $x^0$, $x^1$, $x^2$ and the radial ($\rho$) and
angular coordinates of Eq.~(\ref{polarbackmetric}) as our set of
worldvolume coordinates.

Generically, the embedding of the $D5$-brane probe is then specified
by the values of $x^3$ and $\vec z$ as functions of the $\xi^a$'s. We
will consider static embeddings in which $|\vec z|$ is a fixed
constant, namely $|\vec z|=L$. The simplest of such embeddings is the
one in which the coordinate $x^3$ is also a constant. In this case, it
is clear from (\ref{DBI-D5}) that the WZ coupling will vanish
independently of $F$, since in order to capture the RR potential we
need a nontrivial $x^3$ dependence so that the pull-back in
(\ref{DBI-D5}) does not vanish. Therefore, for constant $x^3$ we can
take the worldvolume gauge field $F$ to vanish. This corresponds to
the Coulomb branch of the dual theory. Since the defect lives at a
fixed $x^3$ position, it represents a domain wall in the
four-dimensional Minkowski.

In turn, if we are to couple the 4-form potential, we need to consider
a nontrivial $x^3$ dependence on the worldvolume coordinates. This
demands to turn a nonzero magnetic $F$ along the two-sphere of its
worldvolume. To be precise, let us assume that $F$ is given by
\begin{eqnarray}	%4.91
F = qVol(S^2) \equiv {\cal F}\,,
\label{wvflux}
\end{eqnarray}
where $q$ is a constant and $Vol(S^2)$ is the volume form of the
worldvolume two-sphere. To understand the implications of having a
magnetic flux across the worldvolume $S^2$, let us look at the form of
the Wess--Zumino term in the action (\ref{DBI-D5}), which will involve
\begin{eqnarray}	%4.92
S_{\rm WZ} \sim \int_{S^2} F \int P[C^{(4)}] \sim q x'\,,
\label{WZbending}
\end{eqnarray}
where $x\equiv x^3$ and the prime denotes the derivative with respect
to the radial coordinate $\rho$. It is clear from (\ref{WZbending})
that the worldvolume flux acts as a source of a nontrivial dependence
of $x$ on the coordinate $\rho$. Assuming that $x$ only depends on
$\rho$, the action (\ref{DBI-D5}) of the probe takes the form:
\begin{eqnarray}
S_{D5}=-4\pi\,T_{5}\,\int\,d^3x\,d\rho\Bigg[\,
\rho^2\,
\sqrt{1\,+\,{(\rho^2+L^2)^2\over R^4}\,x'^{\,2}}\,\,
\sqrt{1\,+\,{(\rho^2+L^2)^2\over R^4}\,
{q^2\over \rho^4}}\,\,-\,\,
{(\rho^2+L^2)^2\over R^4}\,q\,x'\,\Bigg]\,\,,
\label{effeaction}
\end{eqnarray}
where we have assumed that $\vec z$ is constant ($|\hspace{0.5pt}\vec
z\hspace{0.5pt}|=L$) and we have integrated over the coordinates of the
two-sphere. The Euler--Lagrange equation for $x(\rho)$ derived from
(\ref{effeaction}) is quite involved. However, there is a simple
first-order equation for $x(\rho)$ which solves this equation
\cite{ST}, namely
\begin{eqnarray}	%4.94
x'(\rho) = {q\over \rho^2}\,.
\label{first-order}
\end{eqnarray}
Actually, the first-order equation (\ref{first-order}) is a BPS
equation required by supersymmetry, as can be verified by checking the
kappa symmetry of the embedding \cite{ST}. The integration of
Eq.~(\ref{first-order}) is straightforward:
\begin{eqnarray}	%4.95
x(\rho) = x_0 - {q\over \rho}\,,
\label{bending}
\end{eqnarray}
where $x_0$ is an integration constant. The dependence on $\rho$ of
the r.h.s. of Eq.~(\ref{bending}) represents the bending of the
$D5$-brane profile required by supersymmetry when there is a
nonvanishing flux of the worldvolume gauge field. Notice also that now
the probe is located at a fixed value of $x$ only at the asymptotic
value $\rho\to\infty$, whereas when $\rho$ varies the $D5$-brane fills
one-half on the worldvolume of the $D3$-brane (i.e. $x^3\le x_0$ for
$q>0$). Actually, this indicates that this embedding corresponds not
to a deformation of the theory, but rather to a choice of vacuum. The
reason is that both the embedding at constant $x$ and the bended one
share the same asymptotics, so one would expect that they correspond
to different vacua rather than to a deformation of the boundary
theory. We will explicitly see that it is indeed the case when
studying the system from the dual field theory side; where we will
explicitly see, along the lines in the codimension zero case, that the
gauge theory contains both the Coulomb and Higgs branch we studied.

It is interesting to study the modifications of the induced metric
introduced by the bending. Actually, when $q\not=0$ this induced
metric takes the form
\begin{eqnarray}	%4.96
{\cal G}_{ab}\,d\xi^a\, d\xi^b &=& {\rho^2+L^2\over R^2}\,dx^2_{1,2} + 
{R^2\over \rho^2+L^2}\bigg[\bigg(1+{q^2\over R^4}\,
{(\rho^2+L^2)^2\over \rho^4}\bigg)
d\rho^2 + \rho^2\,d\Omega_2^2\bigg]\,.\nonumber \\[-2pt]
\label{ind-met-flux}
\end{eqnarray}

In the $L=0$ case, the metric reduces to an effective $AdS_4$
worldvolume. Therefore, like in the Coulomb phase, the dual theory
enjoys a conformal symmetry even in the Higgs phase. However, even in
the $L\ne 0$ case, the UV metric at $\rho\to\infty$ takes the same
form, since in the UV the quark mass is completely irrelevant and thus
theory asymptotes to a conformal one. Considering the generic massive
case, the worldvolume induced metric is
\begin{eqnarray}	%4.97
AdS_4(R_{\rm eff})\times S^2 (R)\,,
\label{UVind-met-flux}
\end{eqnarray}
where the radius of the $AdS_4$ changes from its fluxless value $R$ to
$R_{\rm eff}$, with the latter given by:
\begin{eqnarray}	%4.98
R_{\rm eff} = \bigg(1 + {q^2\over R^4}\bigg)^{{1\over 2}}\,R\,.
\label{Reff}
\end{eqnarray}
Notice that the radius of the $S^2$ is not affected by the flux, as is
clear from (\ref{ind-met-flux}).

One can understand the appearance of this UV metric as follows. Let
us suppose that we have an $AdS_5$ metric of the form
\begin{eqnarray}	%4.99
ds^2_{AdS_5} = {\rho^2\over R^2}\,dx^2_{1,3} + {R^2\over \rho^2}\,d\rho^2\,.
\label{AdSmetric}
\end{eqnarray}
Let us now change variables from $(\rho, x^3)$ to new coordinates
$(\varrho, \eta)$:
\begin{eqnarray}	%4.100
x^3 = \bar x - {\tanh\eta\over \varrho}\,, \qquad
\rho = R^2\varrho\cosh\eta\,,
\label{changevariablesA}
\end{eqnarray}
where $\bar x$ is a constant. It can be easily seen that the $AdS_5$
metric (\ref{AdSmetric}) in the new variables takes the form
\begin{eqnarray}	%4.101
ds^2_{AdS_5} = R^2(\cosh^2\eta\,ds^2_{AdS_4} + d\eta^2)\,,
\label{foliation1}
\end{eqnarray}
where $ds^2_{AdS_4}$ is the metric of $AdS_4$ with unit radius, given
by
\begin{eqnarray}	%4.102
ds^2_{AdS_4} = \varrho^2\,dx^2_{1,2} + {d\varrho^2\over \varrho^2}\,.
\end{eqnarray}

The first equation in (\ref{changevariablesA}), when written in terms
of $\rho$, reads
\begin{equation}	%4.103
x=\bar{x}-\frac{R^2\sinh\eta}{\rho}\,,
\end{equation}
which is exactly (\ref{bending}) once we identify
$q=R^2\sinh\eta$. Thus, our embedding corresponds to fixed $\eta$
slices of the original $AdS_5$, and thus should correspond to an
$AdS_4$ worldvolume with effective radius
\begin{eqnarray}	%4.104
R_{\rm eff} = R\cosh\eta\,.
\label{sliceradius}
\end{eqnarray}

The worldvolume gauge field (\ref{wvflux}) is constrained by a flux
quantization condition \cite{Flux} which, with our notations, reads
\begin{eqnarray}	%4.105
\int_{S^2}F = {2\pi k\over T_f}\,,\qquad
k\in Z\,,\qquad T_f = {1\over 2\pi\alpha'}\,.
\label{fluxquantization}
\end{eqnarray}
It is now immediate to conclude that the condition
(\ref{fluxquantization}) restricts the constant $q$ to be of the form:
\begin{eqnarray}	%4.106
q = k\pi\alpha'\,,
\label{q-k}
\end{eqnarray}
where $k$ is an integer.

\subsubsection*{A microscopical picture} \label{microD3D7}

The presence of a worldvolume flux as in (\ref{wvflux}) induces,
through the Wess--Zumino term of the action (\ref{DBI-D5}), a
$D3$-brane charge, proportional to $\int_{S^2}\,F$, on the
$D5$-brane. Indeed, we can think again the system as a recombination
of some of the color branes with the flavor ones. For this reason it
is not surprising that this $D5$-brane configuration admits also a
microscopical description in terms of a bound state of coincident
$D3$-branes. Actually, the integer $k$ of the quantization condition
(\ref{fluxquantization}) has the interpretation of the number of
$D3$-branes that build up the $D5$-brane. The dynamics of a stack of
coincident $D3$-branes is determined by the Myers dielectric action
\cite{M} (see Appendix).

The Wess--Zumino term for the $D3$-brane under consideration is
\begin{eqnarray}	%4.107
S_{\rm WZ}^{D3} = T_{3}\int d^{\hspace{0.5pt}4}\xi Str\big[P\big[C^{(4)}\big]\big]\,.
\label{couplingD3D5}
\end{eqnarray}
\vskip-\lastskip
\pagebreak

\noindent
Let us now choose $x^0$, $x^1$, $x^2$ and $\rho$ as our set of
worldvolume coordinates of the $D3$-branes. Moreover, we shall
introduce new coordinates $Y^I(I=1,2,3)$ for the two-sphere of the
metric (\ref{polarbackmetric}). These new coordinates satisfy $\sum_I
Y^I Y^I = 1$ and the line element $d\Omega_2^2$ is given by
\begin{eqnarray}	%4.108
d\Omega_2^2 = \sum_I\, dY^I\,dY^I\,, \qquad \sum_I\,Y^I Y^I = 1\,.
\end{eqnarray}
We will assume that the $Y^I$'s are the only noncommutative scalars.
They will be represented by $k\times k$ matrices. In this case the
matrix $Q$ appearing in (\ref{MyersDBI}) is given~by
\begin{eqnarray}	%4.109
Q_J^I = \delta_{J}^{I} + {i\over 2\pi\alpha'}\,[Y^I,Y^K]G_{KJ}\,.
\end{eqnarray}
Actually, we shall adopt the ansatz in which the $Y^I$'s are constant
and given by
\begin{eqnarray}	%4.110
Y^I = {J^I\over \sqrt{C_2(k)}}\,,
\label{Yansatz}
\end{eqnarray}
where the $k\times k$ matrices $J^I$ correspond to the $k$-dimensional
irreducible representation of the $SU(2)$ algebra:
\begin{eqnarray}	%4.111
[J^I,J^J] = 2i\epsilon_{IJK}J^K\,,
\label{Jcommutator}
\end{eqnarray}
and $C_2(k)$ is the quadratic Casimir of the $k$-dimensional
irreducible representation of $SU(2)$ ($C_2(k)=k^2-1$). Then, the
$Y^I$ satisfy
\begin{equation}	%4.112
Y^IY^I=1
\end{equation}
as a matrix identity, and therefore, the $Y^I$ scalars parametrize a
fuzzy two-sphere. Moreover, let us assume that we consider embeddings
in which the scalars $\vec z$ and $x^3$ are commutative and such that
$|\vec z|=L$ and $x^3=x(\rho)$ (a unit $k\times k$ matrix is
implicit). With these conditions, as the metric
(\ref{polarbackmetric}) does not mix the directions of the two-sphere
with the other coordinates, the matrix $Q^{-1}-\delta$ does not
contribute to the first square root on the r.h.s. of
(\ref{MyersDBI}). Then
\begin{eqnarray}	%4.113
\sqrt{-\det\big[ P[G]\,\big]} = {\rho^2+L^2\over R^2}\,
\sqrt{1 + {(\rho^2+L^2)^2\over R^4}\,x'^{2}}\,.
\end{eqnarray}
Moreover, by using the ansatz (\ref{Yansatz}) and the commutation
relations (\ref{Jcommutator}) we obtain that, for large $k$, the
second square root appearing in (\ref{MyersDBI}) can be\break written~as
\begin{eqnarray}	%4.114
Str\big[\sqrt{\det Q}\big] \approx {R^2\over \pi\alpha'}\,
{\rho^2\over \rho^2+L^2}\,\sqrt{1 + {(\rho^2+L^2)^2\over R^4}\,
{(k\pi\alpha')^2\over \rho^4}}\,.
\end{eqnarray}
\vskip-\lastskip
\pagebreak

Using these results, the DBI part of the $D3$-brane action in this
large $k$ limit takes the form
\begin{eqnarray}	%4.115
S_{\rm BI}^{D3} = -{T_3\over \pi\alpha'} \int d^{\hspace{0.5pt}3}x\,d\rho\,
\rho^2\sqrt{1 + {(\rho^2+L^2)^2\over R^4}\,x'^{2}}
\sqrt{1 + {(\rho^2+L^2)^2\over R^4}\,{q^2\over \rho^4}}\,,\quad
\label{microBI}
\end{eqnarray}
where we have already used (\ref{q-k}) to write the result in terms of
$q$. Due to the relation $T_3 = 4\pi^2 \alpha' T_5$ between the
tensions of the $D3$- and $D5$-branes, one checks by inspection that
the r.h.s. of (\ref{microBI}) coincides with the Born--Infeld term of
the $D5$-brane action (\ref{effeaction}). Notice also that the
quantization integer $k$ in (\ref{fluxquantization}) is identified
with the number of $D3$-branes. Moreover, the Wess--Zumino term
(\ref{couplingD3D5}) becomes
\begin{eqnarray}	%4.116
S_{\rm WZ}^{D3} = kT_3 \int d^{\hspace{0.5pt}3}x\,d\rho\,
{(\rho^2+L^2)^2\over R^4}\,x'\,.
\label{microWZ}
\end{eqnarray}
The factor $k$ in (\ref{microWZ}) comes from the trace of the unit
$k\times k$ matrix.

By comparing (\ref{microWZ}) with the Wess--Zumino term of the
macroscopical action (\ref{effeaction}) one readily concludes that
they coincide; since because of (\ref{q-k}) we have that $4\pi qT_5 =
kT_3$.

\subsubsection*{Field theory analysis}	
\label{fieldtheory}

The field theory dual to the $D3-D5$ intersection has been worked out
by DeWolfe {\it et~al.} in \cite{WFO} (see also
\cite{EGK}). Let us consider for simplicity the massless
case. Then, the theory, which includes ${\mathcal N}=4$ $SU(N)$ $SYM$
in four-dimensional plus an ${\cal N}=4$ hypermultiplet confined to
the defect, has an $SU(2)_H\times SU(2)_V$ R-symmetry. The $SU(2)_H$
($SU(2)_V$) symmetry corresponds to the rotations in the 456 (789)
directions of the array (\ref{DpDp+2intersection}). Written in terms
of ${\mathcal N}=1$ SUSY, this hypermultiplet gives rise to a chiral
($Q$) and an antichiral ($\bar{Q}$) supermultiplet, which are both
doublets under $SU(2)_H$ while being in the fundamental representation
of the gauge group. In addition, the six scalars of the bulk
${\mathcal N}=4$, which are in the adjoint of the gauge group,
naturally split in two sets, the first (which we will call $\phi_H^I$)
forming a vector of $SU(2)_H$ and the second, which we denote by
$\phi_V^A$, a vector of $SU(2)_V$. Thus, the bosonic content of the
theory is as follows:
\begin{center}
\begin{tabular}{|c|c|c|c|}
\hline 
Field & $SU(N)$ & $SU(2)_H$ & $SU(2)_V$\\
\hline
$A_{\mu}$ & adjoint & singlet & singlet\\
\hline
$\phi_H^I$ & adjoint & vector & singlet\\
\hline
$\phi_V^A$ & adjoint & singlet & vector\\
\hline
$q$ & fundamental & doublet & singlet\\
\hline
$\bar{q}$&fundamental& doublet & singlet\\
\hline
\end{tabular}
\label{table}
\end{center}
We will assume that only the fields $\phi_H$, $\phi_V$, $q$ and $\bar
q$ are nonvanishing. The defect action for this theory has a potential
term which can be written as \cite{WFO}
\begin{eqnarray}	%4.117
S_{\rm defect} &=& -\frac{1}{g^2}\int d^{\hspace{0.5pt}3}x
\bigg[\bar{q}^m(\phi_V^A)^2q^m + \frac{i}{2}\,\epsilon_{IJK}
\bar{q}^m\sigma_{mn}^I[\phi_H^J,\phi_H^K]q^n\bigg] \nonumber \\
&&{}-\frac{1}{g^2}\int d^{\hspace{0.5pt}3}x\bigg[\bar{q}^m\sigma_{mn}^I\partial_3
\phi_H^Iq^n + \frac{1}{2}\,\delta(x_3)
(\bar{q}^m\sigma_{mn}^IT^aq^n)^2\bigg]\,,\quad 
\label{actiondefect}
\end{eqnarray}
where the integration is performed over the $x^3=0$ three-dimensional
submanifold and $g$ is the Yang--Mills coupling constant.

For the supersymmetric configurations we are looking for, the
potential term must vanish. One way to achieve this is to consider the
quark fields to zero; and the $\Phi_H$, $\Phi_V$ fields to be
commuting fields. The eigenvalues of the adjoint scalars parametrize,
once again, the Coulomb branch of the theory, which in general
\hbox{involves} a broken gauge group.

However, we can have more involved situations. Focusing on the
equation of motion for the $\tilde{q}$ field, we have that
\begin{eqnarray}	%4.118
\phi_Vq = 0\,.
\end{eqnarray}
We can insure this property by taking $q$ as
\begin{eqnarray}	%4.119
q = \left(\begin{array}{c}
0 \\ 
\vdots \\ 
0 \\ 
\alpha_1 \\ 
\vdots \\ 
\alpha_k\end{array}\right)\,,
\label{qvev}
\end{eqnarray}
and by demanding that $\phi_V$ is of the form
\begin{eqnarray}	%4.120
\phi_V = \left(\arraycolsep5pt\begin{array}{@{\hspace{1pt}}cc@{\hspace{1pt}}}
A & 0 \\
0 & 0 \end{array}\right)\,,
\end{eqnarray}
where $A$ is an $(N-k)\times (N-k)$ traceless matrix. Moreover, we
shall take $\phi_V$, $q$ and $\bar q$ constant, which is enough to
guarantee that their kinetic energy vanishes. Notice that the scalars
$\phi_V$ correspond to the directions 789 in the array
(\ref{DpDp+2intersection}), which are orthogonal to both the $D3$- and
$D5$-brane. Note that, in a similar manner to the $D3-D7$ case, we
have to pick a particular configuration for the transverse scalars to
the system, corresponding in this case to the $\phi_V$. Therefore, we
also need to go to a particular point in the Coulomb branch to enter
the Higgs branch, and indeed, had we chosen a nonzero mass for the
quarks, we would have had that some of the $\Phi_V$ eigenvalues should
have been adjusted to cancel the mass term in order to enter the Higgs
branch.

In order to find the supersymmetric vacua, let us consider the
configurations of $\phi_H$ with vanishing energy. First of all we
will impose that $\phi_H$ is a matrix whose only nonvanishing entries
are in the lower $k\times k$ block. In this way the mixing terms of
$\phi_V$ and $\phi_H$ cancel. Moreover, assuming that $\phi_H$ only
depends on the coordinate $x^3$, the surviving terms in the bulk
action are \cite{WFO}
\begin{eqnarray}	%4.121
S_{\rm bulk} = -{1\over g^2} \int d^{\hspace{0.5pt}4}xTr
\bigg[{1\over 2}\,(\partial_3\phi_H^I)^2 - {1\over 4}\,
[\phi_H^I, \phi_H^J]^2\bigg]\,,
\label{bulkaction}
\end{eqnarray}
where the trace is taken over the color indices. It turns out that the
actions (\ref{actiondefect}) and (\ref{bulkaction}) can be combined in
such a way that their sum can be written as an integral over the
four-dimensional space--time of the trace of a square. In order to
write this expression, let us define the matrix $\alpha^I =
\alpha^{Ia}T^a$, where the $T^a$'s are the generators of the gauge
group and the $\alpha^{Ia}$'s are defined as the following expression
bilinear in $q$ and $\bar q$:
\begin{eqnarray}	%4.122
\alpha^{Ia} \equiv \bar{q}^m \sigma_{mn}^I T^a q^n\,.
\end{eqnarray}
It is now straightforward to check that the sum of
(\ref{actiondefect}) and (\ref{bulkaction}) can be put~as
\begin{eqnarray}	%4.123
S_{\rm defect} + S_{\rm bulk} = -{1\over 2g^2}
\int d^{\hspace{0.5pt}4}x Tr\bigg[\partial_3\phi_H^I + {i\over 2}\,
\epsilon_{IJK}[\phi_H^J, \phi_H^K] + \alpha^I\delta(x^3)\bigg]^2\,,\qquad \
\label{actionsquare}
\end{eqnarray}
where we have used the fact that
$\epsilon_{IJK}Tr\big(\partial_3\phi_H^I[\phi_H^J,\phi_H^K]\big)$ is
a total derivative with respect to $x^3$ and, thus, can be dropped if
we assume that $\phi_H$ vanishes at $x^3=\pm\infty$. It is now clear
from (\ref{actionsquare}) that we must require the Nahm equations
\cite{Neq}:
\begin{eqnarray}	%4.124
\partial_3 \phi_H^I + {i\over 2}\,\epsilon_{IJK}
[\phi_H^J, \phi_H^K] + \alpha^I\delta(x^3) = 0\,.
\label{Nahn}
\end{eqnarray}
(For a nice review of the Nahm construction in string theory see
\cite{Tong} and \cite{Tong2}.)

Notice that when $\alpha^I$ vanishes, Eq.~(\ref{Nahn}) admits the
trivial solution $\phi_H=0$. On the contrary, as shown in
\cite{ARR}, if the fundamentals $q$ and $\bar q$ acquire a
nonvanishing vacuum expectation value as in (\ref{qvev}), $\alpha^I$
is generically nonzero and the solution of (\ref{Nahn}) must be
nontrivial. Actually, it is clear from (\ref{Nahn}) that in this case
$\phi_H$ must blow up at $x^3=0$, which shows how a nonvanishing
vacuum expectation value of the fundamentals acts as a source for the
brane recombination in the Higgs branch of the theory. Actually, away
from $x_3=0$, the $\delta$-function term is zero, so we can consider
just
\begin{eqnarray}	%4.125
\partial_3 \phi_H^I + {i\over 2}\,\epsilon_{IJK}[\phi_H^J, \phi_H^K] = 0\,.
\label{Nahn2}
\end{eqnarray}
We shall adopt the ansatz
\begin{eqnarray}	%4.126
\phi_H^I(x) = f(x)\phi_0^I\,,
\end{eqnarray}
where $x$ stands for $x^3$ and $\phi_0^I$ are constant matrices. The
differential equation (\ref{Nahn2}) reduces to
\begin{eqnarray}	%4.127
{f'\over f^2}\,\phi_0^I + {i\over 2}\,
\epsilon_{IJK}[\phi_0^J,\phi_0^K] = 0\,,
\label{soucelessNahm}
\end{eqnarray}
where the prime denotes derivative with respect to $x$. We shall
solve this equation by first putting
\begin{eqnarray}	%4.128
\phi_0^I = {1\over \sqrt{C_2(k)}} 
\left(\arraycolsep5pt\begin{array}{@{\hspace{2pt}}cc@{\hspace{0.8pt}}}
0 & 0 \\ 
0 & J^I\end{array}\right)\,,
\end{eqnarray}
where the $J^I$ are matrices in the $k$-dimensional irreducible
representation of the $SU(2)$ algebra, which satisfy the commutation
relations (\ref{Jcommutator}), and we have normalized the $\phi_0^I$'s
such that $\phi_0^I\phi_0^I$ is the unit matrix in the $k\times k$
block. By using this representation of the $\phi_0^I$'s,
Eq.~(\ref{soucelessNahm}) reduces to
\begin{eqnarray}	%4.129
{f'\over f^2} = {2\over \sqrt{C_2(k)}}\,,
\end{eqnarray}
which can be immediately integrated, namely
\begin{eqnarray}	%4.130
f = -{\sqrt{C_2(k)}\over 2 x}\,.
\end{eqnarray}
For large $k$, the quadratic Casimir $C_2(k)$ behaves as $k^2$ and
this equation reduces~to
\begin{eqnarray}	%4.131
f = -{k\over 2 x}\,.
\label{Nahnsolution}
\end{eqnarray}
Let us now take into account the standard relation between coordinates
$X_H^I$ and scalar fields $\phi^I_H$, namely
\begin{eqnarray}	%4.132
X_H^I = 2\pi\alpha' \phi^I_H\,,
\end{eqnarray}
and the fact that $\rho^2 \equiv X_H^I X_H^I$. Using these facts we
immediately get the following relation between $\rho$ and $f$:
\begin{eqnarray}	%4.133
\rho = 2\pi\alpha' f\,,
\end{eqnarray}
and the solution (\ref{Nahnsolution}) of the Nahm equation can be
 written as
\begin{eqnarray}	%4.134
\rho = -{\pi k\alpha'\over x}\,,
\end{eqnarray}
which, if we take into account the quantization condition (\ref{q-k}),
is just our embedding (\ref{bending}) for $x_0=0$. As expected,
$\rho$ blows up at $x=0$, while its dependence for $x\not=0$ gives
rise to the same bending as in the brane approach. Now we can
understand the $\delta$ term in (\ref{Nahn}), since it is precisely
this term the one taking care of the blow-up of the solution at
$x=0$. Notice also that, in this field theory perspective, the integer
$k$ is the rank of the gauge theory subgroup in which the Higgs branch
of the theory is realized, which corresponds to the number of
$D3$-branes that recombine into a $D5$-brane.

\subsubsection{Fluctuations in $Dp-D(p+2)$ with flux}
\label{Dp-D(p+2)fluctuations}

Let us now study the fluctuations around the codimension one
defect. As in the codimension 0 case, we can give a systematic
treatment of all the $Dp-D(p+2)$ intersection with flux, which in turn
behave similarly to the $D3-D5$ case studied above. Without loss of
generality we can take the unperturbed configuration as $z^1=L$,
$z^m=0$ $(m>1)$. Next, let us consider a fluctuation of the type:
\begin{eqnarray}	%135
\begin{array}{c}
z^1 = L + \chi^1\,, \qquad z^m = \chi^m \quad (m=2,\ldots,6-p)\,, \\[5pt]
x^p = {\cal X} + x\,,\qquad F = {\cal F} + f\,,
\end{array}
\end{eqnarray}
where the bending ${\cal X}$ and the worldvolume gauge field ${\cal
F}$ are given by Eqs.~(\ref{bending}) and (\ref{wvflux}) respectively
and we assume that $\chi^m$, $x$ and $f$ are small. It is important to
say that even for generic $p$ the bending is that of the $D3-D5$ case,
being the reason that we are always considering the same
codimensionality for the defect. Since this background, in addition to
the presence of the worldvolume gauge field, involves the bending, it
is interesting to consider the whole set of fluctuations.

The induced metric on the $D(p+2)$-brane worldvolume can be written
as
\begin{eqnarray}	%136
g = {\cal G} + g^{(f)}\,,
\end{eqnarray}
with ${\cal G}$ being the induced metric of the unperturbed
configuration:
\begin{eqnarray}	%137
{\cal G}_{ab}\,d\xi^a\,d\xi^b = h^{-1}\,dx_{1,p-1}^2 + 
h\bigg[\bigg(1 + {q^2\over \rho^4 h^2}\bigg)d\rho^2 + 
\rho^2\,d\Omega_2^2\bigg]\,,
\end{eqnarray}
where $h=h(\rho)$ is the function defined in (\ref{h}). Moreover,
$g^{(f)}$ is the part of $g$ that depends on the derivatives of the
fluctuations, namely
\begin{eqnarray}	%138
g^{(f)}_{ab} = {q\over \rho^2 h}\,(\delta_{a\rho}\partial_{b}x + 
\delta_{b\rho}\partial_{a}x) + {1\over h}\,\partial_{a} x \partial_{b}x + 
h\partial_{a}\chi^m\partial_{b}\chi^m\,.
\end{eqnarray}
Let us next rewrite the Born--Infeld determinant as
\begin{eqnarray}	%139
\sqrt{-\det(g+ F)} = \sqrt{-\det({\cal G} + {\cal F})}\sqrt{\det(1+X)}\,,
\label{detX}
\end{eqnarray}
where the matrix $X$ is given in this case by
\begin{eqnarray}	%140
X \equiv ({\cal G} + {\cal F})^{-1}\big(g^{(f)} + f\big)\,.
\label{matrixX}
\end{eqnarray}
We shall evaluate the r.h.s. of (\ref{detX}) by expanding it in powers
of $X$ by means of Eq.~(\ref{expansion}). In order to evaluate more
easily the trace of the powers of $X$ \hbox{appearing} on the
r.h.s. of this equation, let us separate the symmetric and
antisymmetric part in the inverse of the matrix ${\cal G} + {\cal F}$:
\begin{eqnarray}	%141
({\cal G} + {\cal F})^{-1} = \hat{\cal G}^{-1} + {\cal J}\,,
\end{eqnarray}
where
\begin{eqnarray}	%142
\hat{\cal G}^{-1} \equiv {1\over ({\cal G}+{\cal F})_S}\,,\qquad
{\cal J} \equiv {1\over ({\cal G}+{\cal F})_A}\,.
\end{eqnarray}
Notice that $\hat{\cal G}$ is just the open string metric which,
generalizing for any $p$ (\ref{ind-met-flux}), is given by
\begin{eqnarray}	%143
\hat{\cal G}_{ab}\,d\xi^a\,d\xi^b = h^{-1}\,dx_{1,p-1}^2 + h
\bigg(1 + {q^2\over \rho^4 h^2}\bigg)(d\rho^2 + \rho^2\,d\Omega_2^2)\,. 
\label{openstrmetric}
\end{eqnarray}
Moreover, the antisymmetric matrix ${\cal J}$ takes the form
\begin{eqnarray}	%4.144
{\cal J}^{\theta\varphi} = -{\cal J}^{\varphi\theta} = 
-{1\over \sqrt{\tilde g}}\,{q\over q^2 + \rho^4h^2}\,,
\end{eqnarray}
where $\theta$, $\varphi$ are the standard polar coordinates on $S^2$
and $\tilde g=\sin^2\theta$ is the determinant of its round metric.

After some algebra, one has that, dropping constant global factors
that do not affect the equations of motion, the relevant Lagrangian
for the fluctuations is
\begin{eqnarray}	%4.145
{\cal L} &=& -\rho^2\sqrt{\tilde g}\,\bigg[{h\over 2}
\bigg(1 + {q^2\over \rho^4 h^2}\bigg)\hat{\cal G}^{ab}
\partial_a\chi^m\partial_b\chi^m \nonumber \\
&&{}+ {1\over 2 h}\,\hat{\cal G}^{ab}\partial_a x\partial_b x +
{1\over 4}\bigg(1 + {q^2\over \rho^4 h^2}\bigg)f_{ab}f^{ab}\bigg] - 
{C(\rho)\over 2}\,x\epsilon^{ij}f_{ij}\,,
\label{fluct-lag}
\end{eqnarray}
where the indices $a$, $b$ are raised with the open string metric
$\hat{\cal G}$, and where we have made use of the Bianchi identity for
the gauge field fluctuations $\epsilon^{ij}\partial_i f_{j\rho} +\break
{\epsilon^{ij}\over 2}\partial_\rho f_{ij}=0$. Finally, the functions
$A(\rho)$, $C(\rho)$ are
\begin{equation}	%4.146
A(\rho) = \frac{d}{d\rho}\bigg[\frac{q^2}{h^2(q^2+\rho^4h^2)}\bigg]\,,
\qquad C(\rho)=\frac{d}{d\rho}\bigg[\frac{\rho^4}{q^2+\rho^4h^2}\bigg]\,.
\end{equation}

As it is manifest from (\ref{fluct-lag}), the transverse scalars
$\chi$ do not couple to other fields, while the scalar $x$ is coupled
to the fluctuations $f_{ij}$ of the gauge field strength along the
two-sphere. For simplicity we will restrict to the $\chi$ sector from
now on, although a complete analysis can be found in
\cite{ARR} and \cite{ARR2}. For the fluxless case $q=0$
these equations were solved in \cite{AR}, where it was shown
that they give rise to a discrete meson mass spectrum, which can be
computed numerically and, in the case of the $D3-D5$ intersection,
analytically. Let us examine here the situation for $q\not=0$. The
equation of motion for $\chi$ that follow from (\ref{fluct-lag}) is
\begin{eqnarray}	%4.147
\partial_a\bigg[\sqrt{\tilde g}\rho^2 h
\bigg(1 + {q^2\over \rho^4 h^2}\bigg)
\hat{\cal G}^{ab}\partial_b\chi\bigg] = 0\,.
\label{chi-fluct}
\end{eqnarray}
By using the explicit form of the open string metric $\hat{\cal
G}^{ab}$ (Eq.~(\ref{openstrmetric})), we can rewrite
(\ref{chi-fluct}) as
\begin{eqnarray}	%4.148
\partial_{\rho}(\rho^2\partial_\rho \chi) + 
\bigg[\rho^2h^2 + {q^2\over \rho^2}\bigg]
\partial^{\mu}\partial_{\mu}\chi+\nabla^i\nabla_i\chi = 0\,.
\end{eqnarray}
Let us separate variables and write the scalars in terms of the
eigenfunctions of the Laplacian in the Minkowski and sphere parts of
the brane geometry as
\begin{equation}	%4.149
\chi=e^{ikx} Y^l(S^2)\xi(\rho)\, ,
\end{equation}
where the product $kx$ is performed with the Minkowski metric and $l$
is the angular momentum on the $S^2$. The fluctuation equation for
the function $\xi$ is
\begin{equation}	%4.150
\partial_{\rho}(\rho^2\partial_{\rho}\xi) + 
\bigg\{\bigg[R^{4\alpha}\,\frac{\rho^2}{(\rho^2 + L^2)^{2\alpha}} + 
\frac{q^2}{\rho^2}\bigg]M^2 - l(l+1)\bigg\}\xi = 0\,, 
\label{radial-chi-fluct}
\end{equation}
where $M^2=-k^2$ is the mass of the meson. When the distance $L\not=0$
and $q=0$ Eq.~(\ref{radial-chi-fluct}) gives rise to a set of
normalizable solutions that occur for a discrete set of values of $M$
\cite{AR}. As argued in \cite{ARR} for the $D3-D5$ system, the
situation changes drastically when the flux is switched on. Indeed,
let us consider Eq.~(\ref{radial-chi-fluct}) when $L$,
$q\not=0$ in the IR, i.e. when $\rho$ is close to zero. In this case,
for small values of $\rho$, Eq.~(\ref{radial-chi-fluct}) reduces to
\begin{eqnarray}	%4.151
\partial_{\rho}(\rho^2\partial_{\rho}\xi) + 
\bigg[{q^2M^2\over\rho^2} - l(l+1)\bigg]\xi = 0 \quad (\rho\approx 0)\,.
\label{IRfluc}
\end{eqnarray}
Equation~(\ref{IRfluc}) can be solved in terms of Bessel functions,
namely
\begin{eqnarray}	%4.152
\xi = {1\over \sqrt{\rho}}\,J_{\pm \left(l+{1\over 2}\right)}
\bigg({qM\over \rho}\bigg) \quad (\rho\approx 0)\,.
\label{Bessel}
\end{eqnarray}
Near $\rho\approx 0$ the Bessel function (\ref{Bessel}) oscillates
infinitely as
\begin{eqnarray}	%4.153
\xi \approx e^{\pm i{qM\over \rho}} \quad (\rho\approx 0)\,.
\label{IRfluct-behaviour}
\end{eqnarray}
The behavior (\ref{IRfluct-behaviour}) implies that the spectrum of
$M$ is continuous and gapless. Actually, one can understand this
result by rewriting the function (\ref{Bessel}) in terms of the
coordinate $x^p$ by using (\ref{bending}). Indeed, $\rho\approx 0$
corresponds to large $|x^p|$ and $\xi(x^p)$ can be written in this
limit as a simple plane wave:
\begin{eqnarray}	%4.154
\xi \approx e^{\pm iMx^p} \quad (|x^p|\to\infty)\,.
\label{planewave}
\end{eqnarray}
Thus, the fluctuation spreads out of the defect locus at fixed $x^p$,
reflecting the fact that the bending has the effect of recombining,
rather than intersecting, the $Dp$-branes with the $D(p+2)$-branes. We
can understand this result by looking at the IR form of the open
string metric (\ref{openstrmetric}) and (\ref{ind-met-flux}). One
gets
\begin{eqnarray}	%4.155
\hat{\cal G}_{ab}\,d\xi^a\,d\xi^b \approx {L^{2\alpha}\over R^{2\alpha}}
\bigg[dx_{1,p-1}^2 + q^2\bigg({d\rho^2\over \rho^4} + 
{1\over \rho^2}\,d\Omega_2^2\bigg)\bigg] \quad (\rho\approx 0)\,.
\label{IRmetric}
\end{eqnarray}
By changing variables from $\rho$ to $u=q/\rho$, this metric can be
written as
\begin{eqnarray}	%4.156
{L^{2\alpha}\over R^{2\alpha}}
\big[dx_{1,p-1}^2 + du^2 + u^2\,d\Omega_2^2\big]\,,
\label{IRMinkowski}
\end{eqnarray}
which is nothing but the $(p+3)$-dimensional Minkowski space and,
thus, one \hbox{naturally} expects to get plane waves as in
(\ref{planewave}) as solutions of the fluctuation equations. This fact
is generic for all the fluctuations of this system. Although the rest
of the fluctuations in (\ref{fluct-lag}) are coupled, in
\cite{ARR} it is shown that they can be decoupled by
generalizing the results of \cite{WFO} and \cite{AR}. The
decoupled fluctua\-tion equations can actually be mapped \cite{MT} to
that satisfied by the scalars $\chi$. Thus, we conclude that the full
mesonic mass spectrum is continuous and gapless, as a consequence of
the recombination of the color and flavor branes induced by the
worldvolume flux.

\subsubsection{An S-dual picture\/$:$ the $F1-Dp$ intersection}

We would like to gain some more insight about the loss of the discrete
spectrum. In order to analyze in more detail the systems under study,
following \cite{ARR2}, let us consider increasing $g_s$. For
the IIB backgrounds, at some point the $D1$ string, rather than the
fundamental string, starts to be the light object. Upon performing an
S-duality, we can continue the description in terms of the S-dual
backgrounds.

Consider for a moment the particular case of the intersections above
in which $p=3$, corresponding to a $D3-D5$ intersection. In our
approach, the $D5$ is a probe in the background of the
$D3$. Interestingly, the $D3$ is a self S-dual object, and thus the
S-dual background will be once again $AdS_5\times S^5$. In turn, the
flavor $D5$ brane gets mapped to a $NS5$ brane. However, since the
dilaton is zero in this background, at least formally this situation
will be identical to the $D3-D5$ case already studied above. In
particular we will lose again the discrete spectrum.

We can look at the $p=1$ case, whose S-dual version is the $F1-D3$
intersection. In this case the system will not, at least not
trivially, behave as the one so far studied. Since from the gravity
point of view we can treat all the intersections in a generalized way,
we will analyze the more general system corresponding to the $F1-Dp$
intersection, according to the array:
\begin{eqnarray*}	
\arraycolsep5pt\begin{array}{rcccccccccl}
& 1 & 2 & \cdots & p+1 & p+2 & \cdots & 9 & & & \\[2pt]
F1: & \times & - & \cdots & - & - & \cdots & - & & \\[2pt]
Dp: & - & \times & \cdots & \times & - & \cdots & - & & &
\end{array}
\label{F1Dpintersection}
\end{eqnarray*}
The supersymmetry of such configurations can be explicitly seen in
\cite{ARR2}.

As in previous cases, we will consider a stack of $F1$ strings and we
will take the decoupling limit. Then, from the gravity perspective,
the system will be described as the near-horizon region of the $F1$
background, whose metric is given by
\begin{equation}	%4.157
ds^2=H^{-1}\,dx_{1,1}^2+d\vec{r}^{2}\,,
\label{F1metric}
\end{equation}
where, in the near-horizon limit, $H=R^6/r^6$, with $R^6 = 32\pi^2
(\alpha')^3g_s^2N$. The $F1$ background is also endowed with a NSNS
$B$ field and a nontrivial dilaton, given by
\begin{equation}	%4.158
B=H^{-1}\,dx_0\wedge dx_1\,, \qquad e^{-\Phi}=H^{\frac{1}{2}}\,.
\end{equation}
Let us now rewrite this solution in terms of a new coordinate system
more suitable for our probe analysis. First of all, we split the
coordinates transverse to the $F1$ as $\vec r=(\vec y,\vec z)$, where
the $\vec y$ vector corresponds to the directions $2,\ldots,p+1$ and
$\vec z$ refers to the coordinates transverse to both the $F1$ and
$Dp$-brane. Moreover, let us assume that $p>1$ and use spherical
coordinates to parametrize the subspace spanned by the $y$'s, i.e.
$d\vec y^{2} = d\rho^2 + \rho^2\,d\Omega^2_{p-1}$. Then, the metric
(\ref{F1metric}) can be rewritten as
\begin{equation}	%4.159
ds^2=H^{-1}\,dx_{1,1}^2 + d\rho^2 + \rho^2\,d\Omega_{p-1}^2 + d\vec{z}^{2}\,.
\label{F1metric-polar}
\end{equation}
The dynamics of the $Dp$-brane probe is determined by the DBI
Lagrangian, which in this case takes the form
\begin{eqnarray}	%4.160
{\cal L} = -T_p e^{-\phi}\sqrt{-\det(g+{\cal F})}\,,
\label{DBIactionF1Dp}
\end{eqnarray}
where ${\cal F}$ is the following combination of the worldvolume gauge
field strength $F$ and the pullback $P[B]$ of the NSNS $B$ field
\begin{eqnarray}	%4.161
{\cal F} = F - P[B]\,.
\end{eqnarray}
Let us choose $x^0$, $\rho$ and the $p-1$ angles parametrizing the
$S^{p-1}$ sphere as our set of worldvolume coordinates. We will
consider embeddings of the type
\begin{eqnarray}	%4.162
x^1 = x(\rho)\,, \qquad |\hspace{0.5pt}\vec z\hspace{0.5pt}| = L\,.
\label{F1Dp-embed-ansatz}
\end{eqnarray}
Moreover, we will switch on an electric field $F_{0\rho}\equiv F$ in
the worldvolume, such that the only nonvanishing component of ${\cal
F}$ is
\begin{eqnarray}	%4.163
{\cal F}_{0\rho} = F - H^{-1}x'\,,
\label{F1Dp-curlyF}
\end{eqnarray}
where, from now on, $H$ should be understood as the following function
of $\rho$:
\begin{eqnarray}	%4.164
H=H(\rho) = \bigg[{R^2\over \rho^2+L^2}\bigg]^3\,.
\end{eqnarray}
The introduction of the electric field is the counterpart of the
magnetic field we introduced prior to the S-duality accounting for
the dissolved color branes in the flavor one. Consider for simplicity
the $D1-D3$ intersection. As we know, the Higgs branch is achieved, in
the gravity picture, by adding a magnetic worldvolume gauge field
which had the effect of dissolving some of the background $D1$ in the
$D3$. In the S-dual case it is to be expected that we have to
dissolve some of the background $F1$ in the $D3$. However, this is
done by means not of a magnetic worldvolume gauge field, but in terms
of an electric one \cite{CM}.

The form of the Lagrangian density (\ref{DBIactionF1Dp}) for this
ansatz can be straight\-forwardly computed, with the result:
\begin{eqnarray}	%4.165
{\cal L} = -T_p\rho^{p-1}\sqrt{\tilde g}\sqrt{1+2Fx'-HF^2}\,,
\label{F1Dp-L}
\end{eqnarray}
and the equation of motion for the electric field $F$ is
\begin{eqnarray}	%4.166
{\partial\over \partial\rho}
\bigg[{\partial {\cal L}\over \partial F}\bigg] = 0\,.
\end{eqnarray}
This equation can be immediately integrated, namely
\begin{eqnarray}	%4.167
{\rho^{p-1}(HF - x')\over \sqrt{1+2Fx'-HF^2}} = c\,,
\label{electricFeq}
\end{eqnarray}
where $c$ is a constant. Moreover, from (\ref{electricFeq}) we can
obtain $F$ as a function of $x'$ and $\rho$:
\begin{eqnarray}	%4.168
F = H^{-1}\bigg[x' + c{\sqrt{H+(x')^2}\over \sqrt{c^2+\rho^{2(p-1)}H}}\bigg]\,.
\label{F1Dp-F}
\end{eqnarray}
Actually, $F$ can be eliminated in a systematic way by means of a
Legendre transformation. Indeed, let us define the Routhian density
${\cal R}$ as follows:
\begin{eqnarray}	%4.169
{\cal R} = F\,{\partial {\cal L}\over \partial F} - {\cal L}\,.
\end{eqnarray}
By computing the derivative in the explicit expression of ${\cal L}$
in (\ref{F1Dp-L}), and by using (\ref{F1Dp-F}), one can readily show
that ${\cal R}$ can be written as
\begin{eqnarray}	%4.170
{\cal R} = T_p\sqrt{\tilde g}H^{-1}
\big[\sqrt{c^2+\rho^{2(p-1)}H}\sqrt{H+(x')^2} + cx'\big]\,.
\label{F1DpRouthian}
\end{eqnarray}
The equation of motion for $x$ derived from ${\cal R}$ is just
\begin{eqnarray}	%4.171
{\partial\over \partial\rho}
\bigg[{\partial {\cal R}\over \partial x'}\bigg] = 0\,.
\end{eqnarray}
A particular solution of this equation can be obtained by requiring
the vanishing of $\partial {\cal R}/\partial x'$. By computing
explicitly this derivative from the expression of ${\cal R}$ in
(\ref{F1DpRouthian}) one easily shows that the value of $x'$ for this
particular solution is simply
\begin{eqnarray}	%4.172
x' = -{c\over \rho^{p-1}}\,,
\label{F1Dpbending}
\end{eqnarray}
which, for $p\not=2$ can be integrated as
\begin{eqnarray}	%4.173
x(\rho) = {c\over p-2}\,{1\over \rho^{p-2}} + {\rm const} \quad (p\not=2)\,,
\label{integratedF1Dpbending}
\end{eqnarray}
while for $p=2$ the $D2$-brane has a logarithmic bending of the type
$x(\rho)\sim -c\log \rho$.

After substituting (\ref{F1Dpbending}) on the r.h.s. of (\ref{F1Dp-F})
one can see that the world\-volume gauge field $F$ for this
configuration vanishes, i.e.
\begin{eqnarray}	%4.174
F=0\,.
\label{F=0}
\end{eqnarray}
Actually, it is also easy to verify from (\ref{F1Dp-F}) that the
requirement of having vanish\-ing electric gauge field on the
worldvolume is equivalent to having a bending given by
Eq.~(\ref{F1Dpbending}). Notice also that the on-shell Lagrangian
density (\ref{F1Dp-L}) for this configuration becomes ${\cal
L}=-T_p\rho^{p-1}\sqrt{\tilde g}$, which is independent of the
distance $L$. This suggests that the configuration is supersymmetric,
a fact that was explicitly verified in \cite{ARR2}.

Notice that the embedding (\ref{F1Dpbending}) depends on the constant
$c$. This constant is constrained by a flux quantization condition
which, for electric worldvolume gauge fields, was worked out in
\cite{Camino:2001at} and reads
\begin{eqnarray}	%4.175
\int_{S^{p-1}}{\partial {\cal L}\over \partial F} = nT_f\,,\qquad n\in Z\,.
\label{F1Dp-fluxquantization}
\end{eqnarray}
From (\ref{F1Dp-L}) one easily gets
\begin{eqnarray}	%4.176
{\partial {\cal L}\over \partial F}\bigg|_{ F=0} = T_p\sqrt{\tilde g}c\,,
\end{eqnarray}
which allows one to compute the integral on the l.h.s. of
(\ref{F1Dp-fluxquantization}). Let us express the result in terms of
the Yang--Mills coupling. Taking into account that the $Dp$-brane
tension $T_p$ is related to $g_{\rm YM}$ as $T_p=T_f^2/g^2_{\rm YM}$,
one easily arrives at the following expression of $c$ in terms of the
integer $n$:
\begin{eqnarray}	%4.177
c = {\alpha' g^2_{\rm YM}\over \Omega_{p-1}}\,2\pi n\,,
\label{c-quantization}
\end{eqnarray}
where $\Omega_{p-1}$ is the volume of a unit $S^{p-1}$, namely
$\Omega_{p-1}=2\pi^{{p\over 2}}/\Gamma\big({p\over 2}\big)$.
Physically, the integer $n$ represents the number of fundamental
strings that are reconnected to the $Dp$-brane. Notice that for $p=3$
Eq.~(\ref{c-quantization}) reduces to $c=n\pi\alpha' g_s$, to be
compared with the S-dual relation (\ref{q-k}).

\subsubsection*{Fluctuations}

Now we will study the fluctuations around the configuration described
by Eqs.~(\ref{F1Dp-embed-ansatz}) and (\ref{F=0}). We will only
analyze the fluctuations on the transverse $\vec z$ space, which we
will denote by $\chi$. After a straightforward computation, we get
that, up to quadratic order, the Lagrangian density of these
fluctuations is
\begin{equation}	%4.178
{\cal L} = -\rho^{p-1}\sqrt{\tilde{g}}
\bigg(1+\frac{c^2}{\rho^{2(p-1)}H}\bigg)
{\cal{G}}^{\mu\nu}\partial_{\mu}\chi\partial_{\nu}\chi\,,
\label{F1Dp-fluc-lag}
\end{equation}
where the effective metric ${\cal{G}}_{\mu\nu}$ is given by
\begin{equation}	%4.179
{\cal{G}}_{\mu\nu}\,dx^{\mu}\,dx^{\nu} = -H^{-1}(dx^0)^2 + 
\bigg(1+\frac{c^2}{\rho^{2(p-1)}H}\bigg)(d\rho^2+\rho^2\,d\Omega_{p-1}^2)\,.
\label{F1Dp-fluc-metric}
\end{equation}
One can verify that the equation derived from (\ref{F1Dp-fluc-lag})
for $p=3$ (i.e. for the $F1-D3$ intersection) matches precisely that
of the transverse scalar fluctuations of the $D1-D3$ system (i.e.
Eq.~(\ref{chi-fluct}) with $p=1$), once the constants $c$ and $q$ are
identified. This is, of course, expected from S-duality and implies
that the $F1-D3$ spectrum is continuous and gapless. For $p>3$ the
meson spectrum displays the same characteristics as in the $F1-D3$
intersection. However, the $F1-D2$ system behaves differently. Indeed,
for $p=2$ the profile function $x(\rho)$ is logarithmic (see
Eqs.~(\ref{F1Dpbending}) and (\ref{integratedF1Dpbending})). Moreover,
one can check that in this case the effective metric
(\ref{F1Dp-fluc-metric}) in the IR region $\rho\sim 0$ corresponds to
an space of the type ${\rm Min}_{1,1}\times S^1$. Actually, by
studying the fluctuation equation derived from (\ref{F1Dp-fluc-lag})
for $p=2$ and $\rho\sim 0$, one can verify that nonoscillatory
solutions can exist if the KK momentum in the $S^1$ is nonzero. As one
can check by solving numerically the fluctuation equation, in this
case the mass spectrum starts with a finite number of discrete states,
followed by a continuum.

\subsubsection[$M2-M5$ intersection and codimension one defects in\\ M-theory]{$M2-M5$ intersection and codimension one defects in M-theory}
\label{M2M5section}

Let us focus for a while on the $D2-D4$ intersection. In the probe
brane regime we have been considering so far, in which the flavor
brane is treated as a probe, we embed the $D4$ in the $D2$ background
and then take the decoupling limit. As discussed in the first section,
in this case, the gravity approximation is valid for $1\ll g_{\rm
eff}\ll N^{\frac{2}{5}}$; while the field theory description is valid
for $g_{\rm eff}\ll 1$. Since $g_{\rm eff}^2=\lambda\mu^{-1}=g_{\rm
YM}^2N\mu^{-1}$, and the decoupling limit involves fixed $\lambda$,
the regime in which we can trust the field theory is that of large
$\mu$; whereas in the low energy region we cannot trust the field
theory since we need some completion. However, as $\mu$ decreases, we
can enter a regime in which, in the dual gravity side, we have
\begin{equation}	%4.180
e^{\Phi}\sim g_{\rm eff}^{\frac{5}{2}}N^{-1} = 
g_{\rm YM}^{\frac{5}{2}}N^{\frac{1}{4}}\mu^{-\frac{5}{4}} = 
\lambda^{\frac{5}{4}}\mu^{-\frac{5}{4}}N^{-1}\sim 1\,,
\end{equation}
where we open the M-theory circle. From the gravity point of view, we
can uplift the system to eleven dimensions and continue its
description in terms of an 11-dimensional gravity dual. Then, our
system would be mapped to a $M2-M5$ intersection. In this case, the
completion of the field theory, for this energy range, is in terms of
the dual field theory of the M-theory system. Actually, we can give a
gravity description along the lines we have presented as long as we
ensure small curvatures. Then, we can use an 11-dimensional
supergravity approximation and consider, in very much of the same
spirit as we have been doing, the $M5$ brane as a probe in the $M2$
near-horizon.

More explicitly, the $M2-M5$ intersection we will consider is along
one common spatial dimension like:
\begin{eqnarray*}
\arraycolsep5pt\begin{array}{rccccccccccc}
& 1 & 2 & 3 & 4 & 5 & 6 & 7 & 8 & 9 & 10 \\[2pt]
M2: & \times & \times & - & - & - & - & - & - & - & - \\[2pt]
M5: & \times & - & \times & \times & \times & \times & - & - & - & -
\end{array}
\label{M2M5intersection}
\end{eqnarray*}

Since this configuration can be thought as the uplift of the $D2-D4$
intersection to eleven dimensions, we expect a behavior similar to the
one studied in Subsec.~\ref{DpDp+2section}. Indeed, notice that the
$M5$-brane induces a codimension one defect in the $M2$-brane
worldvolume. Considering the same probe-brane approximation as in the
string theory examples, we will treat the highest-dimensional brane
(i.e. the $M5$-brane) as a probe in the background created by the
lower-dimensional object, which in this case is the $M2$-brane. The
near-horizon metric of the $M2$-brane background of 11-dimensional
supergravity is
\begin{equation}	%4.181
ds^2=\frac{r^4}{R^4}\,dx_{1,2}^2 + \frac{R^2}{r^2}\,d\vec{r}^{2}\, ,
\label{M2metric}
\end{equation}
where $R$ is constant, $dx_{1,2}^2$ represents the Minkowski metric in
the directions $x^0$, $x^1$, $x^2$ of the $M2$-brane worldvolume and
$\vec r$ is an eight-dimensional vector transverse to the
$M2$-brane. The metric (\ref{M2metric}) is the one%\forcebreak{} 
of the
$AdS_4\times S^7$ space, where the radius of the $AdS_4$ ($S^7$)
factor is $R/2$ ($R$). The actual value of $R$ for a stack of $N$
coincident $M2$-branes is
\begin{eqnarray}	%4.182
R^6=32\pi^2 l_p^6 N\,,
\label{M2-radius}
\end{eqnarray}
where $l_p$ is the Planck length in eleven dimensions. This background
is also endowed with a three-form potential $C^{(3)}$, whose explicit
expression is
\begin{eqnarray}	%4.183
C^{(3)} = {r^6\over R^6}\,dx^0\wedge dx^1\wedge dx^2\,.
\label{C3M2}
\end{eqnarray}

The dynamics of the $M5$-brane probe is governed by the so-called PST
\hbox{action} \cite{PST,PST2,PST3}. In the PST formalism the worldvolume
fields are a three-form field strength $F$ and an auxiliary scalar
$a$. This action is given by \cite{PST}
\begin{eqnarray}	%4.184
S &=& T_{M5} \int d^{\hspace{0.5pt}6}\xi
\bigg[{-%\ko
}\sqrt{-\det(g_{ij} + \tilde H_{ij})} + 
{\sqrt{-\det g} \over 4\partial a\cdot\partial a}\,
\partial_i a(\star H)^{ijk}H_{jkl}\partial^l a\bigg] \nonumber \\
&&{}+T_{M5}\int\bigg[P\big[C^{(6)}\big] + {1\over 2}\,
F \wedge P\big[C^{(3)}\big]\bigg]\,,
\label{stnueve}
\end{eqnarray}
where $T_{M5}=1/(2\pi)^5l_p^6$ is the tension of the $M5$-brane, $g$
is the induced metric and $H$ is the following combination of the
worldvolume gauge field $F$ and the pullback of the three-form
$C^{(3)}$:
\begin{eqnarray}	%4.185
H = F - P\big[C^{(3)}\big]\,.
\label{stsiete}
\end{eqnarray}
Moreover, the field ${\tilde H}$ is defined as follows:
\begin{eqnarray}	%4.186
{\tilde H}^{ij} = {1\over 3!%\ko
\sqrt{-\det g}}\,
{1\over \sqrt{-(\partial a)^2}}\,\epsilon^{ijklmn}\partial_k a H_{lmn}\,,
\label{stocho}
\end{eqnarray}
and the worldvolume indices in (\ref{stnueve}) are lowered with the
induced metric $g_{ij}$.

In order to study the embedding of the $M5$-brane in the $M2$-brane
background, let us introduce a more convenient set of coordinates. Let
us split the vector $\vec r$ as $\vec r= (\vec y,\vec z)$, where $\vec
y=(y^1,\ldots, y^4)$ is the position vector along the directions
$3,4,5$ and 6 in the array (\ref{M2M5intersection}) and $\vec z =
(z^1,\ldots, z^4)$ corresponds to the directions $7,8,9$ and
$10$. Obviously, if $\rho^2=\vec y\cdot \vec y$, one has that $\vec
r^{2}=\rho^2+\vec z^{2}$ and $d\vec r^{2} = d\rho^2 +
\rho^2\,d\Omega_3^2 + d\vec{z}^{2}$, where $d\Omega_3^2$ is the line
element of a three-sphere. Thus, the metric (\ref{M2metric})\break
becomes
\begin{equation}	%4.187
ds^2 = \frac{(\rho^2+\vec z^{2})^2}{R^4}\,dx_{1,2}^2 + 
\frac{R^2}{\rho^2+\vec z^{2}}\,
(d\rho^2+\rho^2\,d\Omega_3^2+d\vec{z}^{2})\,.
\end{equation}
We will now choose $x^0$, $x^1$, $\rho$ and the three angular
coordinates that parametrize $d\Omega_3^2$ as our worldvolume
coordinates $\xi^i$. Moreover, we will assume that the vector $\vec
z$ is constant and we will denote its modulus by $L$, namely
\begin{eqnarray}	%4.188
|\hspace{0.5pt}\vec z\hspace{0.5pt}| = L\,.
\end{eqnarray}
%\vskip-\lastskip
%\pagebreak

\noindent
To specify completely the embedding of the $M5$-brane we must give the
form of the remaining scalar $x^2$ as a function of the worldvolume
coordinates. For simplicity we will assume that $x^2$ only depends on
the radial coordinate $\rho$, i.e. 
\begin{eqnarray}	%4.189
x^2 = x(\rho)\,.
\end{eqnarray}
Moreover, we will switch on a magnetic field $F$ along the
three-sphere of the $M5$-brane worldvolume, in the form
\begin{equation}	%4.190
F=qVol(S^3)\, ,
\label{M5flux}
\end{equation}
where $q$ is a constant and $Vol(S^3)$ is the volume form of the
worldvolume three-sphere. Notice that the induced metric for this
configuration is given by
\begin{eqnarray}	%4.191
g_{ij}\,d\xi^i\, d\xi^j &=& {(\rho^2+L^2)^2\over R^4}\,dx^{2}_{1,1} + 
{R^2\over \rho^2+L^2}\bigg\{\bigg(1 + {(\rho^2+L^2)^3\over R^6}\,(x')^2\bigg)
d\rho^2 + \rho^2\,d\Omega_3^2\bigg\}\,. \nonumber \\
\label{inducedmetricM5}
\end{eqnarray}

In order to write the PST action for our ansatz we must specify the
value of the PST scalar $a$. As pointed out in \cite{PST} the
field $a$ can be eliminated by gauge fixing, at the expense of losing
manifest covariance. Here we will choose a gauge such that the
auxiliary PST scalar is
\begin{eqnarray}	%4.192
a = x_1\,.
\end{eqnarray}
It is now straightforward to prove that the only nonvanishing
component of the field $\tilde H$ is
\begin{eqnarray}	%4.193
\tilde H_{x^0\rho} = -{i\over R^4}\,{(\rho^2+L^2)^2\over \rho^3}
\bigg(1 + {(\rho^2+L^2)^3\over R^6}\,(x')^2\bigg)^{{1\over 2}}q\,.
\label{tildeH}
\end{eqnarray}
Using these results we can write the PST action (\ref{stnueve}) as
\begin{eqnarray}	%4.194
S &=& -2\pi^2T_{M5}\int d^{\hspace{0.5pt}2}x\,d\rho \nonumber \\
&&{}\times\Bigg[\rho^3\sqrt{1 + {(\rho^2+L^2)^3\over R^6}\,(x')^2}%\ko
\sqrt{1 + {(\rho^2+L^2)^3\over R^6}\,{q^2\over \rho^6}} + 
{(\rho^2+L^2)^3\over R^6}\,qx'\Bigg]\,.\qquad \ \ \quad
\label{PSTansatz}
\end{eqnarray}
Let ${\cal L}$ be the Lagrangian density for the PST action, which we
can take as given by the expression inside the brackets in
(\ref{PSTansatz}). Since $x$ does not appear explicitly in the
action, one can immediately write a first integral of the equation of
motion of $x(\rho)$, namely
\begin{eqnarray}	%4.195
{\partial {\cal L}\over \partial x'} = {\rm const}\,.
\label{PSTcyclic}
\end{eqnarray}
By setting the constant on the r.h.s. of (\ref{PSTcyclic}) equal to
zero, this equation reduces to a simple first-order equation for
$x(\rho)$, i.e.
\begin{equation}	%4.196
x' = -\frac{q}{\rho^3}\,,
\label{BPSx}
\end{equation}
which can be immediately integrated to give
\begin{equation}	%4.197
x(\rho) = \bar x + \frac{q}{2\rho^2}\,,
\label{x-explicit}
\end{equation}
where $\bar x$ is a constant. Notice that the flux parametrized by $q$
induces a bending of the $M5$-brane, which is characterized by the
nontrivial dependence of $x$ on the holographic coordinate
$\rho$. Actually, when the first-order equation (\ref{BPSx}) holds,
the two square roots in (\ref{PSTansatz}) are equal and there is a
cancellation with the last term in (\ref{PSTansatz}). Indeed, the
on-shell action takes the form
\begin{equation}	%4.198
S=-2\pi^2T_5\int d^{\hspace{0.5pt}2}x\,d\rho \,\rho^3\, ,
\end{equation}
which is independent of the $M2-M5$ distance $L$. This is a signal of
supersymmetry and, indeed, as explicitly verified in
\cite{ARR}, the embeddings in which the flux and the bending
are related as in (\ref{BPSx}) are kappa symmetric. Thus,
Eq.~(\ref{BPSx}) can be regarded as the first-order BPS equation
required by supersymmetry. Notice also that the three-form flux
(\ref{M5flux}) induces $M2$-brane charge in the M5-brane worldvolume,
as it is manifest from the form of the PST action (\ref{stnueve}). In
complete analogy with the $Dp-D(p+2)$ system, we can interpret the
present M-theory configuration in terms of $M2$-branes that recombine
with the $M5$-brane. Moreover, in order to gain further insight on the
effect of the bending, let us rewrite the induced metric
(\ref{inducedmetricM5}) when the explicit form of $x(\rho)$ written in
Eq.~(\ref{x-explicit}) is taken into account. One gets
\begin{eqnarray}	%4.199
{(\rho^2+L^2)^2\over R^4}\,dx^{2}_{1,1} + {R^2\over \rho^2+L^2}
\bigg\{\bigg(1 + {q^2\over R^6}\,{(\rho^2+L^2)^3\over \rho^6}\bigg)
d\rho^2 + \rho^2\,d\Omega_3^2\bigg\}\,.\quad \
\label{AdS3-M5metric}
\end{eqnarray}
From (\ref{AdS3-M5metric}) one readily notices that the UV induced
metric at $\rho\to\infty$ takes the form $AdS_3 (R_{\rm eff}/2)\times
S^3 (R)$, where the $AdS_3$ radius $R_{\rm eff}$ depends on the flux
as
\begin{eqnarray}	%4.200
R_{\rm eff} = \bigg(1 + {q^2\over R^6}\bigg)^{1\over 2}\,R\,.
\label{ReffM}
\end{eqnarray}
Clearly, the case $L=0$, corresponding to a massless quark, verifies
that the induced worldvolume metric is of the form $AdS_3\times
S^3$. Thus, in this case, the theory is expected to enjoy a conformal
symmetry. Note that this system can be thought as the strong coupling
completion of the $D2-D4$ system; and therefore it seems that the
system develops a conformal symmetry in this regime.

Therefore, our $M5$-brane is wrapping an $AdS_3$ submanifold of the
$AdS_4$ background. Actually, there are infinite ways of embedding an
$AdS_3$ within an $AdS_4$ space and the bending of the probe induced
by the flux is selecting one particular case of these embeddings. In
order to shed light on this, let us suppose that we have an $AdS_4$
metric of the form
\begin{eqnarray}	%4.201
ds^2_{AdS_4} = {\rho^4\over R^4}\,dx^2_{1,2} + {R^2\over \rho^2}\,d\rho^2\,.
\label{AdS4metric}
\end{eqnarray}
Let us now change variables from $(x^{0,1}, x^2, \rho)$ to $(\hat
x^{0,1}, \varrho, \eta)$, as follows:
\begin{eqnarray}	%4.202
x^{0,1} = 2\hat x^{0,1}\,, \quad 
x^2 = \bar x + {2\over \varrho}\tanh\eta\,, \quad
\rho^2 = {R^3\over 4}\,\varrho\cosh\eta\,,
\label{changevariables}
\end{eqnarray}
where $\bar x$ is a constant. In these new variables the $AdS_4$
metric (\ref{AdS4metric}) can be written as a foliation by $AdS_3$
slices, namely
\begin{eqnarray}	%4.203
ds^2_{AdS_4} = {R^2\over 4}\,(\cosh^2\eta\,ds^2_{AdS_3} + d\eta^2)\,,
\label{foliation}
\end{eqnarray}
where $ds^2_{AdS_3}$ is given by
\begin{eqnarray}	%4.204
ds^2_{AdS_3} = \varrho^2(-(d\hat x^{0})^2 + (d\hat x^{1})^2) + 
{d\varrho ^2\over \varrho^2}\,.
\end{eqnarray}
Clearly the $AdS_3$ slices in (\ref{foliation}) can be obtained by
taking $\eta={\rm const}$. The radius of such $AdS_3$ slice is $R_{\rm
eff}/2$, with:
\begin{eqnarray}	%4.205
R_{\rm eff} = R\cosh\eta\,.
\label{sliceradiusM}
\end{eqnarray}
Moreover, one can verify easily by using the change of variables
(\ref{changevariables}) that our embedding (\ref{x-explicit})
corresponds to one of such $AdS_3$ slices with
\begin{eqnarray}	%4.206
\eta = \eta_q = \sinh^{-1}\bigg({q\over R^3}\bigg)\,.
\end{eqnarray}
Furthermore, one can check that the $AdS_3$ radius $R_{\rm eff}$ of
Eq.~(\ref{sliceradiusM}) reduces to (\ref{ReffM}) when $\eta=\eta_q$.

\subsubsection*{Fluctuations}

Let us now study the fluctuations of the $M2-M5$ intersection. For
simplicity we will focus on the fluctuations of the transverse scalars
which, without loss of generality, we will parametrize as
\begin{equation}	%4.207
z^1=L+\chi^1\,, \qquad z^m=\chi^m \quad (m=2,\ldots,4)\,.
\end{equation}
Let us substitute this ansatz in the PST action and keep up to second
order terms in the fluctuation $\chi$. As the calculation is very
similar to the one performed in Subsec.~\ref{Dp-D(p+2)fluctuations},
we skip the details and give the final result for the effective
Lagrangian of the fluctuations, namely
\begin{eqnarray}	%4.208
{\cal L} = -\rho^3\sqrt{\tilde g}\,{R^2\over \rho^2 + L^2}
\bigg[1 + {q^2\over R^6}\,{(\rho^2+L^2)^3\over \rho^6}\bigg]
\hat{\cal G}^{ij}\partial_i\chi\partial_j\chi\,,
\label{lag-fluc-M2M5}
\end{eqnarray}
where $\tilde g$ is the determinant of the round metric of the $S^3$
and $\hat{\cal G}_{ij}$ is the following effective metric on the
$M5$-brane worldvolume:
\begin{eqnarray}	%4.209
\hat{\cal G}_{ij}\,d\xi^i\,d\xi^j &=& {(\rho^2+L^2)^2\over R^4}\,
dx^{2}_{1,1}\nonumber \\
&&{} + {R^2\over \rho^2+L^2}\bigg(1 + {q^2\over R^6}\,
{(\rho^2+L^2)^3\over \rho^6}\bigg)(d\rho^2 + \rho^2\,d\Omega_3^2)\,.
\label{effect-metricM2M5}
\end{eqnarray}
Notice the close analogy with the $Dp-D(p+2)$ system studied in
Subsec.~\ref{Dp-D(p+2)fluctuations}. Actually
(\ref{effect-metricM2M5}) is the analogue of the open string metric in
this case. The equation of motion for the scalars can be derived
straightforwardly from the Lagrangian density
(\ref{lag-fluc-M2M5}). For $q=0$ this equation was integrated in
\cite{AR}, where the meson mass spectra was also
computed. This fluxless spectra is discrete and displays a mass
gap. As happened with the codimension one defects in type II theory
studied in Subsec.~\ref{DpDp+2section}, the situation changes
drastically when $q\not=0$. To verify this fact let us study the form
of the effective metric (\ref{effect-metricM2M5}) in the UV
($\rho\to\infty$) and in the IR ($\rho\to 0$). After studying this
metric when $\rho\to\infty$, one easily concludes that the UV is of
the form $AdS_3 (R_{\rm eff}/2)\times S^3 (R_{\rm eff})$, where
$R_{\rm eff}$ is just the effective radius with flux written in
(\ref{ReffM}). Thus, the effect of the flux in the UV is just a
redefinition of the $AdS_3$ and $S^3$ radii of the metric governing
the fluctuations. On the contrary, for $q\not=0$ the behavior of this
metric in the IR changes drastically with respect to the fluxless
case. Indeed, for $\rho\approx 0$ the metric (\ref{effect-metricM2M5})
takes the form
\begin{eqnarray}	%4.210
{L^{4}\over R^{4}}\bigg[dx_{1,1}^2 + q^2
\bigg({d\rho^2\over \rho^6} + {1\over \rho^4}\,d\Omega_2^2\bigg)\bigg] 
\quad (\rho\approx 0)\,.
\label{IR-M2M5metric}
\end{eqnarray}
Notice the analogy of (\ref{IR-M2M5metric}) with the IR metric
(\ref{IRmetric}) of the $Dp-D(p+2)$ defects. Actually, the IR limit of
the equation of motion of the fluctuation can be integrated, as in
(\ref{Bessel}), in terms of Bessel functions, which for $\rho\approx
0$ behave as plane waves of the form $e^{\pm i Mx}$, where $x$ is the
function (\ref{x-explicit}). Notice that $\rho\approx 0$ corresponds
to large $x$ in (\ref{x-explicit}). Thus, the fluctuations spread out
of the defect and oscillate infinitely at the IR and, as a
consequence, the mass spectrum is continuous and gapless. In complete
analogy with the $Dp-D(p+2)$ with flux, this is a consequence of the
recombination of the $M2$- and $M5$-branes and should be understood
microscopically in terms of dielectric multiple $M2$-branes polarized
into a $M5$-brane, once such an action is constructed.

\subsection{The codimension two defect} 

We will now analyze the codimension two defect, realized as a $Dp-Dp$
intersection over $p-2$ spatial dimensions. We will consider a single
$Dp'$-brane intersecting a stack of $N$ $Dp$-branes, according to the
array
\begin{eqnarray*}
\arraycolsep5pt\begin{array}{ccccccccccl}
& 1 & \cdots & p-2 & p-1 & p & p+1 & p+2 & \cdots & 9 & \\[2pt]
Dp: & \times & \cdots & \times & \times 
& \times & - & - & \cdots & - & \\[2pt]
Dp': & \times & \cdots & \times & - & - & \times 
& \times & \cdots & - &
\end{array}
\label{DpDpintersection}
\end{eqnarray*}
It is clear from the array (\ref{DpDpintersection}) that the
$Dp'$-brane produces a defect of codimension two in the field theory
dual to the stack of $Dp$-branes.

This intersection is very different from the others we have so far
studied. It was first analyzed along the gauge/gravity duality lines
in \cite{CEGK} and \cite{Erdmenger:2003kn}, and it was
further studied in \cite{ARR2}. For this intersections, for a
start, the two intersecting branes are of the same dimensionality, and
thus we do not have a decoupling of any of the local symmetries on the
branes. Indeed, the field theory dual is a $SU(N)\times SU(N')$ theory
which contains 2 copies of the dimensional reduction to the
worldvolumes of both $Dp$, $Dp'$ of the dimensional ${\mathcal{N}}=1$
Yang--Mills, coupled through some fields living in the common
intersection.

\subsubsection*{Gravity description}

Let us first start with the gravity description of these
intersections. In order to describe the dynamics of the $Dp'$-brane
probe, let us relabel the $x^{p-1}$ and $x^{p}$ coordinates appearing
in the metric (\ref{metric}) as
\begin{eqnarray}	%4.211
\lambda^1 \equiv x^{p-1}\,, \qquad \lambda^2 \equiv x^{p}\,.
\end{eqnarray}
Moreover, we will split the coordinates $\vec r$ transverse to the
$Dp$-branes as $\vec r = (\vec y,\vec z)$, where $\vec y = (y^1,y^2)$
corresponds to the $p+1$ and $p+2$ directions in
(\ref{DpDpintersection}) and $\vec z = (z^1,\ldots, z^{7-p})$ to the
remaining transverse coordinates. With this split of coordinates the
background metric reads
\begin{eqnarray}	%4.212
ds^2 = \bigg[{\vec y^{2} + \vec z^{2} \over R^2}\bigg]^{\alpha}
(dx_{1,p-2}^2 + d \vec\lambda^{2}) + 
\bigg[{R^2\over \vec y^{2} + \vec z^{2}}\bigg]^{\alpha}
(d\vec y^{2} + d \vec z^{2})\,,
\end{eqnarray}
where $dx_{1,p-2}^2$ is the Minkowski metric in the coordinates
$x^0,\ldots, x^{p-2}$ and $\alpha$ has been defined in (\ref{metric}).

To study the embeddings of the $Dp'$-brane probe in this background,
let us consider $\xi^m = (x^0,\ldots,x^{p-2},y^1,y^2)$ as worldvolume
coordinates. In this approach $\vec \lambda$ and $\vec z$ are scalar
fields that characterize the embedding. Actually, we will restrict
ourselves to the case in which $\vec \lambda$ depends only on the
$\vec y$ coordinates (i.e. $\vec\lambda = \vec\lambda(\vec y)$) and
the transverse separation $|\hspace{0.5pt}\vec z\hspace{0.5pt}|$ is constant, i.e.
$|\hspace{0.5pt}\vec z\hspace{0.5pt}| = L$.

Indeed, let us define the following complex combinations of
worldvolume coordinates and scalars:\footnote{The complex worldvolume
coordinate $Z$ should not be confused with the real transverse scalars
$\vec z$. Notice also that $\rho^2=|Z|^2$.}
\begin{eqnarray}	%4.213
Z = y^1 + iy^2\,, \qquad W = \lambda^1 + i\lambda^2\,.
\end{eqnarray}
In addition, if we define the holomorphic and antiholomorphic
derivatives as
\begin{eqnarray}	%4.214
\partial = {1\over 2}\,(\partial_1 - i\partial_2)\,, \qquad
\bar\partial = {1\over 2}\,(\partial_1 + i\partial_2)\,,
\end{eqnarray}
then \cite{ARR} one can see that the supersymmetric intersections can
be written as
\begin{eqnarray}	%4.215
\bar\partial\,W = 0\,,
\end{eqnarray}
whose general solution is an arbitrary holomorphic function of $Z$,
namely
\begin{eqnarray}	%4.216
W = W(Z)\,.
\end{eqnarray}
It is also straightforward to check that for these holomorphic
embeddings the \hbox{induced} metric takes the form
\begin{eqnarray}	%4.217
\bigg[{\rho^2+L^2\over R^2}\bigg]^{\alpha} dx^2_{1,p-2} + 
\bigg[{R^2\over \rho^2+L^2}\bigg]^{\alpha}
\bigg[1 + \bigg[{\rho^2+L^2\over R^2}\bigg]^{2\alpha}
\partial W\bar\partial \bar W\bigg]dZ\,d\bar Z\,,\qquad
\label{DpDp-ind-metric-holo}
\end{eqnarray}
whose determinant is
\begin{eqnarray}	%4.218
\sqrt{-\det (g)} = \bigg[{\rho^2+L^2\over R^2}\bigg]^{{(p-3)\alpha\over 2}}
\bigg[1 + \bigg[{\rho^2+L^2\over R^2}\bigg]^{2\alpha}
\partial W\bar\partial \bar W\bigg]\,.
\end{eqnarray}
Moreover, for these holomorphic embeddings the DBI Lagrangian density
takes the form
\begin{eqnarray}	%4.219
{\cal L}_{\rm DBI} = -T_p e^{-\phi}\sqrt{-\det (g)} = 
-T_p\bigg[1 + \bigg[{\rho^2 + L^2\over R^2}\bigg]^{2\alpha}
\partial W\bar\partial \bar W\bigg]\,,
\label{DpDpLDBI}
\end{eqnarray}
where we have used the value of $e^{-\phi}$ for the $Dp$-brane
background. On the other hand, from the form of the RR potential
$C^{(p+1)}$ written in (\ref{DpRR}) one can readily check that, for
these holomorphic embeddings, the WZ piece of the Lagrangian can be
written as
\begin{eqnarray}	%4.220
{\cal L}_{\rm WZ} = T_p\bigg[{\rho^2+L^2\over R^2}\bigg]^{2\alpha}
\partial W\bar\partial \bar W\,.
\end{eqnarray}
Notice that, for these holomorphic embeddings, the WZ Lagrangian
${\cal L}_{\rm WZ}$ cancels against the second term of ${\cal L}_{\rm
DBI}$ (see Eq.~(\ref{DpDpLDBI})). Thus, once again, the on-shell
action is independent of the distance $L$, a result which is a
consequence of supersymmetry and holomorphicity.

It can be seen \cite{ARR} that, from the point of view of
supersymmetry, any holomorphic curve $W(Z)$ is allowed. Obviously, we
could have $W={\rm const}$. In this case the probe sits at a
particular constant point of its transverse space and does not
recombine with branes of the background. Along the lines in the
previous sections, this corresponds to the Coulomb branch of the
theory. If, on the contrary, $W(Z)$ is not constant, Liouville theorem
ensures that it cannot be bounded in the whole complex plane. The
points at which $|W|$ diverge are spikes of the probe profile, and one
can interpret them as the points where the probe and background branes
merge. Notice that, as opposed to the other cases studied in this
paper, the nontrivial profile of the embedding is not induced by the
addition of any worldvolume field. Thus, we are not dissolving any
further charge in the probe brane and a dielectric interpretation is
not possible now. However, we can still think that this represents a
dissolution of some of the background branes in the ``flavor''
ones. From this perspective, since both ``flavor'' and ``color'' are
of the same type, we clearly do not need any further field. However,
we still have a bending arising from the recombination, which now can
be in any holomorphic way.

\subsubsection*{Field theory dual}

We now turn to the field theory description of the codimension 2
defect, for which we will focus in the particular case of $p=3$. The
field theory dual for this case has been worked out in
\cite{CEGK} and \cite{Kirsch:2004km}. The dual gauge
theory for this $D3-D3$ intersection was shown to correspond to two
${\cal N}=4$ four-dimensional theories living in two different copies
of $R^{1,3}$ which intersect each other along a two-dimensional
common subspace that hosts a bifundamental hypermultiplet. The action
for such a theory is quite involved, since the matter lives confined
to a $1+1$ defect in both of the $R^{1,3}$. For a start it is
necessary a careful embedding of a two-dimensional superspace into the
four-dimensional one. Using complex coordinates in four dimensions as
$(x^0,x^1,\theta, \bar\theta; w,\bar w)$, the two-dimensional
superspace is spanned by $(x^0,x^1,\theta, \bar\theta)$, while $w =
x^2 + ix^3$ should be thought of as a continuous index. Then, each
copy of the bulk ${\mathcal{N}}=4$ theory Lagrangian can be written in
terms of the two-dimensional superspace as
\begin{eqnarray}
S_{D3} &=&  \frac1{g^2}\int d^2x\, d^2w\, d^4\theta  Tr\Big(
\Sigma^\dagger\Sigma+(d w+g\bar\Phi)e^{gV}(d{\bar w} + g\Phi)e^{-gV}\\
&&\qquad\qquad+ \Sigma_{i=1,2}e^{-gV}\bar Q_i e^{gV} Q_i\Big)\\
&&\qquad +\int d^2x\, d^2w\, d^2\theta Tr\left(Q_1[ d{\bar
w}+g\Phi,Q_2]\right) + c.c.\,, \label{action}
\end{eqnarray}
In this notation, the theory contains a vector superfield $V$ and
three chiral super\-fields $\Phi$, $Q_1$, and $Q_2$, all with respect to
the two-dimensional superspace. The four-dimensional gauge vector
splits in two pieces: $A_0$, $A_1$, which are contained in $V$; and
$A_2$, $A_3$, which form the lowest component of the chiral superfield
$\Phi$ as $\phi=A_2 + i A_3$. This chiral superfield transforms
inhomogeneously under $U(N)$ gauge transformations with nontrivial
dependence on the index $w$, which is inherited from the
four-dimensional point of view.

Out of the six original adjoint scalars, two of them are contained in
$V$ through a twisted chiral superfield which in the Abelian case is
$\Sigma = \bar D_+ \bar D_- V$. The four remaining adjoint scalars
comprise the lowest components of the chiral superfields $Q_1$ and
$Q_2$.

As we said, the bulk theory contains two copies of the
${\mathcal{N}}=4$ theory (one for each of the intersecting stacks). In
this bulk theory there is a defect which contains two chiral
superfields $B$ and $\tilde B$ in the $(N, \bar N^{\prime})$ and
$(\bar N, N^{\prime})$ representations of $U(N)\times U(N^{\prime})$,
which represent the $D3-D3'$ strings. The part of the action
containing these fields is
\begin{eqnarray}	%4.222
\label{defct}
S_{D3-D3'} &=& \int d^{\hspace{0.5pt}2}x\, d^{\hspace{0.5pt}4}\theta Tr
\big(e^{-gV'}\bar B e^{gV}B+e^{-gV}\bar{\tilde B}e^{gV'}\tilde B\big) \nonumber \\
&&{} +\frac{ig}2\int d^{\hspace{0.5pt}2}x\, d^{\hspace{0.5pt}2}\theta 
Tr(B\tilde BQ_1-\tilde BBQ_1')\,.
\end{eqnarray}
From now on, we will not write out the explicit dependence on the
coupling constant $g$ anymore, which is easily reintroduced as it
always enters as a prefactor of the $V$ and $\Phi$ superfields.

For this action, the vanishing of the $F$-terms in this theory requires
\begin{eqnarray}	%4.223--4.230
F_{Q_1} &=& d\bar w\,q_2 + [\phi,q_2] + \delta^{(2)}(w)b\tilde b=0\,, 
\label{fq1} \\
F_{Q_2} &=& d\bar w\,q_1+[\phi,q_1] = 0\,, \label{fq2}\\
F_\Phi &=& [q_1,q_2]\,, \label{fphi} \\
F_{Q^{\prime}_1} &=& d\bar y\, q_2^{\prime} + 
[\phi^{\prime},q_2^{\prime}] + \delta^{(2)}(y)\tilde b b=0\,, 
\label{fq1p} \\
F_{Q_2^{\prime}} &=& d\bar y\, q_1^{\prime} + 
[\phi^{\prime},q_1^{\prime}]=0\,, \label{fq2p} \\
F_{\Phi^{\prime}} &=& [q_1^{\prime},q_2^{\prime}]\,, 
\label{fphip} \\
F_B &=& \tilde bq_1\delta^{(2)}(w)-q_1'\tilde b\delta^{(2)}(y)=0\,,
\label{fb} \\
F_{\tilde B} &=& q_1b\delta^{(2)}(w)-bq_1'\delta^{(2)}(y)=0\,,
\label{fbt}
\end{eqnarray}
whereas the $D$-terms require
\begin{equation}	%4.231`
D= d w\phi - d{\bar w}\,\phi^\dagger + [\phi,\phi^\dagger] + 
[q_1,q_1^\dagger]+[q_2,q_2^\dagger] + \delta^{(2)}(w)
(bb^\dagger-\tilde b^\dagger\tilde b)=0\,.
\label{Dterm} 
\end{equation}

Assuming that all the gauge fields vanish, and that the $q$ fields are
regular, we have to impose that the $\delta$ term vanishes, so
\begin{equation}	%4.232
\tilde b\tilde b^\dagger = b^\dagger b\,.\label{backandforth}
\end{equation}

We can simultaneously diagonalize $q_1$ and $q_1'$ at $w=0$ since they
transform under different gauge groups. Then (\ref{fbt}) becomes
\begin{equation}	%4.233
0=b_{i'j}q_{1jj}(0)-q_{1i'i'}'(0)b_{i'j} = 
b_{i'j}(q_{1jj}(0)- q_{1i'i'}'(0))\,,
\label{branches}
\end{equation}
where the indices $i$, $j$ and $i'$, $j'$ denote $SU(N)$ and
$SU(N^{\prime})$ indices, respectively. The expression
(\ref{branches}) is satisfied if $b_{i'j}$ or
$q_{1jj}(0)-q_{1i'i'}'(0)$ vanish (and the same for the $\tilde{b}$)
fields). The vanishing of the $b$ fields corresponds to the Coulomb
branch. Then, $q_{1jj}(0)-q_{1i'i'}'(0)$ parametrize a particular
point of that branch, in which the gauge group will be
broken. However, we can demand that $q_{1jj}(0)-q_{1i'i'}'(0)$
vanishes, which corresponds to the Higgs branch.

Since it is possible to diagonalize simultaneously the $q$ fields,
there is no non-Abelian structure. This is the counterpart of what we
found in the gravity side, namely, that in this case there is no
worldvolume gauge field which could give rise to a microscopical
interpretation along the lines of the rest of the intersections
studied. One can restrict therefore to the Abelian case in which we
have a single $D3$ intersecting another $D3$.

Equation (\ref{fq2}) implies that $q_1$ is a holomorphic function of
$w$, a condition on the embedding coordinates that is well known to be
necessary for a supersymmetric brane configuration. The solution of
(\ref{fq1}) is
\begin{equation}	%4.234
\label{gnrl}
q_2(w) = \frac{b\tilde b}{2\pi i w} + h(w) \, ,
\end{equation}
where $h(w)$ is a holomorphic function of $w$. Assuming that this
function vanishes, we have a unique solution
\begin{equation}	%4.235
q_2(w) = \frac{\tilde b b}{2\pi i w}\,, \qquad 
q_2^{\prime}(y) = \frac{b\tilde b }{2\pi i y}\,. 
\label{solns}
\end{equation}
From the gravity perspective, $q_2$ ($q_2^{\prime}$) describe the
transverse fluctuations of each of the $D3$-branes. The actual
relation involves $q_2 \rightarrow \alpha^{\prime}y$ and $q_2'
\rightarrow \alpha^{\prime}w$ in (\ref{solns}), so finally
\begin{equation}	%4.236
wy = \frac{1}{2\pi i} b\tilde b \alpha^{\prime}\,,
\end{equation}
which is one particular holomorphic curve of the ones obtained above
from the gravity point of view. Interestingly, only for the embeddings
corresponding to the Higgs branch the induced UV metric is of the form
$AdS_3\times S^1$. Indeed, one can check that the metric
(\ref{DpDp-ind-metric-holo}) for $p=3$ (and $\alpha=1$) and for the
profile $W=c/Z$ reduces in the UV to that of the $AdS_3\times S^1$
space, where the two factors have the same radii $R_{\rm
eff}=\sqrt{1+{c^2\over R^4}}\,\,R$. Thus, the constant $c$
parametrizes the particular $AdS_3\times S^1$ slice of the
$AdS_5\times S^5$ space that is occupied by our $D3$-brane probe.

\subsubsection*{Fluctuations of the $Dp-Dp$ intersection}

Let us now study the fluctuations around the configurations above for
a generic curve. We will concentrate on the fluctuations of the
scalars transverse to both types of branes, i.e. those along the
$\vec z$ directions. Let $\chi$ be one of such fields. Expanding the
action up to quadratic order in the fluctuations it is easy to see
that the Lagrangian density for $\chi$ is
\begin{eqnarray}	%4.237
{\cal L} = -\bigg[{R^2\over \rho^2+L^2}\bigg]^{\alpha}
\bigg[1 + \bigg[{\rho^2+L^2\over R^2}\bigg]^{2\alpha}
\partial W\bar\partial \bar W\bigg]
{\cal G}^{mn} \partial_m\chi \partial_n\chi\,,
\end{eqnarray}
where ${\cal G}_{mn}$ is the induced metric
(\ref{DpDp-ind-metric-holo}). Let us parametrize the complex variable
$Z$ in terms of polar coordinates as $Z=\rho e^{i\theta}$ and let us
separate variables in the fluctuation equation as
\begin{eqnarray}	%4.238
\chi = e^{ikx} e^{il\theta} \xi(\rho)\,,
\end{eqnarray}
where the product $kx$ is performed with the Minkowski metric of the
defect. If $M^2=-k^2$, the equation of motion for the radial function
$\xi(\rho)$ takes the form
\begin{eqnarray}	%4.239
\bigg[\bigg[{R^2\over \rho^2+L^2}\bigg]^{2\alpha}
\bigg[1 + \bigg[{\rho^2+L^2\over R^2}\bigg]^{2\alpha}
\partial W\bar\partial \bar W\bigg]M^2 - {l^2\over \rho^2} + 
{1\over \rho}\,\partial_{\rho}(\rho\partial_\rho)\bigg]
\xi(\rho) = 0\,.\qquad\quad
\label{DpDpflucteq}
\end{eqnarray}
For $W={\rm const}$, Eq.~(\ref{DpDpflucteq}) was solved in
\cite{AR}, where it was shown to give rise to a mass gap and a
discrete spectrum of $M$. As in the case of the codimension one
defects, this conclusion changes completely when we go to the Higgs
branch. Indeed, let us consider the embeddings with $W\sim 1/Z$. One
can readily prove that for $\rho\to\infty$ the function $\xi(\rho)$
behaves as $\xi(\rho)\sim c_1\rho^{l}+c_2\rho^{-l}$, which is exactly
the same behavior as in the $W={\rm const}$ case. However, in the
opposite limit $\rho\to 0$ the fluctuation equation can be solved in
terms of Bessel functions which oscillate infinitely as
$\rho\to0$. Notice that, for our Higgs branch embeddings, $\rho\to 0$
means $W\to\infty$ and, therefore, the fluctuations are no longer
localized at the defect, as it happened in the case of the $Dp-D(p+2)$
and $M2-M5$ intersections. Thus we conclude that, also in this case,
the mass gap is lost and the spectrum is continuous.

\section{Perspectives}

Understanding the strong coupling dynamics of gauge theories remains a
\hbox{major} issue. Clearly, the gauge/gravity duality represents a very 
deep and powerful \hbox{approach}. However, a vast number of issues
are still open. Among them, an outstanding problem is that of the
flavor. Including flavors in a fully satisfactory \hbox{manner} is a
very elusive problem, and it is just since very recently that a full
approach to the problem has been started.

The gauge/gravity correspondence can be seen as an open/closed
duality; in which, in the very specific low energy limit known as the
decoupling limit, going from weak to strong coupling takes us from an
open string description in terms of the worldvolume gauge theory on a
stack of branes into a closed string description in terms of strings
propagating in the near-horizon region of the supergravity solution
representing the branes. Since we assumed the branes in flat space, as
we described, the field theory description of the system is in terms
of the dimensional reduction down to $p+1$ dimensions of the maximally
supersymmetric Yang--Mills theory in ten dimensions. Those theories do
not have fundamental matter, and precisely our target was to consider
the inclusion in these theories of a quark sector in a way such that
we have a controlled gravity dual. As we described, this amounts to
bring into the game a new sector of open strings localized in the
common intersection of two stacks of branes. The spirit of the
gauge/gravity duality suggests to find a gravity description in terms
of the near-horizon of the supergravity background corresponding to
the brane intersection. More precisely, the supersymmetric
intersections which we considered are the $Dp-Dp+4$, $Dp-Dp+2$ and
$Dp-Dp$. However, this approach is in general quite involved, and it
is just since very recently that it has been started for some cases
(see
\cite{unquenched,unquenched2,unquenched3,unquenched4, rf60,rf61,unquenched5}). In
turn, we described a simpler approach, in which we consider the flavor
branes as probes in the background of the color ones. Since the
flavors do not backreact in the color, this approximation is some sort
of quenched approach. In addition, the gauge symmetry on the flavor
branes decoupled as a local symmetry, and remained (except for the
$Dp-Dp$ case) as a global flavor symmetry. With the limitations set by
the quenched approximation on mind, we were able to obtain the gravity
duals for a series of gauge theories preserving 8 supercharges
engineered by considering $N_f\ll N_c$ fundamental hypermultiplets
confined into a defect in a bulk Yang--Mills theory which preserves 16
supercharges. For this class of theories, the gravity dual is the near
horizon of the color branes with the flavor branes embedded as probes.

\looseness-1
The theories under study have a somehow rich phase structure, which
should be captured in some way by the supergravity approach. We have
reviewed how gravity beautifully reflects the Coulomb and Higgs
branches of the theories. Heuristically, the Coulomb branch
corresponds to the bare intersection. Motion along the Coulomb branch
is achieved by moving the color branes, as if there were no
\hbox{flavors}. However, in the flavored case, one can separate some of 
the color branes and dissolve them in the flavor branes by means of
turning on a nontrivial world\-volume gauge field. The dissolution
amounts to give a VEV for the open string fields (namely the quarks),
thus entering the Higgs branch. Interestingly, for all cases but the
$Dp-Dp$ intersection, it is possible to describe this process from the
point of view of the \hbox{separated} color branes via the dielectric
effect. This provides a very nice and \hbox{explicit} bridge between the
field theory and the gravity description, which we \hbox{explicitly}
saw from field theory and gravity for the case of color $D3$
branes. The $Dp-Dp$ intersection is somehow special, since there for a
start both the flavor and color branes are of the same dimensionality,
and therefore, the flavor symmetry is still a gauge\break symmetry. In this
case, although the Higgs branch is still in terms of a brane
recom\-bination, it is not possible to pass through the microscopical
description.

Moreover, the gauge/gravity duality allows us to get more knowledge of
the strong coupling dynamics of the gauge theories. Indeed, by
computing the fluctua\-tions of the probe branes, we were able to
compute the meson spectrum of the field theories. From the field
theory point of view, as long as we move from the Coulomb to the Higgs
branch, one would expect a change in the meson spectrum. This is
indeed confirmed from the gravity point of view. Since the embeddings
of the flavor branes are different in the Coulomb and Higgs branches
because of the world\-volume gauge fields, the fluctuations spectrum
changes. We saw that very interesting phenomena, such as the spectral
flow in the $Dp-Dp+4$ case and the loss of the discrete spectrum in
the other cases happened.

We have concentrated in theories preserving eight supercharges
engineered as $N_f\ll N_c$ hypermultiplets confined into a $1/2$-BPS
defect in a bulk gauge \hbox{theory} preserving 16 supercharges. This is a
very small subset of all the flavorings which have been considered in
the literature, which is vast in this topic. We did not \hbox{attempt}
to review all the considered possibilities, and we concentrated in
studying a particular class of theories; for which we found the
gravity dual of their branches. Actually, we expect that other types
of theories (namely different supersymmetries, for example) will
behave in a very different manner. For example, it is well known that
the moduli space of instantons can be mapped with the Higgs branch
just for ${\mathcal{N}}=2$ theories with flavors filling the whole
space. This property was a key issue in understanding the Higgs branch
by passing through the microscopic description in $D3-D7$, and thus,
for the supersymmetries, we expect that a different picture would
emerge.\footnote{Flavorings of generic $AdS_5\times Y^{p,q}$ have been
studied in for example \cite{SEdefects} and
\cite{SEdefects2}. However, a systematic study of these theories
has not been carried out.} Actually, a particularly interesting case
would be to analyze the phases of the flavorings of ${\mathcal{N}}=1$
Yang--Mills obtained by adding probe branes both in Klebanov--Strassler
and in Maldacena--Nu\~nez/Chamseddine--Volkov\hspace{1pt}\footnote{The various
flavorings of the MN/ChV \cite{MN2,MN} have been studied in
\cite{dMN} and \cite{flavoring}.} backgrounds. A nice
warmup for the former case would be to consider the branches of the
flavorings of the Klebanov--Witten background considering for example
the embeddings in \cite{Ouyang}. Actually, since a quark VEV
would break the baryonic symmetry, it might be plausible that in this
case the Higgs branch requires motion along the K\"ahler moduli space of
the Calabi--Yau cone.

To finish, let us mention that recently a considerable effort has been
put in under\-standing those theories at finite temperature (for a
review see \cite{rf107}). Introducing temperature amounts to
considering a black brane \cite{MAGOO}. The corresponding supergravity
background has a horizon and temperature. In turn, the field theory
dual is to be taken at finite temperature, where it behaves as some
kind of plasma. The flavoring goes along the same lines, with the very
important difference that now the embeddings are characterized in
terms of those which do not touch the black hole and those which
penetrate the horizon. This is seen as a phase transition in the field
theory side, and it was studied in \cite{MMT}, \cite{CJ1} and others (in \cite{CJ2, CJ3, CJ4, CJ5} this study has been carried out in the presence of external fields, yielding to a number of interesting phenomena) . In addition,
one can study the open string fluctuations to obtain the meson
spectrum at finite temperature. Interesting things can now happen,
like imaginary masses for the bound states, which are interpreted as
decay modes corresponding to melting mesons
\cite{melting,melting2,melting3,melting4}. Although somehow more
distant to what we considered, namely the physics of the flavor, in
this finite temperature context it is very interesting to analyze not
just open string fluctuations but also closed string fluctuations (and
in general quantities related to the closed string sector), which give
information about the properties of the plasma itself such as
conductivity, viscosity and so on (see for example
\cite{finiteT,finiteT2,finiteT3,finiteT4,finiteT5},
references therein and papers referring to).

\section*{Acknowledgments}

We are grateful to D. Arean, C.~Herzog, C.~Hoyos and S.~Montanez for illuminating
discussions. We would like to thank Oviedo University, and specially
Yolanda Lozano, for warm hospitality while this work was being
completed. We would like to specially thank A.~V. Ramallo for his
unlimited patience and wisdom leading to some of the works on which
this contribution is based. This work has been partially supported by
a MEC-Fulbright fellowship FU-2006-07040.

\appendix
\section[The Action for Coincident Branes and the Dielectric Effect]{The Action for Coincident Branes and the\\ Dielectric Effect}

It is well known that a stack of $N$ coincident branes carries, as low
energy worldvolume theory, a $U(N)$ theory. However, the naive
generalization of the action for a single brane to a non-Abelian gauge
theory does not correctly describe the system, since that would
explicitly violate the T-duality expectation that a T-dualizing a
stack of $Dp$ branes along a worldvolume coordinate would yield to a
stack of coincident $Dp-1$ branes. By demanding consistency with
T-duality, Myers \cite{M} found an action for $N$ coincident branes
which, in its most general form, reads
\begin{equation}
\label{MyersDBIapp}
\tilde{S}_{BI}=-T_p \int d^{p+1}\sigma\,Tr\left(e^{-\phi}\sqrt{-det\left(
P\left[E_{ab}+E_{ai}(Q^{-1}-\delta)^{ij}E_{jb}\right]+
\l\,F_{ab}\right)\,det(Q^i{}_j)}
\right)\ .
\end{equation}
for the DBI, where
\begin{eqnarray}	%A.2
Q^i{}_j \equiv \delta^i{}_j+i2\pi l_s^2[\Phi^i,\Phi^k]E_{kj}\,,\qquad 
E_{\mu\nu}=G_{\mu\nu}+B_{\mu\nu}
\label{extra}
\end{eqnarray}
and
\begin{eqnarray}	%A.3
S_{\rm CS} = \mu_p\int Tr\Big(P\Big[e^{i 2\pi l_s^2i_\Phi i_\Phi} 
\Big(\sum C^{(n)}e^B\Big)\Big]e^{2\pi l_s^2F}\Big)
\label{MyersCS}
\end{eqnarray}
for the CS (or WZ).

The trace is assumed to be a symmetrized trace, which today is known
not to be valid beyond $l_s^6$. However, since we will be interested
in comparing with the macroscopical description, we will be
insensitive to those problems.

Since we have a stack of branes, the transverse positions of the stack
becomes a $U(N)$-valued field. The fields $\Phi^a$ are the worldvolume
adjoint scalars, which, from the target space point of view, have the
interpretation of the transverse positions of the branes
\cite{bound}. Actually, to be precise, the relation between positions
and fields is $X^a=2\pi l_s^2\Phi^a$. The diagonal entries are
interpreted as the positions of each single brane, in such a way that
if we have $\Phi^a$ diagonal, this corresponds to separating each
single brane. Setting all the eigenvalues to the same value amounts to
make all the branes coincide. Furthermore, the off-diagonal entries
are interpreted as the open string interactions between the branes in
the stack.

The action (\ref{MyersDBIapp}) + (\ref{MyersCS}) is valid for coincident
branes, so we have to ensure that the typical distance between the
branes (call it $d$) is always smaller than the typical size of the
object which one would use to prove the system, namely a string. Thus,
$d\ll l_s$.

Interestingly, because of the matrix-valued character, the stack of
branes has a much richer dynamics to that of a single brane. One way
to see this is to consider the WZ action. A single $Dp$ brane couples
at most to a $p+1$ RR potential. Actually, the coupling to $C^{(p+1)}$
simply reflects the fact that the brane is the source of that
field. However, from (\ref{MyersCS}), we see that a stack of branes
can couple higher potentials through the $e^{i2\pi l_s^2
i_{\Phi}i_{\Phi}}$ term. This means that a stack of branes can carry
higher-dimensional brane charge, and thus can enjoy the properties of
a higher-dimensional brane. In particular, it is possible that the
$Dp$ branes polarize, in very much of the same spirit as a dipole in
an electric field, into higher dimensional branes. We have explicitly
seen examples of this in the main body of this paper. However, in
order to illustrate this in an easier setup, consider a toy example
in which we have ten-dimensional Minkowski space with a RR 3-form
given by
\begin{equation}	%A.4
\label{RRtoy}
C^{(3)}=fx^k\epsilon_{ijk}\,dx^0\wedge dx^i\wedge dx^j\,,\quad i,j,k=1,2,3\,.
\end{equation}
A stack of $N$ $D0$ branes would couple to this potential through the
CS action as
\begin{equation}	%A.5
S_{\rm CS}^{D0}=-i\frac{T_0f}{2\pi l_s^2}\int
Tr[[X^i,X^j]X^k\epsilon_{ijk}]\,.
\end{equation}
Assume that the branes polarize to a fuzzy 2-sphere of radius $R$:
\begin{equation}	%A.6
X^i=\frac{R}{\sqrt{C_2(N)}}J^i\,, \qquad \vec{X}^2=R^2\,,
\end{equation}
where $C_2(N)$ is the quadratic Casimir of the $SU(2)$ irreducible
representation whose generators are the $J^i$. Then, we have
\begin{equation}	%A.7
S_{\rm CS}^{D0}=\int \frac{2T_0R^2f}{2\pi l_s^2}\frac{N}{\sqrt{C_2(N)}}\,.
\end{equation}

The dimension of the representation is the number of branes. Thus, the
``density" of branes in the sphere is given by $\frac{N}{R^2}$. If we
want the branes to be effectively coincident, we have to demand that
the distance between them is much smaller than the typical open string
size, so
\begin{equation}	%A.8
\frac{R^2}{N}\ll l_s^2\,,
\end{equation}
which forces us to take a large number of branes. Then
$N/\sqrt{C_2(N)}\sim 1$, so approximately
\begin{equation}	%A.9
\label{SD0}
S_{\rm CS}^{D0}\sim\int \frac{2T_0R^2f}{2\pi l_s^2}\,.
\end{equation}

In the coincident branes limit, the system approximates an $S^2$ with
$C^{(3)}$ charge and dissolved $D0$ branes. We can match those charges
and topology considering a spherical $D2$ brane with $N$ dissolved
$D0$. In order to do that, we have to add a magnetic field so that
\begin{equation}	%A.10
T_2\int F\wedge C^{(1)} = NT_0\int C^{(1)} \rightarrow 
F = \frac{N}{2}dVol(S^2)\,.
\end{equation}
For this brane, it is straightforward to compute the CS action, which
reads
\begin{equation}	%A.11
S_{\rm CS}^{D2}=\int 4\pi T_3R^2 f\,.
\end{equation}
However, note that $T_0=4\pi^2l_s^2$, so
\begin{equation}	%A.12
S_{\rm CS}^{D2} = \int \frac{2T_0R^2 f}{2\pi l_s^2}\,,
\end{equation}
which precisely coincides with (\ref{SD0}), explicitly showing how, in
the limit of coincident branes, both the macroscopic and microscopic
descriptions reflect the same physics.

In the example at hand, the sphere wrapped by the system is not
topologically stable, so it must be ensured that the flux is enough to
overcome the tension tending to make the system collapse to a pointlike
object. In this case, as in many other \hbox{examples} such as the giant
graviton case (see \cite{giants}), the equili\-brium is
\hbox{dynamical}, and the flux plays a key role supporting the brane against
collapse. However, there are other cases in which the flux does not
play a role, and it is just the geometric background the responsible
of the stability \cite{gravdielec,gravdielec2}. Hence the name for
those cases of {\it purely gravitational dielectric effect}. This is
precisely the situation in the examples of the main body of the paper.

\section{Meson Masses in $Dp-Dp+4$}

\begin{figure}
\centerline{\hskip -.1in \epsffile{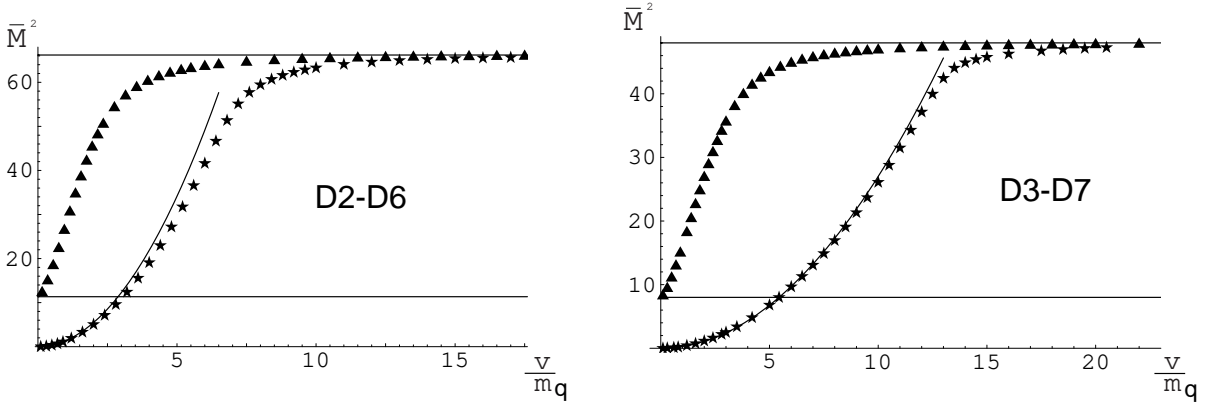}}
\caption{In this figure we  plot the numerical masses for the
first level as a function of 
the instanton size for both the full equation (with stars) and for the
equation obtained in \cite{EGG} (with solid triangles). The quark mass
$m_q$ is such that $g_{eff}(m_q)=1$. The solid line
corresponds to the WKB prediction (\ref{WKB-smallv}) for small $v$. The plot
on the left (right) corresponds to the D2-D6 (D3-D7) intersection.}
\label{massinstanton}
\end{figure}

In order to study the fluctuation equation (\ref{fluctuationsDp+4}) it
is interesting to notice that, after a change of variable,
(\ref{fluctuationsDp+4}) can be converted into a Schr\"odinger
equation. Indeed, let us change from $\varrho$ and $f$ to the new
variables $z$ and $\psi$, defined as
\begin{eqnarray}	%B.1
e^z = \varrho\,, \qquad \psi = \varrho f\,.
\end{eqnarray}
Notice that $\varrho\to\infty$ corresponds to $z\to +\infty$, while
$\varrho=0$ is mapped to $z=-\infty$. Moreover, one can readily prove
that, in terms of $z$ and $\psi$, Eq.~(\ref{fluctuationsDp+4}) can be\break
recast~as
\begin{eqnarray}	%B.2
\partial^2_z \psi - V(z)\psi = 0\,,
\label{Sch-eq}
\end{eqnarray}
where the potential $V(z)$ is given by
\begin{eqnarray}	%B.3
V(z) &=& 1 + \bigg(\frac{v}{m_q}\bigg)^4\frac{8}
{\Big(e^{2z}+ \big(\frac{v}{m_q}\big)^2\Big)^2} \nonumber \\
&&{} - \bar M^2\,{e^{2z}\over (e^{2z}+1)^{{7-p\over 2}}}
\left[1 + c_{p}(v,m_q)\,{(e^{2z}+1)^{{7-p\over 2}} \over 
\Big(e^{2z}+\big(\frac{v}{m_q}\big)^2\Big)^4}\right]\,.
\label{Sch-pot}
\end{eqnarray}
Notice that the reduced mass $\bar M$ is just a parameter in $V(z)$.
Actually, in these new variables the problem of finding the mass
spectrum can be rephrased as that of finding the values of $\bar M$
that allow a zero-energy level for the potential (\ref{Sch-pot}). By
using the standard techniques in quantum mechanics one can convince
oneself that such solutions exist. Indeed, the potential
(\ref{Sch-pot}) is strictly positive for $z\to\pm\infty$ and has some
minima for finite values of $z$. The actual calculation of the mass
spectra must be done by means of numerical techniques. A key
ingredient in this approach is the knowledge of the asymptotic
behavior of the solution when $\varrho\to 0$ and
$\varrho\to\infty$. This behavior can be easily obtained from the form
of the potential $V(z)$ in (\ref{Sch-pot}). Indeed, for
$\varrho\to\infty$, or equivalently for $z\to +\infty$, the potential
$V(z)\to 1$, and the solutions of (\ref{Sch-eq}) behave as $\psi\sim
e^{\pm z}$ which, in terms%\forcebreak{} 
of the original variables,
corresponds to $f={\rm const}$, $\varrho^{-2}$. Similarly for
$\varrho\to0$ (or $z\to-\infty$) one gets that $f=\varrho^{2}$,
$\varrho^{-4}$. Thus, the IR and UV behaviors of the fluctuation are
\begin{eqnarray}	%B.4
\begin{array}{rcll}
f(\varrho) &\approx& a_1\varrho^2 + a_2\varrho^{-4} 
& \quad (\varrho\to 0)\,, \\[5pt]
f(\varrho) &\approx& b_1\varrho^{-2} + b_2\,, 
& \quad (\varrho\to \infty)\,.
\end{array}
\label{UVIRbehaviour}
\end{eqnarray}
\looseness-1
The normalizable solutions are those that are regular at
$\varrho\approx 0$ and decrease at $\varrho\approx \infty$. Thus they
correspond to having $a_2=b_2=0$ in (\ref{UVIRbehaviour}). Upon
applying a shooting technique, we can determine the values of $\bar M$
for which such normalizable solutions exist. Notice that $\bar M$
depends parametrically on the quark mass $m_q$ and on its VEV $v$. In
general, for given values of $m_q$ and $v$, one gets a tower of
discrete values of $\bar M$. In Fig.~\ref{massinstanton} we have
plotted the values of the reduced mass for the first level, as a
function of the quark VEV. For illustrative purposes we have included
the values obtained with the fluctuation equation of
\cite{EGG}. As anticipated in Sec.~4, both results differ
significantly in the region of small $v$ and coincide when
$v\to\infty$. Actually, when $v$ is very large we recover the spectral
flow phenomenon described in \cite{EGG}, i.e. $\bar M$
becomes independent of the instanton size and equals the mass
corresponding to a higher Kaluza--Klein mode on the worldvolume
sphere. However, we see that when $\frac{v}{m_q}$ goes to zero, the
masses of the associated fluctuations also go to zero. Actually, this
limit is pretty singular. Indeed, it corresponds to the small
instanton limit, where it is expected that the moduli space of
instantons becomes effectively noncompact and that extra massless
degrees of freedom show up in the spectrum.

\enlargethispage*{14pt}

It turns out that the mass levels for small $v$ are nicely represented
analytically by means of the WKB approximation for the Schr\"odinger
problem (\ref{Sch-eq}). The WKB method has been very successful
\cite{MInahan,RS} in the calculation of the glueball mass spectra in
the gauge/gravity correspondence and also provides rather reliable
predictions for the mass levels of the mesons \cite{AR}. The WKB
quantization rule is
\begin{eqnarray}	%B.5
\bigg(n+{1\over 2}\bigg)\pi = \int_{z_1}^{z_2}dz\,\sqrt{-V(z)} 
\quad n\ge 0\,,
\label{WKBquantization}
\end{eqnarray}
where $n\in Z$ and $z_1$ and $z_2$ are the turning points of the
potential ($V(z_1)=\break V(z_2)=0$). Following straightforwardly the steps
of \cite{RS} and \cite{AR}, we obtain the following
expression for the WKB values of $\bar M$:
\begin{eqnarray}	%B.6
\bar M_{\rm WKB}^2 = {\pi^2\over \zeta^2}\,(n+1)
\bigg(n + 3 + {2\over 5-p}\bigg)\,,
\label{MWKB}
\end{eqnarray}
where $\zeta$ is the following integral:
\begin{eqnarray}	%B.7
\zeta = \int_{0}^{+\infty} d\varrho
\sqrt{{1\over (1+\varrho^2)^{{7-p\over 2}}} + 
{c_{p}(v,m_q) \over \Big[\big({v\over m_q}\big)^2 + \varrho^2\Big]^4}}\,.
\label{zetaWKB}
\end{eqnarray}
Let us evaluate analytically $\zeta$ when $v$ is small. First of all,
as can be easily checked, we notice that, when $v$ is small, the
second term under the square root in (\ref{zetaWKB}) behaves as
\begin{eqnarray}	%B.8
{1\over \Big[\big({v\over m_q}\big)^2 + \varrho^2\Big]^2} 
\approx {\pi\over 2}\bigg({m_q\over v}\bigg)^3\delta(\varrho)\,,
\quad {\rm as} \quad v\to 0\,.
\label{deltaVEV}
\end{eqnarray}
Then, one can see that this term dominates the integral defining
$\zeta$ around $\varrho\approx 0$ and, for small $v$, one can
approximate $\zeta$ as
\begin{eqnarray}	%B.9
\zeta \approx {\sqrt{c_{p}(v,m_q)}\over 2}\int_{-\epsilon}^{\epsilon}
{d\varrho\over \Big[\big({v\over m_q}\big)^2 + \varrho^2\Big]^2} + 
\int_0^{+\infty}{d\varrho\over (1+\varrho^2)^{{7-p\over 4}}}\,,
\end{eqnarray}
where $\epsilon$ is a small positive number and we have used the fact
that the function in (\ref{zetaWKB}) is an even function of $\varrho$.
Using (\ref{deltaVEV}), one can evaluate $\zeta$ as
\begin{eqnarray}	%B.10
\zeta \approx {\pi\over 4}\bigg({m_q\over v}\bigg)^3
\sqrt{c_{p}(v,m_q)} + {\sqrt{\pi}\over 2}\,
{\Gamma\big({5-p\over 4}\big) \over \Gamma\big({7-p\over 4}\big)}\,.
\label{zeta-approx}
\end{eqnarray}
Clearly, for $v\to 0$, we can neglect the last term in
(\ref{zeta-approx}). Using the expression of $c_{p}(v,m_q)$
(Eq.~(\ref{cp})), we arrive at
\begin{eqnarray}	%B.11
\zeta \approx {\sqrt{3} \cdot 2^{{p-4\over 2}}\pi^{{p+5\over 4}}\over
\sqrt{\Gamma\big({7-p\over 2}\big)}}\,{m_q\over g_{\rm eff}(m_q)v}\,,
\end{eqnarray}
and plugging this result in (\ref{MWKB}), we get the WKB mass of the
ground state ($n=0$) for small $v$:
\begin{eqnarray}	%B.12
\bar M_{\rm WKB}^2 \approx {(17-3p)\Gamma\big({5-p\over 2}\big)\over
3\cdot 2^{p-3}\pi^{{p+1\over 2}}}
\bigg({g_{\rm eff}(m_q) v\over m_q}\bigg)^2\,.
\label{WKB-smallv}
\end{eqnarray}
Thus, we predict that $\bar M^2$ is a quadratic function of $v/m_q$
with the particular coefficient given on the r.h.s. of
(\ref{WKB-smallv}). In Fig.~\ref{massinstanton} we have represented
by a solid line the value of $\bar{M}$ obtained from
Eq.~(\ref{WKB-smallv}). We notice that, for small $v$, this equation
nicely fits the values obtained by the numerical calculation.


\begin{thebibliography}{0}
\bibitem{jm} J.~M.~Maldacena,
``The large $N$ limit of superconformal field
theories and supergravity''
{\it Adv.\ Theor.\ Math.\ Phys.}\  {\bf  2} (1998)
231, {\rm hep-th/9711200}.

\bibitem{MAGOO}O. Aharony, S. Gubser, J. Maldacena,
H. Ooguri and Y. Oz,
``Large $N$ field theories, string theory and gravity",
{\sl Phys.  Rept. } {\bf 323} (2000) 183, {\rm hep-th/9905111}.

\bibitem{Susskind}
  L.~Susskind and E.~Witten,
 ``The holographic bound in anti-de Sitter space,''
 {\rm  arXiv:hep-th/9805114.}

\bibitem{Witten}
  E.~Witten,
  ``Anti-de Sitter space and holography,''
{\it  Adv.\ Theor.\ Math.\ Phys.}\  {\bf 2} (1998) 253
 {\rm arXiv:hep-th/9802150}.
 
 \bibitem{GKP}
   S.~S.~Gubser, I.~R.~Klebanov and A.~M.~Polyakov,
  ``Gauge theory correlators from non-critical string theory,''
  Phys.\ Lett.\  B {\bf 428}, 105 (1998)
  {\rm arXiv:hep-th/9802109}.

\bibitem{Labc1}
 S.~Benvenuti, S.~Franco, A.~Hanany, D.~Martelli and J.~Sparks,
 ``An infinite family of superconformal quiver gauge theories with Sasaki-Einstein duals,''
  {\sl \jhep} {\bf 0506} (2005) 064
  {\rm arXiv:hep-th/0411264}.
 
 \bibitem{Labc2}
    S.~Benvenuti and M.~Kruczenski,
 ``From Sasaki-Einstein spaces to quivers via BPS geodesics: L(p,q|r),''
  {\sl \jhep} {\bf 0604} (2006) 033
  {\rm arXiv:hep-th/0505206}.
  
  \bibitem{Labc3}
    S.~Franco, A.~Hanany, D.~Martelli, J.~Sparks, D.~Vegh and B.~Wecht,
  ``Gauge theories from toric geometry and brane tilings,''
  {\sl \jhep} {\bf 0601} (2006) 128
  {\sl arXiv:hep-th/0505211}.
  
  \bibitem{Labc4}
    A.~Butti, D.~Forcella and A.~Zaffaroni,
  ``The dual superconformal theory for L(p,q,r) manifolds,''
  {\sl \jhep} {\bf 0509} (2005) 018
  {\rm arXiv:hep-th/0505220}.
  
  \bibitem{KW}
  I.~R.~Klebanov and E.~Witten,
  `` Superconformal field theory on threebranes at a Calabi-Yau  
singularity,
 {\sl  Nucl.\ Phys.}\ B {\bf 536}, 199 (1998), {\rm arXiv:hep-th/9905104}.
  
  \bibitem{KM}
  I.~R.~Klebanov and A.~Murugan,
  ``Gauge / gravity duality and warped resolved conifold,''
  {\sl \jhep} {\bf 0703} (2007) 042
  {\rm arXiv:hep-th/0701064}.

\bibitem{KS}
  I.~R.~Klebanov and M.~J.~Strassler,
  ``Supergravity and a confining gauge theory: Duality cascades and chiSB-resolution of naked singularities,''
  {\sl \jhep} {\bf 0008} (2000) 052
  {\rm arXiv:hep-th/0007191}.

\bibitem{BST}
  H.~J.~Boonstra, K.~Skenderis and P.~K.~Townsend,
  ``The domain wall/QFT correspondence,''
  {\sl \jhep} {\bf 9901} (1999) 003
  {\rm arXiv:hep-th/9807137}.

\bibitem{IMSY}
  N.~Itzhaki, J.~M.~Maldacena, J.~Sonnenschein and S.~Yankielowicz,
  ``Supergravity and the large N limit of theories with sixteen supercharges,''
  {\sl \pr} {\bf D 58} (1998) 046004
  {\rm arXiv:hep-th/9802042}.
  
\bibitem{Mateo}
  M.~Bertolini, P.~Di Vecchia, M.~Frau, A.~Lerda, R.~Marotta and I.~Pesando,
  ``Fractional D-branes and their gauge duals,''
  {\sl \jhep} {\bf 0102} (2001) 014
  {\rm arXiv:hep-th/0011077}.
  
  \bibitem{Mateo2}
    M.~Bertolini, P.~Di Vecchia, M.~Frau, A.~Lerda and R.~Marotta,
  ``N = 2 gauge theories on systems of fractional D3/D7 branes,''
  Nucl.\ Phys.\  B {\bf 621} (2002) 157
  {\rm arXiv:hep-th/0107057}.
  
  \bibitem{Mateo3}
    M.~Grana and J.~Polchinski,
  ``Gauge / gravity duals with holomorphic dilaton,''
  Phys.\ Rev.\  D {\bf 65} (2002) 126005
  {\rm arXiv:hep-th/0106014}.

\bibitem{KR}A. Karch and L. Randall,
``Locally localized gravity",
{\sl \jhep} {\bf 0105 } (2001) 008, {\rm hep-th/0011156}.

\bibitem{KR2}
``Open and closed string
interpretation of SUSY CFT's on branes with boundaries",
{\sl \jhep} {\bf 0106 } (2001) 063, {\rm hep-th/0105132}.

\bibitem{KKW} A. Karch and E. Katz,
``Adding flavor to AdS/CFT",
{\sl \jhep} {\bf 0206 }(2002) 043, {\rm hep-th/0205236}.

\bibitem{KKW2}
A. Karch, E. Katz and N. Weiner,
``Hadron masses and screening from AdS Wilson loops",
{\sl \prl} {\bf 90 }(2003) 091601, {\rm hep-th/0211107}.

\bibitem{KMMW}M. Kruczenski, D. Mateos, R. Myers and D. Winters,
``Meson spectroscopy in AdS/CFT with flavour",
{\sl \jhep} {\bf 0307 }(2003) 049, {\rm hep-th/0304032}

\bibitem{Sonnen} T. Sakai and J. Sonnenschein, ``Probing flavored 
mesons of confining
gauge theories by supergravity",
{\sl \jhep} {\bf 0309 }(2003) 047, {\rm hep-th/0305049}.

\bibitem{Johana} J. Babington, J. Erdmenger, N. Evans, Z. Guralnik 
and I. Kirsch,
``Chiral symmetry breaking and pions in non-supersymmetric 
gauge/gravity duals", 
{\sl \pr} {\bf D69} (2004) 066007, {\rm hep-th/0306018}.

\bibitem{Johana2}
R. Apreda, J. Erdmenger and N. Evans, ``Scalar effective potential for
D7-brane probes which break chiral symmetry", {\rm hep-th/0509219}.

\bibitem{Johana3}
R. Apreda, J. Erdmenger, N. Evans, J. Grosse and Z. Guralnik,
``Instantons on D7 brane probes and AdS/CFT with flavour", 
{\rm hep-th/0601130}.

\bibitem{KMMW-two}M. Kruczenski, D. Mateos, R. Myers and D. Winters,
`Towards a holographic dual of large-$N_c$ QCD",
{\sl \jhep} {\bf 0405 }(2004) 041, {\rm hep-th/0311270}.

\bibitem{Carlos}
J. L. F. Barbon, C. Hoyos, D. Mateos and R. C. Myers,
``The holographic life of the eta'", 
{\sl \jhep} {\bf 0410 }(2004) 029, {\rm hep-th/0404260}.

\bibitem{Carlos2}
A.~Armoni, ``Witten-Veneziano from Green-Schwarz'',
{\sl \jhep} {\bf 0406}, 019 (2004), 
{\rm hep-th/0404248}.

\bibitem{Carlos3}
 J.~L.~Hovdebo, M.~Kruczenski, D.~Mateos, R.~C.~Myers and D.~J.~Winters,
``Holographic mesons: Adding flavor to the AdS/CFT duality,''
{\sl Int.\ J.\ Mod.\ Phys.}\ A {\bf 20} (2005) 3428.

\bibitem{Ouyang}P. Ouyang,
``Holomorphic D7-branes and flavored N=1 gauge dynamics", 
{\sl \np} {\bf B699 }(2004) 207, {\rm hep-th/0311084}.

\bibitem{Ouyang2}
T.~S.~Levi and P.~Ouyang, 
``Mesons and flavor on the conifold'', {\rm hep-th/0506021}.

\bibitem{WH}X.-J. Wang and S. Hu, ``Intersecting branes and 
adding flavors to the Maldacena-N\'u\~nez background",
{\sl \jhep} {\bf 0309 }(2003) 017  {\rm hep-th/0307218}.

\bibitem{flavoring}C. N\'u\~nez, A. Paredes and A. V. Ramallo,
``Flavoring the gravity dual of ${\cal N}=1$ Yang-Mills with probes",
{\sl \jhep} {\bf 0312 }(2003) 024, {\rm hep-th/0311201}.

\bibitem{Hong}S. Hong, S. Yoon, M. J. Strassler, 
``Quarkonium from the fifth dimension", 
{\sl \jhep} {\bf 0404 }(2004) 046, {\rm hep-th/0312071}.

\bibitem{Evans}N. Evans, J. P. Shock, ``Chiral dynamics from AdS space",
{\sl \pr} {\bf D70} (2004) 046002,{\rm hep-th/0403279}.

\bibitem{Evans2}
N. Evans, J. P. Shock and T. Waterson, ``D7 brane embeddings and chiral
symmetry breaking", {\sl \jhep} {\bf 0503 }(2005) 005, {\rm hep-th/0502091}.

\bibitem{Evans3}
J. P. ~Shock, ``Canonical coordinates and meson spectra for scalar deformed
${\cal N}=4$ SYM from the AdS/CFT correspondence", {\rm hep-th/0601025}.

\bibitem{Ghoroku} K. Ghoroku, M. Yahiro, ``Chiral symmetry breaking driven by
the dilaton", {\sl \pl} {\bf B604 }(2004) 235, {\rm hep-th/0408040}.

\bibitem{Ghoroku1}
``Holographic models for mesons at finite temperature", 
{\rm hep-ph/0512289}.

\bibitem{Ghoroku2}
K. Ghoroku, T. Sakaguchi, N. Uekusa and M. Yahiro, ``Flavor quark at high
temperature from a holographic model", 
{\sl \pr} {\bf D71} (2005) 106002, {\rm hep-th/0502088}.

\bibitem{Ghoroku3}
I. Brevik, K. Ghoroku and  A. Nakamura, ``Meson mass and confinement force
driven by the dilaton", {\rm hep-th/0505057}. 

\bibitem{melting}  K.~Peeters, J.~Sonnenschein and M.~Zamaklar,
``Holographic melting and related properties of mesons in a quark
gluon plasma", 
{\sl  Phys.\ Rev.} \  D {\bf 74},106008 (2006), {\rm hep-th/0606195}.

\bibitem{melting2}
S.~Kobayashi, D.~Mateos, S.~Matsuura, R.~C.~Myers and R.~M.~Thomson,
``Holographic phase transitions at finite baryon density",
{\sl \jhep} {\bf 0702}, 016 (2007),  {\rm hep-th/0611099}.

\bibitem{melting3}
 C.~Hoyos, K.~Landsteiner and S.~Montero,
``Holographic meson melting", {\rm hep-th/0612169}.

\bibitem{melting4}
 D.~Mateos, R.~C.~Myers and R.~M.~Thomson,
``Thermodynamics of the brane", {\rm  hep-th/0701132}.
 
\bibitem{conifold}D. Arean, D. Crooks and A. V. Ramallo, 
``Supersymmetric probes on the conifold", 
{\sl \jhep} {\bf 0411 }(2004) 035, {\rm hep-th/0408210}.

\bibitem{Kuper} S. Kuperstein, ``Meson spectroscopy from holomorphic probes
on the warped deformed conifold", 
{\sl \jhep} {\bf 0503 }(2005) 014, {\rm hep-th/0411097}.

\bibitem{Sakai}T. Sakai and S. Sugimoto, ``Low energy hadron physics in
holographic QCD", {\sl \ptp} {\bf 113 }(2005) 843, {\rm hep-th/0412141}.

\bibitem{Sakai1}
``More on a holographic dual of QCD", 
{\sl \ptp} {\bf 114 }(2006) 1083, {\rm hep-th/0507073};

\bibitem{APR}
  D.~Arean, A.~Paredes and A.~V.~Ramallo,
  ``Adding flavor to the gravity dual of non-commutative gauge theories,''
{\sl \jhep} {\bf 0508} (2005) 017, {\rm hep-th/0505181}.

\bibitem{AR}D.~Arean and A.~V.~Ramallo,
``Open string modes at brane intersections",
{\sl \jhep} {\bf 0604} (2006) 037, {\rm hep-th/0602174}.

\bibitem{MT} R.C.Myers, R.M.Thompson, 
"Holographic mesons in various dimensions", {\sl\jhep}{\bf 0609} (2006) 066,
{\rm hep-th/0605017}.

\bibitem{Apreda:2006bu}
 R.~Apreda, J.~Erdmenger, D.~Lust and C.~Sieg,
``Adding flavour to the Polchinski-Strassler background", 
{\sl \jhep} {\bf 0701}, 079 (2007), 
{\rm  hep-th/0610276}.

\bibitem{R} A. V. Ramallo,
``Adding open string modes to the gauge/gravity correspondence",
{\sl \mpl} {\bf A21 }(2006) 1, {\rm hep-th/0605261}.

\bibitem{unquenched}
 R.~Casero, C.~Nunez and A.~Paredes,
  `` Towards the string dual of N = 1 SQCD-like theories,''
  Phys.\ Rev.\ D {\bf 73}, 086005 (2006),
{\rm hep-th/0602027}.

\bibitem{unquenched2}
  A.~Paredes,
  `` On unquenched N = 2 holographic flavor,''
{\sl \jhep} {\bf 0612 }(2006) 032, {\rm hep-th/0610270}.

\bibitem{unquenched3}
R.~Casero and A.~Paredes,
``A note on the string dual of N = 1 SQCD-like theories,''
{\rm hep-th/0701059}.

\bibitem{unquenched4}
 F. Benini, F. Canoura, S. Cremonesi, C. N\'u\~nez and A. V. Ramallo,
``Unquenched flavors in the Klebanov-Witten model", 
{\sl \jhep} {\bf 0702 }(2007) 090, {\rm hep-th/0612118}.

\bibitem{rf60}
  F.~Benini, ``A chiral cascade via backreacting D7-branes with flux,''
  arXiv:0710.0374 [hep-th].

\bibitem{rf61}
 R.~Casero, C.~Nunez and A.~Paredes, ``Elaborations on the String Dual to N=1 SQCD,''
  arXiv:0709.3421 [hep-th].

\bibitem{unquenched5}
 F.~Benini, F.~Canoura, S.~Cremonesi, C.~Nunez and A.~V.~Ramallo,
 ``Backreacting Flavors in the Klebanov-Strassler Background,''
  arXiv:0706.1238 [hep-th].


\bibitem{EGG} J.Erdmenger, J. Grosse, Z. Guralnick, 
"Spectral flow on the Higgs branch and $AdS/CFT$ duality", 
{\sl \jhep} {\bf 0506 }(2005) 052, {\rm hep-th/0502224}.

\bibitem{Guralnik:2004ve}
  Z.~Guralnik, S.~Kovacs and B.~Kulik,
``Holography and the Higgs branch of N = 2 SYM theories", 
{\sl \jhep} {\bf 0503}, 063 (2005),  {\rm hep-th/0405127}.

\bibitem{ARR2}
  D.~Arean, A.~V.~Ramallo and D.~Rodriguez-Gomez,
  ``Holographic flavor on the Higgs branch,''
  {\sl \jhep} {\bf 0705} (2007) 044
  {\rm arXiv:hep-th/0703094}.

\bibitem{kutasov}
  A.~Giveon and D.~Kutasov, ``Brane dynamics and gauge theory,''
Rev.\ Mod.\ Phys.\  {\bf 71}, 983 (1999)
{\rm hep-th/9802067}.

\bibitem{Aharony}
  O.~Aharony,
``A note on the holographic interpretation of string theory backgrounds  with varying flux,''
  {\sl \jhep} {\bf 0103} (2001) 012
  {\rm arXiv:hep-th/0101013}.

\bibitem{M}R. C. Myers, ``Dielectric branes",
{\sl \jhep} {\bf 9912 }(1999) 022, {\rm hep-th/9910053}

\bibitem{WFO}O. DeWolfe, D. Z. Freedman and H. Ooguri, 
``Holography and defect conformal field theories", 
{\sl \pr} {\bf D66} (2002) 025009, {\rm hep-th/0111135}.

\bibitem{EGK}J. Erdmenger, Z. Guralnik and I. Kirsch,
``Four-dimensional superconformal theories with interacting boundaries
 or defects",
{\sl \pr} {\bf D66} (2002) 025020, {\rm hep-th/0203020}.

\bibitem{ARR} D.Arean, A.V.Ramallo, D.Rodriguez-Gomez, 
"Mesons and Higgs branch in defect theories", 
{\sl\pl} {\bf B641} (2006) 393, {\rm hep-th/0609010}.

\bibitem{ST} K.Skenderis, M.Taylor, 
"Branes in $AdS$ and $PP$-wave spacetimes",
 {\sl\jhep} {\bf 0206} (2002) 025, {\rm hep-th/0204054}.

\bibitem{CEGK} N.Constable, Z.Guralnik, J.Erdmenger, I.Kirch, 
"Intersecting D3-branes and holography",  {\sl\pr} {\bf D68} (2003) 106007,
{\rm hep-th/0211222}.

\bibitem{Erdmenger:2003kn}
J.~Erdmenger, Z.~Guralnik, R.~Helling and I.~Kirsch,
``A world-volume perspective on the recombination of intersecting
   branes,''
{\sl \jhep} {\bf 0404 }(2004) 064, {\rm hep-th/0309043}.

\bibitem{Kirsch:2004km}
  I. Kirsch,
  ``Generalizations of the AdS/CFT correspondence,''
  Fortsch.\ Phys.\  {\bf 52} (2004) 727,
{\rm hep-th/0406274}.

\bibitem{Igor}
  I.~R.~Klebanov and E.~Witten,
  ``AdS/CFT correspondence and symmetry breaking,''
  Nucl.\ Phys.\  B {\bf 556} (1999) 89
  {\rm arXiv:hep-th/9905104}.

\bibitem{crosssection}
 I.~R.~Klebanov,
 ``World-volume approach to absorption by non-dilatonic branes,''
  Nucl.\ Phys.\  B {\bf 496} (1997) 231
  {\rm arXiv:hep-th/9702076}.
  
 \bibitem{crosssection2}
   S.~S.~Gubser, I.~R.~Klebanov and A.~A.~Tseytlin,
  ``String theory and classical absorption by three-branes,''
  Nucl.\ Phys.\  B {\bf 499} (1997) 217
  {\rm arXiv:hep-th/9703040}.
  
  \bibitem{crosssection3}
    S.~S.~Gubser and I.~R.~Klebanov,
  ``Absorption by branes and Schwinger terms in the world volume theory,''
  Phys.\ Lett.\  B {\bf 413} (1997) 41
  {\rm arXiv:hep-th/970800}\\

\bibitem{DF}
  E.~D'Hoker and D.~Z.~Freedman,
 ``Supersymmetric gauge theories and the AdS/CFT correspondence,''
 {\rm arXiv:hep-th/0201253}.

\bibitem{holren}
  K.~Skenderis,
  ``Lecture notes on holographic renormalization,''
  Class.\ Quant.\ Grav.\  {\bf 19} (2002) 5849
  {\rm arXiv:hep-th/0209067}.

\bibitem{intersections}
K.~Behrndt, E.~Bergshoeff and B.~Janssen,
``Intersecting $D$--Branes in Ten and Six Dimensions,''
 {\sl \pr} {\bf D55} (1997) 
 {\rm arXiv:hep-th/9604168}.
 
 \bibitem{intersections2}
   E.~Bergshoeff, M.~de Roo, E.~Eyras, B.~Janssen and J.~P.~van der Schaar,
  ``Multiple intersections of D-branes and M-branes,''
 {\sl  Nucl.\ Phys.}\ B {\bf 494} (1997) 119
 {\rm arXiv:hep-th/9612095}.

\bibitem{MRD1} M.R.Douglas, "Branes within branes", {\rm hep-th/9512077}.

\bibitem{MRD2} M.R.Douglas, "Gauge Fields and D-branes", {\sl\jgp}{\bf 28}, 
255 (1998), {\rm hep-th/9604198}.

\bibitem{W} E.Witten, "Sigma models and the $ADHM$ construction of instantons",
 {\sl\jgp}{\bf 15}, 215 (1995), {\rm hep-th/9410052}.

\bibitem{Tong} D. Tong, ``TASI lectures on solitons: instantons,
monopoles, vortices and kinks", {\rm hep-th/0509216}.

\bibitem{Tong2}
  N.~Dorey, T.~J.~Hollowood, V.~V.~Khoze and M.~P.~Mattis,
 ``The calculus of many instantons,''
  Phys.\ Rept.\  {\bf 371} (2002) 231
 {\rm arXiv:hep-th/0206063}.\\ 
 
 \bibitem{ND}
   M.~R.~Douglas and N.~A.~Nekrasov,
  ``Noncommutative field theory,''
  Rev.\ Mod.\ Phys.\  {\bf 73} (2001) 977
  {\rm arXiv:hep-th/0106048}.

\bibitem{SW}
N. Seiberg and E. Witten, ``String theory and non-commutative geometry",
{\sl \jhep} {\bf 9909} (1999) 032,  {\rm hep-th/9908142}.

\bibitem{gravdielec}
D.Rodriguez-Gomez, {\sl jhep} vol{{\bf 0601}}, 
079 (2006), hep-th/0509228.

\bibitem{gravdielec2}
  B.~Janssen, Y.~Lozano and D.~Rodriguez-Gomez, ``The baryon vertex with magnetic flux,''
  {\sl jhep} {\bf 0611} (2006) 082, arXiv:hep-th/0606264.

\bibitem{HW}
  A.~Hanany and E.~Witten,
  ``Type IIB superstrings, BPS monopoles, and three-dimensional gauge dynamics,''
  Nucl.\ Phys.\  B {\bf 492} (1997) 152
  {\rm arXiv:hep-th/9611230}.

\bibitem{Flux} C. Bachas, M. Douglas and C. Schweigert,
``Flux stabilization of D-branes", 
{\sl \jhep} {\bf 0005 }(2000) 048, {\rm hep-th/0003037}. 

\bibitem{Neq}
W.Nahm in
 N.~S.~Craigie, P.~Goddard and W.~Nahm,
 ``Monopoles In Quantum Field Theory. Proceedings, Monopole Meeting, Trieste,
 Italy, December 11-15, 1981,''

\bibitem{CM}
  C.~G.~.~Callan and J.~M.~Maldacena,
  ``Brane dynamics from the Born-Infeld action,''
  Nucl.\ Phys.\  B {\bf 513} (1998) 198
  {\rm arXiv:hep-th/9708147}.

\bibitem{PST} P.Pasti, D.Sorokin, M.Tonin, "Covariant action for a "d=11" 
five-brane with the chiral field", {\sl \pl} {\bf B398} (1997) 41, {\rm
hep-th/9701037}.

\bibitem{PST2}
I. Bandos, K. Lechner, A. Nurmagambetov, P.
Pasti, D. Sorokin and M. Tonin, 
{\sl \prl} {\bf 78} (1997) 4332, {\rm hep-th/9701149}.

\bibitem{PST3}
D. Sorokin, ``On some features of the M5-brane", {\rm hep-th/9807050}.

\bibitem{Camino:2001at}
  J.~M.~Camino, A.~Paredes and A.~V.~Ramallo,
  ``Stable wrapped branes,''
{\sl \jhep} {\bf 0105 }(2001) 011, {\rm hep-th/0104082}. 

\bibitem{SEdefects}
  F.~Canoura, J.~D.~Edelstein, L.~A.~P.~Zayas, A.~V.~Ramallo and D.~Vaman,
``Supersymmetric branes on AdS(5) x Y**(p,q) and their field theory
duals,''
{\sl \jhep} {\bf 0603 }(2006) 101, {\rm hep-th/0512087}.

\bibitem{SEdefects2}
F.~Canoura, J.~D.~Edelstein and A.~V.~Ramallo,
  ``D-brane probes on L(a,b,c) superconformal field theories,''
  {\sl \jhep} {\bf 0609} (2006) 038
  {\rm arXiv:hep-th/0605260}.

\bibitem{MN}
 J.~M.~Maldacena and C.~Nunez,
  ``Towards the large N limit of pure N = 1 super Yang Mills,''
  Phys.\ Rev.\ Lett.\  {\bf 86} (2001) 588
  {\rm arXiv:hep-th/0008001}.
  
  \bibitem{MN2}
  A.~H.~Chamseddine and M.~S.~Volkov,
  ``Non-Abelian BPS monopoles in N = 4 gauged supergravity,''
  Phys.\ Rev.\ Lett.\  {\bf 79} (1997) 3343
  {\rm arXiv:hep-th/9707176}.

\bibitem{dMN}
  D.~Arean, A.~Paredes and A.~V.~Ramallo,
  ``Adding flavor to the gravity dual of non-commutative gauge theories,''
  {\sl \jhep} {\bf 0508} (2005) 017
  {\rm arXiv:hep-th/0505181}.

  \bibitem{rf107}
D. Mateos, String theory and quantum chromodynamics, arXiv:0709.1523.

\bibitem{MMT}
 D.~Mateos, R.~C.~Myers and R.~M.~Thomson,
 ``Holographic phase transitions with fundamental matter,''
  Phys.\ Rev.\ Lett.\  {\bf 97} (2006) 091601
  {\rm arXiv:hep-th/0605046}.
  
  \bibitem{CJ1}
  T.~Albash, V.~G.~Filev, C.~V.~Johnson and A.~Kundu,
  ``A topology-changing phase transition and the dynamics of flavour,''
  arXiv:hep-th/0605088

\bibitem{CJ2}
  V.~G.~Filev, C.~V.~Johnson, R.~C.~Rashkov and K.~S.~Viswanathan,
  ``Flavoured large N gauge theory in an external magnetic field,''
  JHEP {\bf 0710} (2007) 019
  [arXiv:hep-th/0701001].
  
  \bibitem{CJ3}
  V.~G.~Filev,
  ``Criticality, Scaling and Chiral Symmetry Breaking in External Magnetic
  Field,''
  arXiv:0706.3811 [hep-th].
  
  \bibitem{CJ4}
  T.~Albash, V.~G.~Filev, C.~V.~Johnson and A.~Kundu,
  ``Finite Temperature Large N Gauge Theory with Quarks in an External Magnetic
  Field,''
  arXiv:0709.1547 [hep-th].
  
  \bibitem{CJ5}
  T.~Albash, V.~G.~Filev, C.~V.~Johnson and A.~Kundu,
  ``Quarks in an External Electric Field in Finite Temperature Large N Gauge
  Theory,''
  arXiv:0709.1554 [hep-th].
  
  
  
 \bibitem{finiteT}
  A.~O.~Starinets,
  ``Quasinormal modes of near extremal black branes,''
  Phys.\ Rev.\  D {\bf 66} (2002) 124013
  {\rm arXiv:hep-th/0207133}.
  
  \bibitem{finiteT2}
  P.~Kovtun, D.~T.~Son and A.~O.~Starinets,
  ``Viscosity in strongly interacting quantum field theories from black hole physics,''
  Phys.\ Rev.\ Lett.\  {\bf 94} (2005) 111601
  {\rm arXiv:hep-th/0405231}.
  
  \bibitem{finiteT3}
  C.~P.~Herzog, A.~Karch, P.~Kovtun, C.~Kozcaz and L.~G.~Yaffe,
  ``Energy loss of a heavy quark moving through N = 4 supersymmetric
  Yang-Mills plasma,''
  {\sl \jhep} {\bf 0607} (2006) 013
  {\rm arXiv:hep-th/0605158}.
  
  \bibitem{finiteT4}
  S.~S.~Gubser,
 ``Drag force in AdS/CFT,''
  Phys.\ Rev.\  D {\bf 74} (2006) 126005
  {\rm arXiv:hep-th/0605182}.
  
  \bibitem{finiteT5}
     H.~Liu, K.~Rajagopal and U.~A.~Wiedemann,
  ``Calculating the jet quenching parameter from AdS/CFT,''
  Phys.\ Rev.\ Lett.\  {\bf 97} (2006) 182301
  {\rm arXiv:hep-ph/0605178}.

\bibitem{MInahan}
J. A. Minahan, ``Glueball mass spectra and other issues for 
supergravity duals of QCD models",
{\sl \jhep} {\bf 9901 }(1999) 020, {\rm hep-th/9811156}.


\bibitem{RS}J. G. Russo and K. Sfetsos,
``Rotating D3-branes and QCD in three dimensions",
{\sl \atmp} {\bf 3}(1999) 131, {\rm hep-th/9901056}.

\bibitem{bound}
  E.~Witten,
  ``Bound states of strings and p-branes,''
  Nucl.\ Phys.\  B {\bf 460} (1996) 335
  {\rm arXiv:hep-th/9510135}.
  
\bibitem{giants}
See  B.~Janssen, Y.~Lozano and D.~Rodriguez-Gomez,
 ``Giant gravitons as fuzzy manifolds,''
  {\rm arXiv:hep-th/0412037} and references therein.  


\end{thebibliography}
\end{document}